\documentclass[12pt,english]{article}
\usepackage{times}
\usepackage[T1]{fontenc}
\usepackage{geometry}
\usepackage{rotating}
\usepackage{graphicx}
\usepackage{setspace}
\usepackage{amssymb}

\makeatletter

\providecommand{\tabularnewline}{\\}

\usepackage{bm}

\usepackage{babel}
\makeatother
\begin{document}

\section*{A Physics-Driven AI \emph{}Approach for Microwave Imaging of Breast
Tumors}

\noindent ~

\noindent \vfill

\noindent F. Zardi,$^{^{\left(1\right)\left(2\right)}}$ \emph{Member,
IEEE}, L. Tosi,$^{^{\left(1\right)\left(2\right)}}$ \emph{Member,
IEEE}, M. Salucci,$^{^{\left(1\right)\left(2\right)}}$ \emph{Senior
Member}, \emph{IEEE}, and A. Massa,$^{^{\left(1\right)\left(2\right)\left(3\right)\left(4\right)\left(5\right)}}$
\emph{Fellow, IEEE}

\noindent \vfill

\noindent {\footnotesize ~}{\footnotesize \par}

\noindent {\footnotesize $^{(1)}$} \emph{\footnotesize ELEDIA Research
Center} {\footnotesize (}\emph{\footnotesize ELEDIA}{\footnotesize @}\emph{\footnotesize UniTN}
{\footnotesize - University of Trento)}{\footnotesize \par}

\noindent {\footnotesize DICAM - Department of Civil, Environmental,
and Mechanical Engineering}{\footnotesize \par}

\noindent {\footnotesize Via Mesiano 77, 38123 Trento - Italy}{\footnotesize \par}

\noindent \textit{\emph{\footnotesize E-mail:}} {\footnotesize \{}\emph{\footnotesize francesco.zardi,
luca.tosi, marco.salucci, andrea.massa}{\footnotesize \}@}\emph{\footnotesize unitn.it}{\footnotesize \par}

\noindent {\footnotesize Website:} \emph{\footnotesize www.eledia.org/eledia-unitn}{\footnotesize \par}

\noindent {\footnotesize ~}{\footnotesize \par}

\noindent {\footnotesize $^{(2)}$} \emph{\footnotesize CNIT - \char`\"{}University
of Trento\char`\"{} ELEDIA Research Unit }{\footnotesize \par}

\noindent {\footnotesize Via Sommarive 9, 38123 Trento - Italy}{\footnotesize \par}

\noindent {\footnotesize Website:} \emph{\footnotesize www.eledia.org/eledia-unitn}{\footnotesize \par}

\noindent {\small ~}{\small \par}

\noindent {\footnotesize $^{(3)}$} \emph{\footnotesize ELEDIA Research
Center} {\footnotesize (}\emph{\footnotesize ELEDIA}{\footnotesize @}\emph{\footnotesize UESTC}
{\footnotesize - UESTC)}{\footnotesize \par}

\noindent {\footnotesize School of Electronic Science and Engineering,
Chengdu 611731 - China}{\footnotesize \par}

\noindent \textit{\emph{\footnotesize E-mail:}} \emph{\footnotesize andrea.massa@uestc.edu.cn}{\footnotesize \par}

\noindent {\footnotesize Website:} \emph{\footnotesize www.eledia.org/eledia}{\footnotesize -}\emph{\footnotesize uestc}{\footnotesize \par}

\noindent {\footnotesize ~}{\footnotesize \par}

\noindent {\footnotesize $^{(4)}$} \emph{\footnotesize ELEDIA Research
Center} {\footnotesize (}\emph{\footnotesize ELEDIA@TSINGHUA} {\footnotesize -
Tsinghua University)}{\footnotesize \par}

\noindent {\footnotesize 30 Shuangqing Rd, 100084 Haidian, Beijing
- China}{\footnotesize \par}

\noindent {\footnotesize E-mail:} \emph{\footnotesize andrea.massa@tsinghua.edu.cn}{\footnotesize \par}

\noindent {\footnotesize Website:} \emph{\footnotesize www.eledia.org/eledia-tsinghua}{\footnotesize \par}

\noindent {\small ~}{\small \par}

\noindent {\small $^{(5)}$} {\footnotesize School of Electrical Engineering}{\footnotesize \par}

\noindent {\footnotesize Tel Aviv University, Tel Aviv 69978 - Israel}{\footnotesize \par}

\noindent \textit{\emph{\footnotesize E-mail:}} \emph{\footnotesize andrea.massa@eng.tau.ac.il}{\footnotesize \par}

\noindent {\footnotesize Website:} \emph{\footnotesize https://engineering.tau.ac.il/}{\footnotesize \par}

\noindent \vfill

\noindent \textbf{\emph{This work has been submitted to the IEEE for
possible publication. Copyright may be transferred without notice,
after which this version may no longer be accessible.}}

\noindent \vfill

\newpage
\section*{A Physics-Driven AI \emph{}Approach for Microwave Imaging of Breast
Tumors}

~

\noindent ~

\noindent ~

\begin{flushleft}F. Zardi, L. Tosi, M. Salucci, and A. Massa\end{flushleft}

\noindent \vfill

\begin{abstract}
\noindent In this paper, an innovative microwave imaging (\emph{MI})
approach for breast tumor diagnosis is proposed that employs a differential
formulation of the inverse scattering problem (\emph{ISP}) at hand
to exploit arbitrary-fidelity \emph{priors} on the inhomogeneous reference/healthy
tissues. The quantitative imaging of the unknown tumor is then rephrased
into a global optimization problem, which is efficiently solved with
an \emph{ad-hoc} physics-driven artificial intelligence (\emph{AI})
strategy inspired by the concepts and guidelines of the System-by-Design
(\emph{SbD}) paradigm. The effectiveness, the robustness, the reliability,
and the efficiency of the proposed method are assessed against both
synthetic and experimental data.
\end{abstract}
\noindent \vfill

\noindent \textbf{Key words}: Inverse Scattering Problem (\emph{ISP}),
Microwave Imaging (\emph{MI}), Breast Imaging, Artificial Intelligence
(\emph{AI}), System-by-Design (\emph{SbD}), Learning-by-Examples (\emph{LBE}),
Evolutionary Algorithms (\emph{EA}s).

\newpage
\section{Introduction }

\noindent As claimed by the World Health Organization, breast tumor
is nowadays the most diffused cancer in the world with over 2 million
cases reported during the year 2020 \cite{WHO 2022}. Recent studies
have shown that early diagnosis through mammography screening leads
to a substantial decrease in mortality \cite{Tabar 2019}. However,
the screening systems currently-in-use, which are based on magnetic
resonance imaging (\emph{MRI}) and X-ray technology, present several
drawbacks such as the need for complex and expensive equipment with
limited availability that hinders regular mass screening programs.
In addition, the use of either injected contrast liquids or ionizing
radiation raises non-negligible concerns on the potential health implications
of a regular/repeated examination with these modalities \cite{Nikolova 2011}. 

\noindent Alternatively, microwave imaging (\emph{MI}) is an emerging
and promising solution to overcome such limitations \cite{Mirbeik-Sabzevari 2019}.
In \emph{MI}, the breast is illuminated with non-ionizing electromagnetic
(\emph{EM}) waves and the arising scattering response is processed
to determine qualitative (e.g., the presence, the position, and the
shape) and/or quantitative (e.g., the permittivity and the conductivity
distributions) information on the diagnosed tissues. From a practical
point of view, the equipment for collecting the microwave data is
relatively inexpensive and it can be designed with compact (i.e.,
portable) form factors \cite{Bassi 2013} or even embedded in wearable
devices \cite{Porter 2016 b}. Moreover, current \emph{MI} systems
support quite expedite data collection processes, in the range of
a few seconds \cite{O'Loughlin 2018}. On the other hand, detecting
and retrieving the features of abnormal (tumoral) tissues from \emph{EM}
field measurements requires the solution of an inverse scattering
problem (\emph{ISP}) and, therefore, to effectively/efficiently cope
with several paramount challenges \cite{Chen 2018}.

\noindent As a matter of fact, \emph{ISPs} are generally ill-posed,
which means that (\emph{i}) more than one solution exists since different
dielectric distributions could match the same scattering data and
that (\emph{ii}) the \emph{MI} solution does not continuously depend
on these latters \cite{Chen 2018}. Such an issue can be mitigated
by including some \emph{a-priori} information on the imaged scenario,
on its underlying physics, and on the type and nature of the unknowns
to integrate the information content of the scattering data. For instance,
a differential formulation of the \emph{ISP} is an effective regularization
recipe since the unknown is no more the complete distribution of the
investigation domain, but only the difference with respect to a known
dielectric profile \cite{Caorsi 2004}\cite{Xu 2018b}. As for the
\emph{MI} of the breast, such a reference scenario corresponds to
the normal/healthy tissues, which knowledge can be gathered from previous
analyses performed with microwave \cite{Kurrant 2013}-\cite{Smith 2022},
\emph{MRI} \cite{Gubern-Merida 2015}\cite{Golnabi 2019}, or ultrasound
\cite{Abdollahi 2019}-\cite{Qin 2020} investigations.

\noindent Another fundamental challenge of \emph{MI} comes from the
non-linear relationship between the \emph{ISP} unknowns and the scattered
data \cite{Chen 2018}. Such an issue can be overcome by adopting
linear formulations aimed at retrieving a map of the reflected energy
from the breast, rather than its permittivity and conductivity distributions,
by means of radar-based strategies \cite{Lim 2008}-\cite{Casu 2017}.
However, the result is a qualitative imaging of the breast that often
does not provide sufficient information to discriminate between benign
and malign abnormalities \cite{Poplack 2007} and to monitor the follow-up
of an ongoing medical treatment \cite{Meaney 2013}. Similar limitations
may occur when counteracting the non-linearity by means of linearized
formulations of the scattering equations (e.g., Born's and/or Rytov's
approximations \cite{Miao 2017}-\cite{Gao.2015}).

\noindent Otherwise, fully non-linear quantitative solutions of the
\emph{ISP} can be found by means of properly customized deterministic
(\emph{DA}s) or stochastic (\emph{SA}s) algorithms. The former class
includes local search techniques such as the Conjugate Gradient (\emph{CG})
method \cite{Hosseinzadegan 2021}\cite{Liu 2002}. They are usually
very fast, but they may be trapped into local minima unless the starting
solution already belongs to the {}``attraction basin'' of the global
optimum. Conversely, \emph{SA}s (e.g., Evolutionary Algorithms (\emph{EA})
\cite{Rocca 2009}-\cite{Goudos 2021}) perform a global search and
they are more resilient to the initialization problem thanks to their
hill-climbing properties. Unfortunately, the repeated full-wave evaluation
of the forward scattering operators to assess the fitness of the trial
guess solutions \cite{Rocca 2009} could result in a computation burden
not compatible with the goal of a fast medical screening.

\noindent Recently, artificial intelligence (\emph{AI}) has emerged
as a powerful framework for solving \emph{ISP}s with unprecedented
computational efficiency leveraging the Learning-by-Examples (\emph{LBE})
and Deep Learning (\emph{DL}) paradigms \cite{Salucci 2022b}-\cite{Zhou 2021}.
Such techniques exploit a training set of properly-chosen examples
of the input-output relation between unknowns and \emph{ISP} data
to build a surrogate model (\emph{SM}) significantly faster than a
traditional full-wave solver in predicting the output/data \cite{Salucci 2022b}.
Clearly, the success of such strategies depends on the ability to
generate a minimum-cardinality set of representative training examples,
sampling a minimum-dimensionality solution space to cope with the
{}``curse-of-dimensionality'' \cite{Salucci 2022b}\cite{Friedman 2001},
thus minimizing the computational effort of the off-line training
phase. 

\noindent In such a context, this paper presents a novel physics-driven
\emph{AI} method for reliably solving fully non-linear microwave breast
\emph{ISP}s with a high computational efficiency. Thanks to a differential
formulation of the underlying scattering phenomena, which enables
the exploitation of arbitrary-fidelity \emph{priors} on the inhomogeneous
reference/healthy tissues, a minimum-dimensionality search space is
defined and adaptively sampled by means of an \emph{EA}-based global
search engine that leverages a \emph{SM} of the cost function to speed-up
the inversion, while keeping the same prediction accuracy. The arising
inversion strategy, inspired by the concepts and guidelines of the
System-by-Design (\emph{SbD}) framework \cite{Massa 2022}, is validated
on both synthetic data based on \emph{MRI}-derived numerical phantoms
\cite{Burfeindt 2012} and experimental data \cite{Solis-Nepote 2019}. 

\noindent The outline of the work is the following. The formulation
of the differential breast \emph{ISP} is detailed in Sect. \ref{sec:mathematical-formulation},
whereas the \emph{AI}-based \emph{MI} approach is described in Sect.
\ref{sec:solution_approach}. An extensive validation considering
different operative conditions and benchmark problems of increasing
complexity is reported in Sect. \ref{sec:results}. Finally, some
conclusions and final remarks are drawn (Sect. \ref{sec:Conclusions}).

\section{\noindent Mathematical Formulation of the Breast \emph{ISP}\label{sec:mathematical-formulation}}

\noindent Let us consider a \emph{2-D} square investigation domain
$\Omega$ of side $L$ embedding a coronal slice $\Lambda$ parallel
to the chest wall of the patient's breast {[}Fig. 1(\emph{a}){]}.
The normal (healthy) tissue within $\Lambda$ is modeled by means
of the complex permittivity distribution $\widetilde{\varepsilon}_{\mathcal{N}}\left(\mathbf{r}\right)$
($\widetilde{\varepsilon}_{\mathcal{N}}\left(\mathbf{r}\right)\triangleq\varepsilon_{0}\varepsilon_{\mathcal{N}}\left(\mathbf{r}\right)-j\frac{\sigma_{\mathcal{N}}\left(\mathbf{r}\right)}{2\pi f}$),
$\varepsilon_{\mathcal{N}}\left(\mathbf{r}\right)$ and $\sigma_{\mathcal{N}}\left(\mathbf{r}\right)$
being the relative permittivity and the conductivity at the position
$\mathbf{r}=\left(x,\, y\right)$, respectively. Moreover, $\varepsilon_{0}$
is the vacuum permittivity and $f$ is the working frequency %
\footnote{\noindent To simplify the notation, the time dependence $\exp\left(j2\pi ft\right)$
is omitted hereinafter.%
}. Under the hypothesis that $\Lambda$ includes a tumor, which is
modeled as a region with bounded support $\Psi\subset\Lambda$ of
abnormal permittivity $\widetilde{\varepsilon}_{\mathcal{A}}\left(\mathbf{r}\right)$
{[}$\widetilde{\varepsilon}_{\mathcal{A}}\left(\mathbf{r}\right)\triangleq\varepsilon_{0}\varepsilon_{\mathcal{A}}\left(\mathbf{r}\right)-j\frac{\sigma_{\mathcal{A}}\left(\mathbf{r}\right)}{2\pi f}${]},
the dielectric profile of the region $\Lambda$ turns out to be

\noindent \begin{equation}
\widetilde{\varepsilon}\left(\mathbf{r}\right)=\left\{ \begin{array}{ll}
\widetilde{\varepsilon}_{\mathcal{A}}\left(\mathbf{r}\right) & \textrm{if }\mathbf{r}\in\Psi\\
\widetilde{\varepsilon}_{\mathcal{N}}\left(\mathbf{r}\right) & \textrm{if }\mathbf{r}\in\left(\Lambda-\Psi\right)\\
\widetilde{\varepsilon}_{\mathcal{B}} & \mathrm{otherwise,}\end{array}\right.\label{eq:complex-permittivity}\end{equation}
$\widetilde{\varepsilon}_{\mathcal{B}}$ being the permittivity of
the homogeneous background medium surrounding the breast {[}$\widetilde{\varepsilon}_{\mathcal{B}}\triangleq\varepsilon_{0}\varepsilon_{\mathcal{B}}-j\frac{\sigma_{\mathcal{B}}}{2\pi f}$
- Fig. 1(\emph{a}){]}. 

\noindent The acquisition system is a transverse magnetic (\emph{TM})
multi-view/multi-static setup composed by a set of $V$ antennas located
at the positions $\left\{ \mathbf{r}^{v}\in\Theta;\, v=1,...,V\right\} $
in an external circular observation domain $\Theta$ of radius $\rho_{\Theta}$
{[}$\Theta\cap\Omega=\emptyset$ - Fig. 1(\emph{a}){]}. Such antennas
are alternatively operated in transmit or receive mode so that, when
the $v$-th ($v=1,...,V$) one is transmitting, the remaining $M=\left(V-1\right)$
ones collect the electric field in $M$ different positions $\left\{ \mathbf{r}_{m}^{v}\in\Theta;\, m=1,...,M\right\} $.
In the following, let us denote with $E_{inc}^{v}\left(\mathbf{r}\right)$
the incident field associated to the $v$-th ($v=1,...,V$) probing
source radiating in free-space, while $E^{v}\left(\mathbf{r}\right)$
stands for the total field due to the \emph{EM} interactions between
the $v$-th ($v=1,...,V$) source and the inhomogeneous distribution
(\ref{eq:complex-permittivity}). These latters can be modeled by
means of the \emph{State},\begin{equation}
E_{inc}^{v}\left(\mathbf{r}\right)=E^{v}\left(\mathbf{r}\right)-\int_{\Omega}G_{\mathcal{B}}\left(\mathbf{r},\,\mathbf{r}^{\prime}\right)\tau\left(\mathbf{r}^{\prime}\right)E^{v}\left(\mathbf{r}^{\prime}\right)d\mathbf{r}^{\prime};\,\,\,\,\,\mathbf{r}\in\Omega;\, v=1,...,V,\label{eq:LSIE state total}\end{equation}
and the \emph{Data},\begin{equation}
S^{v}\left(\mathbf{r}_{m}^{v}\right)=\int_{\Omega}G_{\mathcal{B}}\left(\mathbf{r}_{m}^{v},\,\mathbf{r}^{\prime}\right)\tau\left(\mathbf{r}^{\prime}\right)E^{v}\left(\mathbf{r}^{\prime}\right)d\mathbf{r}^{\prime};\,\,\,\,\,\mathbf{r}_{m}^{v}\in\Theta;\, v=1,...,V;\, m=1,...,M,\label{eq:LSIE data total}\end{equation}
equations, where $\tau\left(\mathbf{r}\right)$ {[}$\tau\left(\mathbf{r}\right)\triangleq\frac{\widetilde{\varepsilon}\left(\mathbf{r}\right)}{\widetilde{\varepsilon}_{\mathcal{B}}}-1${]}
is the actual contrast, $S^{v}\triangleq\left(E^{v}-E_{inc}^{v}\right)$
is the $v$-th ($v=1,...,V$) scattered field, and $G_{\mathcal{B}}$
is the Green's function of the homogeneous background. 

\noindent By observing that the field contribution, $E_{\mathcal{N}}^{v}\left(\mathbf{r}\right)$,
due to the normal tissue, $\tau_{\mathcal{N}}\left(\mathbf{r}\right)$
{[}$\tau_{\mathcal{N}}\left(\mathbf{r}\right)\triangleq\left[\widetilde{\varepsilon}_{\mathcal{N}}\left(\mathbf{r}\right)/\widetilde{\varepsilon}_{\mathcal{B}}\right]-1${]}
can be predicted starting from the available \emph{prior} knowledge
of a \emph{Reference} \emph{Scenario} not including the tumor {[}Fig.
1(\emph{b}){]}, it turns out the total, $E_{\mathcal{N}}^{v}$, and
the scattered, $S_{\mathcal{N}}^{v}\triangleq\left(E_{\mathcal{N}}^{v}-E_{inc}^{v}\right)$,
fields still comply with the Data and the State equations in the following
form\begin{equation}
E_{inc}^{v}\left(\mathbf{r}\right)=E_{\mathcal{N}}^{v}\left(\mathbf{r}\right)-\int_{\Omega}G_{\mathcal{B}}\left(\mathbf{r},\,\mathbf{r}^{\prime}\right)\tau_{\mathcal{N}}\left(\mathbf{r}^{\prime}\right)E_{\mathcal{N}}^{v}\left(\mathbf{r}^{\prime}\right)d\mathbf{r}^{\prime};\,\,\,\,\,\mathbf{r}\in\Omega;\, v=1,...,V\label{eq:LSIE state ref}\end{equation}
\begin{equation}
S_{\mathcal{N}}^{v}\left(\mathbf{r}_{m}^{v}\right)=\int_{\Omega}G_{\mathcal{B}}\left(\mathbf{r}_{m}^{v},\,\mathbf{r}^{\prime}\right)\tau_{\mathcal{N}}\left(\mathbf{r}^{\prime}\right)E_{\mathcal{N}}^{v}\left(\mathbf{r}^{\prime}\right)d\mathbf{r}^{\prime};\,\,\,\,\,\mathbf{r}_{m}^{v}\in\Theta;\, v=1,...,V;\, m=1,...,M.\label{eq:LSIE data ref}\end{equation}
 Accordingly, the contribution due to an unknown abnormal tissue can
be isolated and made explicit by defining a fictitious \emph{Differential}
\emph{Scenario} as the difference between the actual and the reference
ones {[}Fig. 1(\emph{b}){]}. In such a scenario, the $v$-th ($v=1,...,V$)
differential field, $D^{v}\triangleq\left(E^{v}-E_{\mathcal{N}}^{v}\right)=\left(S^{v}-S_{\mathcal{N}}^{v}\right)$,
is computed by subtracting (\ref{eq:LSIE state ref}) from (\ref{eq:LSIE state total})
to yield\begin{equation}
D^{v}\left(\mathbf{r}\right)=\int_{\Omega}G_{\mathcal{N}}\left(\mathbf{r},\,\mathbf{r}^{\prime}\right)\tau_{\Delta}\left(\mathbf{r}^{\prime}\right)E^{v}\left(\mathbf{r}^{\prime}\right)d\mathbf{r}^{\prime};\,\,\,\,\,\mathbf{r}\in\Omega;\, v=1,...,V\label{eq:LSIE state diff}\end{equation}
where $G_{\mathcal{N}}\left(\mathbf{r},\,\mathbf{r}^{\prime}\right)$
is the inhomogeneous Green's function \cite{Chen 2018} of the reference
distribution and $\tau_{\Delta}$ is the differential contrast $\tau_{\Delta}\left(\mathbf{r}\right)\triangleq\left[\tau\left(\mathbf{r}\right)-\tau_{\mathcal{N}}\left(\mathbf{r}\right)\right]=\frac{\widetilde{\varepsilon}_{\mathcal{A}}\left(\mathbf{r}\right)-\widetilde{\varepsilon}_{\mathcal{N}}\left(\mathbf{r}\right)}{\widetilde{\varepsilon}_{\mathcal{B}}}$,
$\mathbf{r}\in\Psi$ {[}$\tau_{\Delta}\left(\mathbf{r}\right)\neq0$
$\mathbf{r}\in\Psi${]}.

\noindent To numerically deal with the \emph{ISP} at hand, the domain
$\Omega$ is discretized into $N$ square sub-domains $\left\{ \Omega_{n}\in\Omega;\, n=1,...,N\right\} $
centered at $\left\{ \mathbf{r}_{n};\, n=1,...,N\right\} $ so that
the discretized form of equation (\ref{eq:LSIE state diff}) turns
out to be\begin{equation}
\underline{D}^{\Omega,v}=\underline{\underline{G}}_{\mathcal{N}}^{\Omega}\,\underline{J}_{\Delta}^{v};\,\,\,\,\, v=1,...,V\label{eq:discrete state diff}\end{equation}
where $\underline{D}^{\Omega,v}=\left\{ D^{v}\left(\mathbf{r}_{n}\right);\, n=1,...,N\right\} $,
the entries of the vector $\underline{J}_{\Delta}^{v}=\left\{ J_{\Delta}^{v}\left(\mathbf{r}_{n}\right);\, n=1,...,N\right\} $
are the samples of the $v$-th ($v=1,...,V$) differential equivalent
current {[}$J_{\Delta}^{v}\left(\mathbf{r}_{n}\right)\triangleq\tau_{\Delta}\left(\mathbf{r}_{n}\right)E^{v}\left(\mathbf{r}_{n}\right)${]},
and $\underline{\underline{G}}_{\mathcal{N}}^{\Omega}\in\mathbb{C}^{N\times N}$
is the internal inhomogeneous Green's matrix given by (see Appendix
A)\begin{equation}
\underline{\underline{G}}_{\mathcal{N}}^{\Omega}=\left[\underline{\underline{\mathbb{I}}}-\underline{\underline{G}}_{\mathcal{B}}^{\Omega}\,\underline{\underline{\tau}}_{\mathcal{N}}\right]^{-1}\underline{\underline{G}}_{\mathcal{B}}^{\Omega}.\label{int green inhomog}\end{equation}

\noindent In (\ref{int green inhomog}), $\underline{\underline{\mathbb{I}}}$
is the $\left(N\times N\right)$ identity matrix, $\underline{\underline{\tau}}_{\mathcal{N}}$
is a square diagonal matrix which diagonal entries fill the vector
$\underline{\tau}_{\mathcal{N}}=\left\{ \tau_{\mathcal{N}}\left(\mathbf{r}_{n}\right);\, n=1,...,N\right\} $,
and $\underline{\underline{G}}_{\mathcal{B}}^{\Omega}\in\mathbb{C}^{N\times N}$
is the internal Green's matrix of the homogeneous background, which
$\left(n,\, s\right)$-th ($n,\, s=1,...,N$) element is given by
$\left.\underline{\underline{G}}_{\mathcal{B}}^{\Omega}\right\rfloor _{n,s}\triangleq j\frac{\kappa_{\mathcal{B}}^{2}}{4}\int_{\Omega_{s}}\mathcal{H}_{0}^{\left(2\right)}\left(\kappa_{\mathcal{B}}\left\Vert \mathbf{r}_{n}-\mathbf{r}^{\prime}\right\Vert \right)d\mathbf{r}^{\prime}$,
$\kappa_{\mathcal{B}}$ being the wavenumber, while $\mathcal{H}_{0}^{\left(2\right)}$
is the zero-th order Hankel function of the second kind and $\left\Vert \cdot\right\Vert $
indicates the $\ell_{2}$-norm.

\noindent On the other hand, the discretized data equation for the
\emph{Differential Scenario} reads as (see Appendix A)\begin{equation}
\underline{D}^{\Theta,v}=\underline{\underline{G}}_{\mathcal{N}}^{\Theta,v}\,\underline{J}_{\Delta}^{v};\,\,\,\,\, v=1,...,V\label{eq:discrete data diff}\end{equation}
where the vector $\underline{D}^{\Theta,v}=\left\{ D^{v}\left(\mathbf{r}_{m}^{v}\right);\, m=1,...,M\right\} $
contains the samples of the $v$-th ($v=1,...,V$) differential field
computed in $\Theta$ and $\underline{\underline{G}}_{\mathcal{N}}^{\Theta,v}\in\mathbb{C}^{M\times N}$
is the $v$-th external inhomogeneous Green's matrix\begin{equation}
\underline{\underline{G}}_{\mathcal{N}}^{\Theta,v}=\underline{\underline{G}}_{\mathcal{B}}^{\Theta,v}\left[\underline{\underline{\mathbb{I}}}+\underline{\underline{\tau}}_{\mathcal{N}}\,\underline{\underline{G}}_{\mathcal{N}}^{\Omega}\right],\label{ext green inhomog}\end{equation}
 $\underline{\underline{G}}_{\mathcal{B}}^{\Theta,v}\in\mathbb{C}^{M\times N}$
being the $v$-th ($v=1,...,V$) external Green's matrix for the homogeneous
background which $\left(m,\, n\right)$-th ($m=1,...,M$, $n=1,...,N$)
element is given by $\left.\underline{\underline{G}}_{\mathcal{B}}^{\Theta,v}\right\rfloor _{m,n}\triangleq j\frac{\kappa_{\mathcal{B}}^{2}}{4}\int_{\Omega_{n}}\mathcal{H}_{0}^{\left(2\right)}\left(\kappa_{\mathcal{B}}\left\Vert \mathbf{r}_{m}^{v}-\mathbf{r}^{\prime}\right\Vert \right)d\mathbf{r}^{\prime}$.
It should be highlighted that, since $\underline{\underline{G}}_{\mathcal{N}}^{\Omega}$
and $\underline{\underline{G}}_{\mathcal{N}}^{\Theta,v}$ only depend
on $\Theta$, $\Omega$, $\underline{\tau}_{\mathcal{N}}$, and $\widetilde{\varepsilon}_{\mathcal{B}}$,
they need to be computed only once by exploiting the available \emph{a-priori}
information on the \emph{Reference Scenario}. 

\noindent According to the previous formulation, the \emph{MI} problem
at hand can be rephrased as follows:

\begin{quotation}
\noindent \textbf{Breast Tumor} \textbf{\emph{ISP}}---Starting from
the knowledge of the \emph{reference} contrast, $\underline{\tau}_{\mathcal{N}}$,
the incident field within $\Omega$, $\underline{E}_{inc}^{\Omega,v}=\left\{ E_{inc}^{v}\left(\mathbf{r}_{n}\right);\, n=1,...,N\right\} $
($v=1,...,V$), and the differential field samples in $\Theta$, $\underline{D}^{\Theta,v}$
($v=1,...,V$), retrieve the unknown contrast distribution in $\Lambda$,
$\underline{\tau}=\left\{ \tau\left(\mathbf{r}_{n}\right)=\left[\tau_{\mathcal{N}}\left(\mathbf{r}_{n}\right)+\tau_{\Delta}\left(\mathbf{r}_{n}\right)\right];\, n=1,...,N\right\} $
by determining the differential term $\underline{\tau}_{\Delta}=\left\{ \tau_{\Delta}\left(\mathbf{r}_{n}\right);\, n=1,...,N\right\} $.
\end{quotation}

\section{\noindent Physics-Driven \emph{AI} Solution Approach\label{sec:solution_approach}}

The {}``\emph{Breast Tumor ISP}{}`` is then reformulated into the
following global optimization task\begin{equation}
\underline{\alpha}^{opt}=\arg\left\{ \min_{\underline{\alpha}}\left[\Phi\left(\underline{\alpha}\right)\right]\right\} \label{eq:global-optimization-problem}\end{equation}
where $\underline{\alpha}=\left\{ \alpha_{k};\, k=1,...,K\right\} $
is a set of properly-defined descriptors of the solution $\tau_{\Delta}$
and $\Phi\left(\underline{\alpha}\right)$ is the differential data
mismatch cost function\begin{equation}
\Phi\left(\underline{\alpha}\right)=\frac{\sum_{v=1}^{V}\left\Vert \underline{D}^{\Theta,v}\left(\underline{\alpha}\right)-\underline{D}_{meas}^{\Theta,v}\right\Vert ^{2}}{\sum_{v=1}^{V}\left\Vert \underline{D}_{meas}^{\Theta,v}\right\Vert ^{2}}\label{eq:cost-function}\end{equation}
where $\underline{D}^{\Theta,v}\left(\underline{\alpha}\right)$ and
$\underline{D}_{meas}^{\Theta,v}$ are the differential field associated
to the solution $\tau_{\Delta}$ defined by $\underline{\alpha}$
and the one obtained from the measurements in $\Theta$, respectively. 

\noindent It is worth noticing that the efficient solution of (\ref{eq:global-optimization-problem})
poses one paramount challenge to tackle, which directly comes from
the high computational burden of an accurate full-wave evaluation
of (\ref{eq:cost-function}). As a matter of fact, while global optimization
solvers, such as those based on \emph{EA}s, effectively explore the
solution space without being trapped into false solutions (i.e., local
minima) \cite{Rocca 2009}, they also require the iterated ($i=1,...,I$,
$I$ being the number of iterations) evaluation of (\ref{eq:cost-function})
for a population of $P$ ($P\propto K$ \cite{Rocca 2009}) agents
(trial solutions). Therefore, it is mandatory to reduce the overall
computational cost of the solution process to yield a solution as
much as possible close to the actual one, $\underline{\alpha}^{act}$
within a reasonable (i.e., compatible with the clinical application)
time frame. 

\noindent Towards this end, a physics-driven \emph{AI} solution strategy,
inspired by the concepts/guidelines of the \emph{SbD} paradigm \cite{Massa 2022}
and leveraging the integration of a three-steps \emph{LBE} strategy
\cite{Salucci 2022b}\cite{Massa 2018} with an \emph{EA}-based global
search engine \cite{Rocca 2009}, is adopted. More specifically, at
each $i$-th ($i=1,...,I$) iteration a \emph{SM} of $\Phi\left(\underline{\alpha}\right)$,
$\widehat{\Phi}_{i}\left(\underline{\alpha}\right)$, is exploited
to make reliable-but-fast predictions of (\ref{eq:cost-function})
starting from the information embedded in a training set $\mathbb{O}_{i}$
of $B_{i}$ previously-computed {}``examples''\begin{equation}
\mathbb{O}_{i}=\left\{ \left[\underline{\alpha}_{i}^{b},\,\Phi\left(\underline{\alpha}_{i}^{b}\right)\right];\, b=1,...,B_{i}\right\} .\label{eq:}\end{equation}
The three \emph{logical} steps of the implemented \emph{LBE} strategy
are briefly summarized in the following and detailed in the next Sections: 

\begin{enumerate}
\item \emph{Physics-Driven Dimensionality Reduction} (Sect. \ref{sub:Dimensionality-Reduction})
- Because of the {}``curse of dimensionality'' \cite{Friedman 2001}
the number of training samples for an accurate prediction of (\ref{eq:cost-function})
exponentially grows with the dimensionality $K$ of the solution space.
For such a reason, a traditional pixel-based expansion of the unknown
dielectric profile in $\Omega$ is unfeasible since it would require
a huge number of unknowns (i.e., $K=2\times N$) to code the permittivity
and the conductivity in each discretization sub-domain $\left\{ \Omega_{n};\, n=1,...,N\right\} $.
Accordingly, the goal of this step is to seek a more convenient/efficient
definition of $\underline{\alpha}$ (i.e., $K\ll2N$), still guaranteeing
a high flexibility in modeling the unknown target and allowing to
embed \emph{a-priori} information on the underlying physics;
\item \emph{Adaptive Solution Space Sampling} (Sect. \ref{sub:Solution-Space-Sampling})
- Once the low-cardinality set of descriptors $\underline{\alpha}$
has been defined (Step 1), the goal of this step is the proper selection
of the minimum set of representative input-output pairs to build the
\emph{SM}. This task is performed by first generating an initial ($i=0$)
training set $\mathbb{O}_{0}=\left\{ \left[\underline{\alpha}_{0}^{b},\,\Phi\left(\underline{\alpha}_{0}^{b}\right)\right];\, b=1,...,B_{0}\right\} $
of $B_{0}$ samples uniformly spread over the $K$-dimensional search
space. Successively, additional samples are progressively ($i=1,...,I$)
generated and added to the training set ($\mathbb{O}_{0}\subseteq...\subseteq\mathbb{O}_{i}\subseteq...\subseteq\mathbb{O}_{I}$)
to adaptively enhance the accuracy of the \emph{SM} only in proximity
of the actual solution/global optimum $\underline{\alpha}^{true}$.
Such an adaptive sampling is performed by globally exploring the solution
space with an \emph{EA}-based strategy driven by the predictions and
the associated degree of uncertainty outputted by the \emph{SM};
\item \emph{Definition of the Learning Strategy} (Sect. \ref{sub:Definition-of-the-LBE})
- This step is aimed at defining the most suitable \emph{LBE} technique
for exploiting the gathered information within $\mathbb{O}_{i}$ to
build the \emph{SM} $\widehat{\Phi}_{i}\left(\underline{\alpha}\right)$
($i=1,...,I$). Such a choice must take into account both the deterministic
nature of (\ref{eq:cost-function}) and the need to predict uncertainty
measures supporting the \emph{EA}-driven adaptive sampling (Step 2). 
\end{enumerate}

\subsection{Physics-Driven Dimensionality Reduction \label{sub:Dimensionality-Reduction}}

The definition of the minimum-cardinality set $\underline{\alpha}$
of $K$ descriptors leverages the \emph{prior} knowledge on the class
of imaged targets as well as the underlying physics of the differential
scattering formulation (Sect. \ref{sec:mathematical-formulation}).
Since the \emph{ISP} unknowns are expressed in terms of a physically-bounded
difference with respect to a known reference \emph{EM} scenario {[}Fig.
1(\emph{b}){]}, the breast tumor is modeled as a relatively small
region $\Psi$ with homogeneous permittivity/conductivity values (i.e.,
$\varepsilon_{\mathcal{A}}\left(\mathbf{r}\right)=\varepsilon_{\Psi}$
and $\sigma_{\mathcal{A}}\left(\mathbf{r}\right)=\sigma_{\Psi}$,
$\mathbf{r}\in\Psi$ \cite{Golnabi 2019}\cite{Solis-Nepote 2019})
and it is described with the vector $\underline{\alpha}$\begin{equation}
\underline{\alpha}\triangleq\left\{ \varepsilon_{\Psi},\,\sigma_{\Psi},\, x_{\Psi},\, y_{\Psi},\,\underline{d}\right\} .\label{eq:DoFs}\end{equation}
In (\ref{eq:DoFs}), $x_{\Psi}$ and $y_{\Psi}$ {[}$\mathbf{r}_{\Psi}=\left(x_{\Psi},\, y_{\Psi}\right)\in\Lambda${]}
are the coordinates of the barycenter of $\Psi$, while $\underline{d}=\left\{ d_{c}>0;\, c=1,...,C\right\} $
is a set of real values that univocally identifies its external contour
$\partial\Psi$ (i.e., $\partial\Psi=\partial\Psi\left(\underline{d}\right)$).
More specifically, $\partial\Psi$ is defined as the union of $C$
quadratic curves, $\partial\Psi=\bigcup_{c=1}^{C}\partial\Psi_{c}$,
each $c$-th ($c=1,...,C$) curve being expressed as the set of points
complying with

\begin{equation}
\begin{array}{r}
\partial\Psi_{c}=\left\{ \forall\mathbf{r}\in\Lambda\,:\,\mathbf{r}=\mathbf{r}_{c}\times\left(\frac{1}{2}+l-l^{2}\right)+\mathbf{r}_{c-1}\times\left(\frac{1}{2}-l+\frac{l^{2}}{2}\right)+\right.\\
\left.+\mathbf{r}_{c+1}\times\left(\frac{l^{2}}{2}\right);\, l\in\left[0,\,1\right]\right\} \end{array}\label{eq:tumor-shape}\end{equation}

\noindent where\begin{equation}
\mathbf{r}_{c}=\mathbf{r}_{\Psi}+d_{c}\times\bm{\beta}_{c};\,\,\,\,\, c=1,...,C.\label{eq:control-points}\end{equation}
In (\ref{eq:control-points}) $\bm{\beta}_{c}=\left[\cos\left(\varphi_{c}\right),\,\sin\left(\varphi_{c}\right)\right]$
and $\varphi_{c}=2\pi\left(c-1\right)/C$ ($c=1,...,C$), while $\mathbf{r}_{0}=\mathbf{r}_{C}$
and $\mathbf{r}_{C+1}=\mathbf{r}_{1}$. Accordingly, the dimensionality
of the search space turns out to be significantly shrinked and equal
to $K=\left(4+C\right)$.

\subsection{Adaptive Solution Space Sampling\label{sub:Solution-Space-Sampling}}

The sampling of the $K$-dimensional solution space is aimed at three
indivisible goals: (\emph{i}) adaptively enhancing the accuracy of
the \emph{SM} within the \emph{attraction basin} of the global optimum,
(\emph{ii}) effectively exploring the multi-minima landscape of the
differential cost function (\ref{eq:cost-function}), and (\emph{iii})
minimizing the number of time-consuming full-wave evaluations. Such
goals are achieved by integrating the \emph{exploration} and the \emph{exploitation}
capabilities of both the \emph{SM} and a multi-agent nature-inspired
global optimizer. On the one hand, the \emph{SM} exploits progressively-acquired
information from the training samples to {}``understand'' where
new full-wave evaluations (i.e., new training samples) should be performed.
This means identifying portions of the search space that should be
further explored because of their higher probability of including
$\underline{\alpha}^{true}$. On the other hand, the optimization
engine exploits both the outputs of the \emph{SM} and the acquired
cognitive/social knowledge to smartly explore the search space by
generating new trial guesses.

\noindent Such an adaptive sampling starts from the generation of
$B_{0}$ solutions $\left\{ \underline{\alpha}_{0}^{b};\, b=1,...,B_{0}\right\} $,
uniformly spread within the solution space by means of the Latin Hypercube
Sampling (\emph{LHS}) method \cite{Garud 2017}. 

\noindent The computation of the corresponding cost function values,
$\left\{ \Phi\left(\underline{\alpha}_{0}^{b}\right);\, b=1,...,B_{0}\right\} $,
is then performed after mapping each $b$-th ($b=1,...,B_{0}$) sample
into the corresponding dielectric map of $\Lambda$. With reference
to a generic guess $\underline{\alpha}$, the following decoding rule
is used

\noindent \begin{equation}
\widetilde{\varepsilon}\left(\mathbf{r}_{n}\left|\underline{\alpha}\right.\right)=\left\{ \begin{array}{ll}
\varepsilon_{0}\varepsilon_{\Psi}-j\frac{\sigma_{\Psi}}{\omega} & \textrm{if }\mathbf{r}_{n}\in\Psi\left(\underline{\alpha}\right)\\
\widetilde{\varepsilon}_{\mathcal{N}}\left(\mathbf{r}_{n}\right) & \textrm{if }\mathbf{r}_{n}\in\left(\Lambda-\Psi\left(\underline{\alpha}\right)\right)\end{array}\right.;\,\,\,\,\, n=1,...,N\label{eq:decode-epsilon}\end{equation}
where $\Psi\left(\underline{\alpha}\right)$ is the tumor support
described by $\underline{\alpha}$ (Sect. \ref{sub:Dimensionality-Reduction}).
Afterwards, the differential equivalent currents are determined as
follows\begin{equation}
\underline{J}_{\Delta}^{v}\left(\underline{\alpha}\right)=\underline{\underline{\tau}}_{\Delta}\left(\underline{\alpha}\right)\,\left[\underline{\underline{\mathbb{I}}}-\underline{\underline{G}}_{\mathcal{B}}^{\Omega}\,\underline{\underline{\tau}}\left(\underline{\alpha}\right)\right]^{-1}\,\underline{E}_{inc}^{v};\,\,\,\,\, v=1,...,V\label{eq:retrieved-differential-currents}\end{equation}
where $\underline{\underline{\tau}}\left(\underline{\alpha}\right)$
and $\underline{\underline{\tau}}_{\Delta}\left(\underline{\alpha}\right)$
are diagonal matrices which diagonal entries are given by $\underline{\tau}\left(\underline{\alpha}\right)=\left\{ \tau\left(\mathbf{r}_{n}\left|\underline{\alpha}\right.\right)=\left[\widetilde{\varepsilon}\left(\mathbf{r}_{n}\left|\underline{\alpha}\right.\right)/\widetilde{\varepsilon}_{\mathcal{B}}-1\right];\, n=1,...,N\right\} $
and $\underline{\tau}_{\Delta}\left(\underline{\alpha}\right)=\left\{ \left[\tau\left(\mathbf{r}_{n}\left|\underline{\alpha}\right.\right)-\tau_{\mathcal{N}}\left(\mathbf{r}_{n}\right)\right];\, n=1,...,N\right\} $,
respectively. Finally, the differential field $\underline{D}^{\Theta,v}\left(\underline{\alpha}\right)$
($v=1,...,V$) radiated by (\ref{eq:retrieved-differential-currents})
is computed using (\ref{eq:discrete data diff}) and then inputted
to (\ref{eq:cost-function}) to compute $\Phi\left(\underline{\alpha}\right)$.

\noindent Such $B_{0}$ input-output examples are stored within $\mathbb{O}_{o}$
and used to train an initial \emph{SM} $\widehat{\Phi}_{0}\left(\underline{\alpha}\right)$.
An iterative ($i=1,...,I$) adaptive sampling is then repeated to
explore the solution space by means of an \emph{EA}-based strategy
based on the Particle Swarm Optimization (\emph{PSO}) operators \cite{Rocca 2009}.
More in detail, the following steps are performed:

\begin{enumerate}
\item \emph{Initialization} ($i=0$) - Randomly pick $P\leq B_{0}$ samples
from $\mathbb{O}_{o}$ to initialize a swarm of trial solutions $\Gamma_{0}=\left\{ \underline{\alpha}_{0}^{p};\, p=1,...,P\right\} $
and randomly set their velocities $\Upsilon_{0}=\left\{ \underline{\upsilon}_{0}^{p};\, p=1,...,P\right\} $.
Initialize the personal best positions, $\underline{\vartheta}_{0}^{p}\leftarrow\underline{\alpha}_{0}^{p}$
($p=1,...,P$);
\item \emph{Adaptive Sampling Loop} ($i=1,...,I$)

\begin{enumerate}
\item \emph{Population Prediction} - Use the $i$-th \emph{SM} $\widehat{\Phi}_{i}\left(\,.\,\right)$
to predict the cost function associated to all agents of the population
$\Gamma_{i}$, $\left\{ \widehat{\Phi}_{i}\left(\underline{\alpha}_{i}^{p}\right);\, p=1,...,P\right\} $,
and compute the corresponding uncertainties $\left\{ \varsigma_{i}\left(\underline{\alpha}_{i}^{p}\right);\, p=1,...,P\right\} $;
\item \emph{Agent Selection} - Determine the best agent of the $i$-th population
$\Gamma_{i}$ as\begin{equation}
\underline{\alpha}_{i}^{*}=\arg\left\{ \min_{\underline{\alpha}\in\Gamma_{i}}\left[\widehat{\Phi}_{i}\left(\underline{\alpha}\right)-\varsigma_{i}\left(\underline{\alpha}\right)\right]\right\} ;\label{eq:}\end{equation}

\item \emph{Training Set Updating} - If $\left[\widehat{\Phi}_{i}\left(\underline{\alpha}_{i}^{*}\right)-\varsigma_{i}\left(\underline{\alpha}_{i}^{*}\right)\right]>\min_{\underline{\alpha}\in\mathbb{O}_{i}}\left\{ \Phi\left(\underline{\alpha}\right)\right\} $,
then let $B_{i}\leftarrow B_{i-1}$, $\mathbb{O}_{i}\leftarrow\mathbb{O}_{i-1}$,
$\widehat{\Phi}_{i}\left(\,.\,\right)\leftarrow\widehat{\Phi}_{i-1}\left(\,.\,\right)$
and jump to the next step. Otherwise, decode $\underline{\alpha}_{i}^{*}$
(\ref{eq:decode-epsilon}) and compute, through (\ref{eq:retrieved-differential-currents}),
the corresponding actual cost function value $\Phi\left(\underline{\alpha}_{i}^{*}\right)$.
Then, let $B_{i}\leftarrow\left(B_{i-1}+1\right)$, $\mathbb{O}_{i}\leftarrow\mathbb{O}_{i-1}\cup\left\{ \underline{\alpha}_{i}^{*},\,\Phi\left(\underline{\alpha}_{i}^{*}\right)\right\} $,
and update the \emph{SM} $\widehat{\Phi}_{i}\left(\,.\,\right)$ using
$\mathbb{O}_{i}$;
\item \emph{Personal Best Updating} - Update the personal best of each $p$-th
($p=1,...,P$) agent \begin{equation}
\underline{\vartheta}_{i}^{p}\leftarrow\mathcal{R}\left\{ \vartheta_{i-1}^{p},\,\underline{\alpha}_{i}^{p}\right\} \end{equation}
where $\mathcal{R}\left\{ \underline{\alpha}^{\prime},\,\underline{\alpha}^{\prime\prime}\right\} $
is a ranking operator selecting the {}``most promising'' solution
among the two inputs $\underline{\alpha}^{\prime}$ and $\underline{\alpha}^{\prime\prime}$
by taking into account whether their cost function has been evaluated
with the full-wave solver or predicted. More in detail, when both
agents belong to the training set (i.e., $\left\{ \underline{\alpha}^{\prime},\,\underline{\alpha}^{\prime\prime}\right\} \in\mathbb{O}_{i}$)
or not (i.e., $\left\{ \underline{\alpha}^{\prime},\,\underline{\alpha}^{\prime\prime}\right\} \notin\mathbb{O}_{i}$)
the selection is based on their predicted cost function values along
with the associated degree of reliability\begin{equation}
\mathcal{R}\left\{ \underline{\alpha}^{\prime},\underline{\alpha}^{\prime\prime}\right\} \equiv\arg\left\{ \min_{\underline{\alpha}\in\left\{ \underline{\alpha}^{\prime},\underline{\alpha}^{\prime\prime}\right\} }\left[\widehat{\Phi}_{i}\left(\underline{\alpha}\right)-\varsigma_{i}\left(\underline{\alpha}\right)\right]\right\} \end{equation}
knowing that $\widehat{\Phi}_{i}\left(\underline{\alpha}\right)=\Phi\left(\underline{\alpha}\right)$
and $\varsigma_{i}\left(\underline{\alpha}\right)=0$ (i.e., exact
knowledge and null uncertainty) if $\underline{\alpha}$ has been
evaluated with the full-wave solver (i.e., $\underline{\alpha}\in\mathbb{O}_{i}$).
Otherwise, when only one agent belongs to the set of full-wave simulated
samples (e.g., $\underline{\alpha}^{\prime}\notin\mathbb{O}_{i}$
and $\underline{\alpha}^{\prime\prime}\in\mathbb{O}_{i}$), the predicted
cost is {}``penalized'' by the uncertainty, letting\begin{equation}
\mathcal{R}\left\{ \underline{\alpha}^{\prime},\underline{\alpha}^{\prime\prime}\right\} \equiv\left\{ \begin{array}{ll}
\underline{\alpha}^{\prime} & \textrm{if }\left[\widehat{\Phi}_{i}\left(\underline{\alpha}^{\prime}\right)+\varsigma_{i}\left(\underline{\alpha}^{\prime}\right)\right]<\Phi\left(\underline{\alpha}^{\prime\prime}\right)\\
\underline{\alpha}^{\prime\prime} & \textrm{otherwise};\end{array}\right.\end{equation}

\item \emph{Global Best Updating -} Update the global best\begin{equation}
\underline{\alpha}_{i}^{opt}\leftarrow\arg\left\{ \min_{\underline{\alpha}\in\Pi_{i}}\left[\widehat{\Phi}_{i}\left(\underline{\alpha}\right)+\varpi\times\varsigma_{i}\left(\underline{\alpha}\right)\right]\right\} \end{equation}
where the value of the constant $\varpi$ depends on whether the global
best at the $\left(i-1\right)$-th iteration was full-wave evaluated
($\underline{\alpha}_{i-1}^{opt}\in\mathbb{O}_{i-1}\Rightarrow\varpi=+1$)
or not ($\underline{\alpha}_{i-1}^{opt}\not\in\mathbb{O}_{i-1}\Rightarrow\varpi=-1$),
while $\Pi_{i}$ comprises both $\underline{\alpha}_{i-1}^{opt}$
and all personal best positions at the current iteration\begin{equation}
\Pi_{i}\triangleq\underline{\alpha}_{i-1}^{opt}\cup\left\{ \underline{\vartheta}_{i}^{p};\, p=1,...,P\right\} ;\end{equation}

\item \emph{Population Updating} - Apply the \emph{PSO} operators \cite{Rocca 2009}
to update the agents velocities $\Upsilon_{i+1}=\left\{ \underline{\upsilon}_{i+1}^{p};\, p=1,...,P\right\} $
by exploiting the acquired cognitive (i.e., $\left\{ \underline{\vartheta}_{i}^{p};\, p=1,...,P\right\} $)
and social (i.e., $\underline{\alpha}_{i}^{opt}$) information, then
update the $p$-th agent of the population\begin{equation}
\underline{\alpha}_{i+1}^{p}\leftarrow\left(\underline{\alpha}_{i}^{p}+\underline{\upsilon}_{i+1}^{p}\right);\, p=1,...,P.\end{equation}
Let $p\leftarrow\left(p+1\right)$ and repeat from Step 2(\emph{a});
\end{enumerate}
\item \emph{Output Phase} - The global best at the last iteration ($i=I$)
is selected as the result of the inversion ($\underline{\alpha}^{opt}\leftarrow\underline{\alpha}_{i=I}^{opt}$)
and the reconstructed permittivity profile is computed through (\ref{eq:decode-epsilon}).
\end{enumerate}

\subsection{Definition of the Learning Strategy\label{sub:Definition-of-the-LBE}}

At each $i$-th ($i=1,...,I$) iteration of the adaptive sampling
(Sect. \ref{sub:Solution-Space-Sampling}), the training samples stored
in $\mathbb{O}_{i}$ are treated as they were observations of a normally
distributed stochastic process. Accordingly, the correlation between
any pair of samples $\left(\underline{\alpha}_{i}^{a},\,\underline{\alpha}_{i}^{b}\right)\in\mathbb{O}_{i}$
is expressed in terms of Kriging basis functions \cite{Forrester 2008},
\begin{equation}
w_{i}\left(\underline{\alpha}_{i}^{a},\,\underline{\alpha}_{i}^{b}\right)\triangleq\exp\left(-\sum_{k=1}^{K}\theta_{i,k}\left|\alpha_{i,k}^{a}-\alpha_{i,k}^{b}\right|^{\nu_{i}}\right).\label{eq:correlation}\end{equation}
From (\ref{eq:correlation}), the $\left(B_{i}\times B_{i}\right)$
correlation matrix $\underline{\underline{W}}_{i}$ of $\mathbb{O}_{i}$
is computed by setting its $\left(a,\, b\right)$-th entry ($a,\, b=1,...,B_{i}$)
to $\left.\underline{\underline{W}}_{i}\right\rfloor _{a,b}=w_{i}\left(\underline{\alpha}_{i}^{a},\,\underline{\alpha}_{i}^{b}\right)$.
As for the real-valued hyper-parameters $\underline{\theta}_{i}=\left\{ \theta_{i,k};\, k=1,...,K\right\} $
and $\nu_{i}$ in (\ref{eq:correlation}), they are found by maximizing
the {}``joint likelihood'' function \cite{Lophaven 2002}

\begin{equation}
\mathcal{L}\left(\underline{\theta}_{i},\nu_{i}\right)=\frac{1}{\left(\sqrt{2\pi}\xi_{i}\right)^{B_{i}}}\exp\left[-\sum_{b=1}^{B_{i}}\frac{\left(\Phi\left(\underline{\alpha}_{i}^{b}\right)-\gamma_{i}\right)^{2}}{2\xi_{i}^{2}}\right],\end{equation}
where $\gamma_{i}$ and $\xi_{i}^{2}$ are the maximum-likelihood
estimates of the process average\begin{equation}
\gamma_{i}=\frac{\sum_{a=1}^{B_{i}}\sum_{b=1}^{B_{i}}\left.\underline{\underline{U}}_{i}\right\rfloor _{a,b}\Phi\left(\underline{\alpha}_{i}^{b}\right)}{\sum_{a=1}^{B_{i}}\sum_{b=1}^{B_{i}}\left.\underline{\underline{U}}_{i}\right\rfloor _{a,b}},\end{equation}
 and its variance\begin{equation}
\xi_{i}^{2}=\frac{1}{B_{i}}\sum_{a=1}^{B_{i}}\sum_{b=1}^{B_{i}}\left(\Phi\left(\underline{\alpha}_{i}^{a}\right)-\gamma_{i}\right)\left.\underline{\underline{U}}_{i}\right\rfloor _{a,b}\left(\Phi\left(\underline{\alpha}_{i}^{b}\right)-\gamma_{i}\right)\end{equation}
respectively being $\underline{\underline{U}}_{i}\triangleq\underline{\underline{W}}_{i}^{-1}$.
Given the above formulation, the prediction of the cost function in
correspondence with an arbitrary solution $\underline{\alpha}$ is
given by \cite{Forrester 2008}\begin{equation}
\widehat{\Phi}_{i}\left(\underline{\alpha}\right)=\gamma_{i}+\sum_{a=1}^{B_{i}}\sum_{b=1}^{B_{i}}w_{i}\left(\underline{\alpha},\,\underline{\alpha}_{i}^{a}\right)\left.\underline{\underline{U}}_{i}\right\rfloor _{a,b}\left(\Phi\left(\underline{\alpha}_{i}^{b}\right)-\gamma_{i}\right),\label{eq:surrogate}\end{equation}
while the associated prediction uncertainty is derived as \begin{equation}
\varsigma_{i}\left(\underline{\alpha}\right)=2\xi_{i}\sqrt{1-\sum_{a=1}^{B_{i}}\sum_{b=1}^{B_{i}}w_{i}\left(\underline{\alpha},\,\underline{\alpha}_{i}^{a}\right)\left.\underline{\underline{U}}_{i}\right\rfloor _{a,b}w_{i}\left(\underline{\alpha},\,\underline{\alpha}_{i}^{b}\right)}.\end{equation}

\section{\noindent Numerical and Experimental Results\label{sec:results}}

\noindent A set of representative benchmarks, concerned with both
synthetic (Sects. \ref{sub:Preliminary-Analysis}-\ref{sub:Analysis-with-Full-Numerical-Phantom})
and experimental (Sect. \ref{sub:Analysis-with-Experimental}) data,
is presented to assess the performance of the proposed \emph{SbD}
inversion methodology in scenarios of increasing complexity/realism.
Quantitatively, the reconstruction {}``quality'' has been evaluated
with the integral error metric\begin{equation}
\Xi\left\{ R\right\} \triangleq\frac{1}{N_{R}}\sum_{n=1}^{N_{R}}\frac{\left|\tau_{\Delta}\left(\mathbf{r}_{n}\right)-\tau_{\Delta}^{true}\left(\mathbf{r}_{n}\right)\right|}{\left|\tau_{\Delta}^{true}\left(\mathbf{r}_{n}\right)+1\right|},\end{equation}
where $\tau_{\Delta}$ and $\tau_{\Delta}^{true}$ are the retrieved
and the actual differential contrasts, respectively, while $R\subseteq\Omega$
indicates the region comprising $N_{R}\leq N$ pixels where the computation
has been performed (i.e., $\Xi_{tot}\triangleq\Xi\left\{ R=\Omega\right\} $,
$\Xi_{int}\triangleq\Xi\left\{ R=\Psi\right\} $, and $\Xi_{ext}\triangleq\Xi\left\{ R=\left(\Omega\smallsetminus\Psi\right)\right\} $).

\noindent The numerical analysis has dealt with an imaging setup at
$f=1.3$ {[}GHz{]} comprising $V=16$ line currents uniformly-distributed
in a circle with radius $\rho_{\Theta}=7.6$ {[}cm{]} from the barycenter
of $\Omega$. The synthetic data have been computed by solving the
forward scattering equations discretized with a dense grid of cells
$\lambda_{\min}/20$-sided, $\lambda_{\min}$ being the wavelength
in the breast tissue with the highest permittivity. To avoid the {}``inverse
crime'' \cite{Chen 2018}, a different (coarser) $\lambda_{\min}/10$
discretization has been adopted in the \emph{ISP}. To emulate the
measurement noise in a breast \emph{MI} setup, the total field has
been blurred with an additive white Gaussian noise characterized by
a signal-to-noise ratio (\emph{SNR}) value.

\subsection{Preliminary Analysis with {}``Ideal'' Breast Phantom \label{sub:Preliminary-Analysis}}

\noindent The first illustrative example considers an {}``ideal''
breast phantom immersed in a background matching medium of permittivity
$\varepsilon_{\mathcal{B}}=22.4$ and conductivity $\sigma_{\mathcal{B}}=1.26$
{[}S/m{]} \cite{Golnabi 2019} with circular coronal section $\Lambda$
and homogeneous characteristics {[}i.e., $\varepsilon_{\mathcal{N}}\left(\mathbf{r}\right)=\varepsilon_{\mathcal{N}}=16.5$
and $\sigma_{\mathcal{N}}\left(\mathbf{r}\right)=\sigma_{\mathcal{N}}=0.60$
{[}S/m{]} for $\mathbf{r}\in\Lambda$ \cite{Golnabi 2019} - Figs.
2(\emph{a})-2(\emph{b})%
\footnote{\noindent For visualization purposes, only the color-scale permittivity/conductivity
distributions within $\Lambda$ are plotted.%
}{]}. A circular tumor of radius $\rho_{\Psi}=0.5$ {[}cm{]}, permittivity
$\varepsilon_{\Psi}=59.3$, and conductivity $\sigma_{\Psi}=1.54$
{[}S/m{]} \cite{Golnabi 2019} has been supposed to be present within
the breast {[}Figs. 2(\emph{c})-2(\emph{d}){]}. For such a scenario,
the performance of the proposed \emph{SbD} approach have been compared
to those of two standard inversion strategies formulated either in
the \emph{DA} or the \emph{EA} frameworks by processing the numerically-computed
data with $SNR=100$ {[}dB{]} \cite{Golnabi 2019}. More specifically,
a state-of-the-art \emph{CG} exploiting pixel basis functions has
been chosen as the \emph{DA} strategy and it has been run for $I_{DA}=400$
iterations to reach convergence \cite{van den Berg 1997}. Concerning
the \emph{EA} approach, a standard \emph{PSO} algorithm \cite{Rocca 2009}
has been used to explore the solution space defined by $\underline{\alpha}$.
The tumor shape has been modeled by setting $C=4$ in (\ref{eq:tumor-shape}),
so that $K=8$ unknowns. The number of agents and initial training
samples have been set to $P=\left(2\times K\right)=16$ and $B_{0}=\left(5\times K\right)=40$,
respectively, according to the literature guidelines \cite{Salucci 2022}\cite{Massa 2022}.
Finally, to guarantee a fair comparison between the \emph{SbD} and
the \emph{DA} in terms of total computational cost, the number of
both \emph{SbD} and \emph{EA} iterations has been set to match the
time elapsed by one \emph{DA} minimization (i.e., $\Delta t_{DA}\approx\Delta t_{SbD}\Rightarrow I=200$). 

\noindent The behavior of the cost function versus the iteration number
is shown in Fig. 3(\emph{a}). As it can be observed, the \emph{SbD}
remarkably minimizes the cost function until a value at the convergence
($i=I$) very close to that yielded by the \emph{EA} {[}i.e., $\left|\left.\Phi_{I}\right|_{SbD}-\left.\Phi_{I}\right|_{EA}\right|<3\times10^{-5}$
- Fig. 3(\emph{a}){]}. The corresponding total reconstruction errors
and execution times are given in Fig. 3(\emph{b}) to point out that
the \emph{SbD} is able to provide almost the same solution {}``quality''
of the \emph{EA} {[}$\left.\Xi_{tot}\right|_{SbD}=2.01\times10^{-5}$
vs. $\left.\Xi_{tot}\right|_{EA}=1.46\times10^{-5}$ - Fig. 3(\emph{b}){]}
with a significant reduction of the overall computational time {[}i.e.,
$\left.\Delta t\right|_{SbD}=7.38$ {[}min{]} vs. $\left.\Delta t\right|_{EA}=98.4$
{[}min{]}, being $\left.\Delta t\right|_{DA}=7.40$ {[}min{]}%
\footnote{\noindent These CPU times refer to non-optimized FORTRAN codes on
a standard laptop with $16$ {[}GB{]} of RAM memory and an Intel(R)
Core(TM) i5-8250U CPU @ 1.60 {[}GHz{]}.%
} - Fig. 3(\emph{b}){]}. By defining the time saving of the \emph{SbD}
over the \emph{EA} as\begin{equation}
\eta=\frac{\left.\Delta t\right|_{EA}-\left.\Delta t\right|_{SbD}}{\left.\Delta t\right|_{EA}}\approx1-\frac{B_{I}}{T},\label{eq:}\end{equation}
$T=\left(P\times I\right)$ being the number of full-wave evaluations
performed by this latter, it turns out that $\eta=93\%$. Such a speed-up
enhancement is even more impressive when observing that almost identical
reconstructions \emph{}have been yielded by both methods as pictorially
confirmed by the permittivity (left column) and conductivity (right
column) maps in Fig. 4. Indeed, it turns out that the \emph{SbD} inherits
the \emph{hill-climbing} features of a nature-inspired multi-agent
exploration \cite{Rocca 2009}, escaping from the multiple local minima
(false solutions) and properly guessing the position, the size, the
shape, and the \emph{EM} composition of the unknown abnormal tissue
{[}Figs. 4(\emph{a})-4(\emph{b}) vs. Figs. 2(\emph{c})-2(\emph{d}){]}.
On the other hand, while being as fast as the \emph{SbD}, the \emph{DA}
only roughly detects the presence of the tumor without accurate information
on its size and composition {[}Figs. 4(\emph{e})-4(\emph{f}){]}. Such
outcomes are also quantitatively confirmed by the value of the internal
error {[}i.e., $\left.\Xi_{int}\right|_{DA}=6.18\times10^{-1}$ -
Figs. 4(\emph{e})-4(\emph{f}){]}, which is two orders in magnitude
higher than that for the \emph{SbD} {[}$\left.\Xi_{int}\right|_{SbD}=9.75\times10^{-3}$
- Figs. 4(\emph{a})-4(\emph{b}){]} and the \emph{EA} {[}$\left.\Xi_{int}\right|_{EA}=7.08\times10^{-3}$
- Figs. 4(\emph{c})-4(\emph{d}){]}. Similar outcomes hold true for
the external error because of the noticeable artifacts generated within
$\Lambda$, as well (i.e., $\left.\Xi_{ext}\right|_{DA}=1.24\times10^{-2}$
vs. $\left.\Xi_{ext}\right|_{SbD}=\left.\Xi_{ext}\right|_{EA}=0.0$).

\subsection{Analysis with Segmented Breast Phantom \label{sub:Analysis-with-Segmented-Phantom}}

\noindent While the {}``ideal'' phantom of Sect. \ref{sub:Preliminary-Analysis}
is an over-simplification of the actual breast structure, which generally
includes several tissues with inhomogeneous \emph{EM} characteristics,
it was an useful benchmark to test the features of the \emph{SbD}
approach. However, let us consider a more faithful/realistic representation
of the breast anatomy. More specifically, the numerical phantoms from
the \emph{UWCEM} repository, an open-access dataset of \emph{MRI}-derived
breast models with realistic geometrical and \emph{EM} features \cite{Burfeindt 2012},
have been used. Such a biomedical database includes phantoms belonging
to all four categories of breast composition identified by the American
College of Radiology \cite{D'Orsi 2013} {[}i.e., (\emph{a}) {}``fatty'',
(\emph{b}) {}``scattered'', (\emph{c}) {}``heterogeneously dense''
(\emph{HD}), and (\emph{d}) {}``extremely dense'' (\emph{XD}){]}
that correspond to increasing concentrations of high-permittivity
tissues. 

\noindent Dealing with piece-wise (i.e., segmented) homogeneous breast
models, a phantom belonging to the \emph{XD} category (\emph{UWCEM
ID} $012304$, coronal slice $\Lambda$ at $z_{\Lambda}=6.65$ {[}cm{]}
\cite{Burfeindt 2012}) has been considered first. Following the guidelines
in \cite{Golnabi 2019}, each cell of the reference breast has been
binary classified into two types of tissue, namely {}``adipose''
and {}``fibroglandular'', which dielectric properties are reported
in Tab. I. Moreover, the skin layer has been approximated as {}``adipose''
tissue \cite{Golnabi 2019} since the breast has been assumed to be
immersed in a bath of matching medium ($\varepsilon_{\mathcal{B}}=22.4$,
$\sigma_{\mathcal{B}}=1.26$ {[}S/m{]} \cite{Golnabi 2019} - Tab.
I) and the thickness of such a layer is generally electrically negligible
($t_{skin}=1.5$ {[}mm{]} $\approx5$ \% of $\lambda_{\min}$ \cite{Golnabi 2019}).
As for the tumor, it has been modeled as an ellipse which shape and
size match one of the realistic phantoms described in \cite{O'Loughlin 2019}
{[}Figs. 5(\emph{a})-5(\emph{b}){]}. The \emph{SbD} reconstruction
in Figs. 5(\emph{c})-5(\emph{d}) positively compares with the actual
configuration {[}Figs. 5(\emph{a})-5(\emph{b}){]} even though the
value of the total error (i.e., $\Xi_{tot}=3.21\times10^{-3}$) increases
with respect to the ideal case in Sect. \ref{sub:Preliminary-Analysis}
because of the increased complexity of the scenario at hand.

\noindent To investigate on the reliability and the robustness of
the \emph{SbD} in case of higher levels of noise, an analysis has
been performed by varying the \emph{SNR} from $SNR=100$ {[}dB{]}
down to $SNR=10$ {[}dB{]}. The inversion results are summarized in
Fig. 6 in terms of total {[}Fig. 6(\emph{a}){]}, internal {[}Fig.
6(\emph{b}){]}, and external {[}Fig. 6(\emph{c}){]} errors. As it
can be observed, the \emph{SbD} accuracy is always comparable to that
of the \emph{EA} and remarkably higher than the \emph{DA} whatever
the blurring of the processed data. As a representative example, Figure
7 shows the breast profiles reconstructed by the three algorithms
when $SNR=20$ {[}dB{]}. Despite the harsh noise conditions, the \emph{SbD}
map is still highly similar to the actual one {[}$\left.\Xi_{tot}\right|_{SbD}=4.38\times10^{-3}$
- Figs. 7(\emph{a})-7(\emph{b}) vs. Figs. 5(\emph{a})-5(\emph{b}){]},
with the same reconstruction accuracy of the \emph{EA} inversion {[}$\left.\Xi_{tot}\right|_{EA}=3.65\times10^{-3}$
- Figs. 7(\emph{c})-7(\emph{d}){]} and visibly better than the \emph{DA}
one {[}$\left.\Xi_{tot}\right|_{DA}=9.22\times10^{-2}$ - Figs. 7(\emph{e})-7(\emph{f}){]}.
Additionally, the \emph{SbD} is always significantly faster than the
\emph{EA} with a time saving of $\eta=93\%$ because of the reduction
of the calls to the full-wave solver thanks to the exploitation of
the \emph{SM} {[}Fig. 6(\emph{d}){]}.

\noindent The \emph{SbD} accuracy has then been assessed by considering
breasts with different densities, as well. Towards this end, three
additional \emph{UWCEM} phantoms have been selected either belonging
to the fatty, the scattered, and the \emph{HD} categories (Tab. I).
Figure 8 compares the \emph{SbD} errors for each considered class,
while Figure 9 shows the corresponding reconstructions. It turns out
that the \emph{SbD} performance are quite similar in all benchmark
scenarios, even though the total error slightly increases when the
breast density changes from the \emph{XD} to the {}``fatty'' one
(i.e., $\left.\Xi_{tot}\right|_{Fatty}/\left.\Xi_{tot}\right|_{XD}\approx2.0$
- Fig. 8). This is not a surprise, since a reduced breast density
translates into lower values of the reference permittivity, $\varepsilon_{\mathcal{N}}\left(\mathbf{r}\right)$,
and conductivity, $\sigma_{\mathcal{N}}\left(\mathbf{r}\right)$,
distributions (Tab. I), thus a higher differential contrast, $\tau_{\Delta}\left(\mathbf{r}\right)$,
have to be imaged, which result in a higher non-linearity of the corresponding
\emph{ISP}. Nevertheless, the \emph{SbD} inversion always provides
a precise indication of both the tumor location and its shape, as
pictorially pointed out in Fig. 9.

\noindent The proposed method has been also tested against the variations
of the \emph{EM} properties of the tumor. Towards this aim, the complex
permittivity of the abnormal tissue has been varied according to the
following rule\begin{equation}
\widetilde{\varepsilon}_{\Psi}\left(\delta\right)=\left(1+\delta\right)\widetilde{\varepsilon}_{\Psi}^{nom}\end{equation}
where $\widetilde{\varepsilon}_{\Psi}^{nom}$ is the nominal tumor
permittivity ($\widetilde{\varepsilon}_{\Psi}^{nom}=59.3\varepsilon_{0}-j\frac{1.54}{2\pi f}$)
while $\delta$ has been set within the range $-10\%\leq\delta\leq10\%$
according to the observed variability in \emph{EM} properties of \emph{ex-vivo}
breast cancer biological samples \cite{Martellosio 2017}. The plots
of the error metrics versus the $\delta$ value are shown in Fig.
10. Such results verify the robustness and general effectiveness of
the proposed inversion method. As it can be observed, the error values
are almost constant and the retrieved images of $\Lambda$ turn out
to be accurate even in the most challenging case with the highest
differential contrast {[}i.e., $\delta=+10\%$ $\rightarrow$ $\left.\Xi_{tot}\right|_{\delta=+10\%}=3.1\times10^{-3}$
- Fig. 11(\emph{d}) vs. Fig. 11(\emph{b}) and Fig. 11(\emph{h}) vs.
Fig. 11(\emph{f}){]}.

\subsection{Analysis with Fully Inhomogeneous Breast Phantom \label{sub:Analysis-with-Full-Numerical-Phantom}}

\noindent In this section, the most complex case where the complex
permittivity values are available for each pixel belonging to $\Lambda$
\cite{Zastrow 2008} is taken into account. More specifically, a {}``full''
breast model has been derived from the \emph{UWCEM} database {[}\emph{ID}
$012204$ - Figs. 12(\emph{a})-12(\emph{b}){]} to provide a more detailed
representation of the internal breast structure than its segmented
{[}Figs. 12(\emph{c})-12(\emph{d}){]} and constant {[}Figs. 12(\emph{e})-12(\emph{f}){]}
versions that have been obtained by averaging the permittivity of
each tissue type or of the entire coronal slice, respectively. Regardless
of the more complex reference scenario, the \emph{SbD} faithfully
guesses the abnormal tissue distribution {[}Figs. 13(\emph{c})-13(\emph{d})
vs. Figs. 13(\emph{a})-13(\emph{b}){]} by keeping the same features
shown in the previous test cases when compared to other \emph{SoA}
inversion techniques (Tab. II).

\noindent It is now worth noticing that the benchmarks discussed so
far considered the full \emph{prior} (\emph{FP}) knowledge of the
reference breast tissues distribution. However, only partial/simplified
information on the patient's breast is generally \emph{a-priori} available
in practical biomedical imaging. Accordingly, the following analysis
investigates the \emph{SbD} performance with \emph{prior} knowledge
on the reference scenario of diminishing accuracy. By considering
the same actual scenario of Figs. 13(\emph{a})-13(\emph{b}) for the
forward scattering data computation, the segmented \emph{prior} {[}\emph{SP}
- Figs. 12(\emph{c})-12(\emph{d}){]} or the constant one {[}\emph{CP}
- Figs. 12(\emph{e})-12(\emph{f}){]} have been inputted to the \emph{SbD}
to perform the inversion. The reconstructions starting from a \emph{SP}
{[}Figs. 13(\emph{e})-13(\emph{f}){]} are slightly less accurate than
those for the \emph{FP} case, but they still are of good quality in
terms of estimated size and complex permittivity of the tumor. As
a matter of fact, the total error increases by $35\%$ between the
\emph{SP} and \emph{FP} cases {[}Figs. 13(\emph{e})-13(\emph{f}) vs.
Figs. 13(\emph{c})-13(\emph{d}) - Tab. III{]}. 

\noindent In the most challenging case of an over-simplified \emph{CP},
totally neglecting the internal structure of the actual breast tissues
{[}Figs. 12(\emph{e})-12(\emph{f}) vs. Figs. 12(\emph{a})-12(\emph{b}){]},
there is still a proper detection and localization of the tumor, even
though its size is over-estimated as quantitatively pointed out by
the increase of the external error {[}$\left.\Xi_{ext}\right|_{CP}=3.52\times10^{-2}$
vs. $\left.\Xi_{ext}\right|_{FP}=1.72\times10^{-3}$ - Figs. 13(\emph{g})-13(\emph{h})
vs. 13(\emph{c})-13(\emph{d}) and Tab. III{]}.

\subsection{Analysis with Experimental Phantoms \label{sub:Analysis-with-Experimental}}

The final assessment is concerned with real scattering data and the
inversion of experimental measurements at $f=2$ {[}GHz{]} from the
University of Manitoba Breast Microwave Imaging Dataset (\emph{UM-BMID},
\emph{}$3^{rd}$ generation \cite{Reimer 2021}\cite{Reimer 2020}).
The \emph{UM-BMID} contains data from various 3D-printed breast phantoms
in air ($\varepsilon_{\mathcal{B}}=1$, $\sigma_{\mathcal{B}}=0$
{[}S/m{]}). The measurements have been collected with a bi-static
setup consisting of two \emph{LB-20200-SF} horn antennas \cite{A-Info 2022}
located at $\rho_{\Theta}=18$ {[}cm{]} from the center of the investigation
domain and $72$ {[}deg{]} apart from each other. A multi-view/illumination
imaging system has been implemented by rotating the system in $V=72$
different angular positions with $5$ {[}deg{]} steps \cite{Reimer 2021}\cite{Reimer 2020}.

\noindent The first experiment is related to a tumor of circular section
and radius $\rho_{\Psi}^{true}=1.5$ {[}cm{]} embedded within a breast
phantom of {}``fatty'' density. The coronal section of the actual
permittivity and conductivity distributions are shown in Figs. 14(\emph{a})-14(\emph{b}).
Once again, the \emph{SbD} reconstruction {[}Figs. 14(\emph{c})-14(\emph{d}){]}
well resembles the one outputted by the \emph{EA} {[}Figs. 14(\emph{e})-14(\emph{f}){]}
by yielding, unlike the \emph{DA} {[}Figs. 14(\emph{g})-14(\emph{h}){]},
accurate qualitative and quantitative information on the tumor. There
is only a slight over/under estimation of the relative permittivity/conductivity
of the anomalous tissue as also indicated by the corresponding error
indexes (e.g., $\left.\Xi_{int}\right|_{\rho_{\Psi}^{true}=1.5\textrm{ [cm]}}^{EA}/\left.\Xi_{int}\right|_{\rho_{\Psi}^{true}=1.5\textrm{ [cm]}}^{SbD}\approx1.1$
and $\left.\Xi_{int}\right|_{\rho_{\Psi}^{true}=1.5\textrm{ [cm]}}^{DA}/\left.\Xi_{int}\right|_{\rho_{\Psi}^{true}=1.5\textrm{ [cm]}}^{SbD}\approx2.8$
- Tab. IV), the speed-up still being $\left.\Delta t\right|_{EA}/\left.\Delta t\right|_{SbD}\approx14.5$
times with respect to the standard global optimization ($\rightarrow\eta=93\%$
- Tab. IV). 

\noindent To give some insights towards the clinical exploitation
of the proposed imaging approach, let us analyze the spatial resolution
of the tumor localization by evaluating the following key-performance
indicator \cite{Reimer 2021}\begin{equation}
\zeta\triangleq\left\Vert \mathbf{r}_{\Psi}-\mathbf{r}_{\Psi}^{true}\right\Vert \end{equation}
$\mathbf{r}_{\Psi}$/$\mathbf{r}_{\Psi}^{true}$ being the retrieved/actual
tumor barycenter, respectively. Moreover, let the tumor detection
be successful (i.e., marked as {}``\emph{Yes}'' in Tab. V) when
the value of $\zeta$ complies with\begin{equation}
\zeta\leq\left(\rho_{\Psi}^{true}+\chi\right),\label{eq:detection rule}\end{equation}
$\chi=0.5$ {[}cm{]} being the measurement uncertainty in the position
of the tumor within the experimental breast phantom \cite{Reimer 2021}. 

\noindent For comparison purposes, Table V resumes the results from
the inversions with the \emph{SbD}, the \emph{EA}, and the \emph{DA}
along with those of other competitive alternatives in the available
literature (i.e., the \emph{DAS}, \emph{DMAS}, and \emph{ORR} methods
\cite{Reimer 2021}). Although all approaches positively operate since
(\ref{eq:detection rule}) is always verified, the \emph{SbD} and
the \emph{EA} provide the lowest errors (i.e., $\left.\zeta\right|_{SbD}=0.48\,[\mathrm{cm}]\simeq\left.\zeta\right|_{EA}=0.45\,[\mathrm{cm}]$
- Tab. V).

\noindent The last scenario is very challenging since it considers
the diagnosis of a significantly smaller tumor {[}i.e., $\rho_{\Psi}^{true}=0.75$
{[}cm{]} - Figs. 15(\emph{a})-15(\emph{b}){]}. Such a benchmark is
of particular biomedical interest since a $7$-year survival rate
of $84\%$ is expected for patients that are diagnosed in the early
stages of breast cancer (i.e., $\rho_{\Psi}^{true}\leq0.75$ {[}cm{]}
\cite{Veronesi 1990}), but unfortunately tumors of such a small size
are quite hard to detect because of their weak scattering signature
\cite{Reimer 2022 a}. The \emph{SbD} reconstruction in Figs. 15(\emph{c})-15(\emph{d})
allows one to suitably estimate both the tumor size and shape, but
there are some inaccuracies in its localization and composition that
are quantified by an increment of both $\Xi_{int}$ and $\zeta$ with
respect to the previous test case (i.e., $\left.\Xi_{ext}\right|_{\rho_{\Psi}^{true}=0.75\textrm{ [cm]}}^{SbD}/\left.\Xi_{ext}\right|_{\rho_{\Psi}^{true}=1.5\textrm{ [cm]}}^{SbD}\approx1.6$
and $\left.\zeta\right|_{\rho_{\Psi}^{true}=0.75\textrm{ [cm]}}^{SbD}/\left.\zeta\right|_{\rho_{\Psi}^{true}=1.5\textrm{ [cm]}}^{SbD}\approx2.1$
- Figs. 15(\emph{c})-15(\emph{d}) vs. Figs. 14(\emph{c})-14(\emph{d})
and Tabs. IV-V). On the other hand, only the \emph{SbD} and the \emph{EA}
prove to be able to detect the unknown pathology according to  (\ref{eq:detection rule})
(Tab. V).

\section{\noindent Conclusions\label{sec:Conclusions}}

\noindent A new physics-driven \emph{AI} approach for a reliable and
time-effective \emph{MI} of breast tumors has been presented\emph{.}
Based on a proper differential formulation of the biomedical \emph{ISP}
at hand to enable the exploitation of \emph{a-priori} physical knowledge
on the imaged domain as well as a compact yet flexible description
of the unknown pathology, an \emph{AI}-based strategy \cite{Salucci 2022b}
inspired by the concepts and guidelines of the \emph{SbD} paradigm
\cite{Salucci 2022}\cite{Massa 2022} has been proposed by reformulating
the inversion of differential scattering data as a global optimization
problem efficiently and reliably solved thanks to an adaptively-refined
\emph{SM} of the cost function. 

\noindent The numerical and experimental assessments (Sect. \ref{sec:results})
have shown that the proposed \emph{MI} method

\begin{itemize}
\item is able to retrieve faithful guesses of the breast with a degree of
accuracy comparable to a traditional global search based on \emph{EA}s,
but at a remarkably reduced computational cost being fast as a \emph{DA}-based
local search;
\item exhibits a remarkable robustness against the noise blurring the differential
scattering data;
\item reliably retrieves tumors embedded within breasts of different densities,
showing almost independent performance on the level of contrast between
abnormal and normal tissues;
\item properly performs also in case of an imperfect/approximated knowledge
of the reference (healthy) breast tissues and internal structure;
\item shows promising performance also when applied to real experimental
data by overcoming competitive approaches even more when imaging small
(i.e., early stages) tumoral pathologies. 
\end{itemize}
Future works, beyond the scope of this article, will be aimed at extending
the proposed imaging method to fully \emph{3-D} geometries as well
as to investigate proper customizations to other biomedical applications
of undoubted interest (e.g., brain stroke imaging).

\section*{\noindent Appendix A\label{sec:Appendix-A}}

\noindent To derive the expression for the inhomogeneous internal
Green's matrix $\underline{\underline{G}}_{\mathcal{N}}^{\Omega}$,
equation (\ref{eq:LSIE state ref}) is first subtracted from (\ref{eq:LSIE state total}).
Then, by exploiting the definition of $D^{v}\left(\mathbf{r}\right)$
and $\tau_{\Delta}\left(\mathbf{r}\right)$, one obtains\begin{equation}
\begin{array}{r}
D^{v}\left(\mathbf{r}\right)=\int_{\Omega}G_{\mathcal{B}}\left(\mathbf{r},\,\mathbf{r}^{\prime}\right)\left[\tau_{\Delta}\left(\mathbf{r}^{\prime}\right)E^{v}\left(\mathbf{r}^{\prime}\right)+\tau_{\mathcal{N}}\left(\mathbf{r}^{\prime}\right)D^{v}\left(\mathbf{r}^{\prime}\right)\right]d\mathbf{r}^{\prime}\\
\mathbf{r}\in\Omega;\, v=1,...,V\end{array}\end{equation}
that in matrix notation becomes\begin{equation}
\underline{D}^{\Omega,v}=\underline{\underline{G}}_{\mathcal{B}}^{\Omega}\,\underline{J}_{\Delta}^{v}+\underline{\underline{G}}_{\mathcal{B}}^{\Omega}\,\underline{\underline{\tau}}_{\mathcal{N}}\,\underline{D}^{\Omega,v};\,\,\,\,\, v=1,...,V.\label{eq:D_Omega_v}\end{equation}
By simple algebraic manipulations on (\ref{eq:D_Omega_v}), it turns
out that \begin{equation}
\underline{D}^{\Omega,v}=\left\{ \left[\underline{\underline{\mathbb{I}}}-\underline{\underline{G}}_{\mathcal{B}}^{\Omega}\,\underline{\underline{\tau}}_{\mathcal{N}}\right]^{-1}\underline{\underline{G}}_{\mathcal{B}}^{\Omega}\right\} \,\underline{J}_{\Delta}^{v};\,\,\,\,\, v=1,...,V\label{eq:explicit discrete state diff}\end{equation}
so that it is possible to derive the definition of $\underline{\underline{G}}_{\mathcal{N}}^{\Omega}$
in (\ref{int green inhomog}).

\noindent Similarly, equation (\ref{eq:LSIE data ref}) is subtracted
from (\ref{eq:LSIE data total}) to yield\begin{equation}
\begin{array}{r}
D^{v}\left(\mathbf{r}_{m}^{v}\right)=\int_{\Omega}G_{\mathcal{B}}\left(\mathbf{r}_{m}^{v},\,\mathbf{r}^{\prime}\right)\left[\tau_{\Delta}\left(\mathbf{r}^{\prime}\right)E^{v}\left(\mathbf{r}^{\prime}\right)+\tau_{\mathcal{N}}\left(\mathbf{r}^{\prime}\right)D^{v}\left(\mathbf{r}^{\prime}\right)\right]d\mathbf{r}^{\prime}\\
\mathbf{r}_{m}^{v}\in\Theta;\, v=1,...,V;\, m=1,...,M\end{array}\end{equation}
that after discretization becomes\begin{equation}
\underline{D}^{\Theta,v}=\underline{\underline{G}}_{\mathcal{B}}^{\Theta,v}\,\left[\underline{J}_{\Delta}^{v}+\underline{\underline{\tau}}_{\mathcal{N}}\,\underline{D}^{\Omega,v}\right];\,\,\,\,\, v=1,...,V.\label{eq:D_Theta_v}\end{equation}
Finally, by substituting $\underline{D}^{\Omega,v}$ with (\ref{eq:discrete state diff})
in (\ref{eq:D_Theta_v}), we obtain\begin{equation}
\underline{D}^{\Theta,v}=\left\{ \underline{\underline{G}}_{\mathcal{B}}^{\Theta,v}\,\left[\underline{\underline{\mathbb{I}}}+\underline{\underline{\tau}}_{\mathcal{N}}\,\underline{\underline{G}}_{\mathcal{N}}^{\Omega}\right]\right\} \,\underline{J}_{\Delta}^{v};\,\,\,\,\, v=1,...,V\end{equation}
to deduce the definition of the inhomogeneous external Green's matrix
$\underline{\underline{G}}_{\mathcal{N}}^{\Theta}$ in (\ref{ext green inhomog}).

\section*{\noindent Acknowledgments}

\noindent This work benefited from the networking activities carried
out within the Project {}``ICSC National Centre for HPC, Big Data
and Quantum Computing (CN HPC)'' funded by the European Union - NextGenerationEU
within the PNRR Program (CUP: E63C22000970007), the Project ''AURORA
- Smart Materials for Ubiquitous Energy Harvesting, Storage, and Delivery
in Next Generation Sustainable Environments'' funded by the Italian
Ministry for Universities and Research within the PRIN-PNRR 2022 Program
(CUP: E53D23014760001), the Project DICAM-EXC (Grant L232/2016) funded
by the Italian Ministry of Education, Universities and Research (MUR)
within the {}``Departments of Excellence 2023-2027'' Program (CUP:
E63C22003880001), and the Project {}``SPEED'' (Grant No. 6721001)
funded by National Science Foundation of China under the Chang-Jiang
Visiting Professorship Program. Views and opinions expressed are however
those of the author(s) only and do not necessarily reflect those of
the European Union or the European Research Council. Neither the European
Union nor the granting authority can be held responsible for them.
A. Massa wishes to thank E. Vico for her never-ending inspiration,
support, guidance, and help.

\newpage
\section*{FIGURE CAPTIONS}

\begin{itemize}
\item \textbf{Figure 1.} Sketch of the geometry of (\emph{a}) the \emph{2-D}
microwave breast imaging setup and (\emph{b}) the \emph{Actual}, the
\emph{Reference}, and the \emph{Differential} scattering scenarios.
\item \textbf{Figure 2.} \emph{Illustrative Example} (\emph{''Ideal''
Breast Phantom}, $\varepsilon_{\mathcal{B}}=22.4$, $\sigma_{\mathcal{B}}=1.26$
{[}S/m{]}, $\varepsilon_{\mathcal{N}}=16.5$, $\sigma_{\mathcal{N}}=0.60$
{[}S/m{]}, $\varepsilon_{\Psi}=59.3$, $\sigma_{\Psi}=1.54$ {[}S/m{]})
- Distributions of the relative permittivity (\emph{a})(\emph{c})
and the conductivity (\emph{b})(\emph{d}) for (\emph{a})(\emph{b})
the \emph{Reference} and (\emph{c})(\emph{d}) the \emph{Actual} scenarios.
\item \textbf{Figure 3.} \emph{Illustrative Example} (\emph{''Ideal''
Breast Phantom}, $f=1.3$ {[}GHz{]}, $\varepsilon_{\mathcal{B}}=22.4$,
$\sigma_{\mathcal{B}}=1.26$ {[}S/m{]}, $\varepsilon_{\mathcal{N}}=16.5$,
$\sigma_{\mathcal{N}}=0.60$ {[}S/m{]}, $\varepsilon_{\Psi}=59.3$,
$\sigma_{\Psi}=1.54$ {[}S/m{]}, $SNR=100$ {[}dB{]}) - Plots of (\emph{a})
the evolution of the differential data cost function, $\Phi_{i}$,
versus the iteration index, $i$, and (\emph{b}) the total reconstruction
error, $\Xi_{tot}$ vs. execution time, $\Delta t$.
\item \textbf{Figure 4.} \emph{Illustrative Example} (\emph{''Ideal''
Breast Phantom}, $f=1.3$ {[}GHz{]}, $\varepsilon_{\mathcal{B}}=22.4$,
$\sigma_{\mathcal{B}}=1.26$ {[}S/m{]}, $\varepsilon_{\mathcal{N}}=16.5$,
$\sigma_{\mathcal{N}}=0.60$ {[}S/m{]}, $\varepsilon_{\Psi}=59.3$,
$\sigma_{\Psi}=1.54$ {[}S/m{]}, $SNR=100$ {[}dB{]}) - Profiles of
(\emph{a})(\emph{c})(\emph{e}) the relative permittivity and (\emph{b})(\emph{d})(\emph{f})
the conductivity retrieved by (\emph{a})(\emph{b}) the \emph{SbD},
(\emph{c})(\emph{d}) the \emph{EA}, and (\emph{e})(\emph{f}) the \emph{DA}
methods.
\item \textbf{Figure 5.} \emph{Numerical Results} (\emph{{}``Segmented''
Breast Phantom}, \emph{XD} \emph{Tissues} (Tab. I), $f=1.3$ {[}GHz{]},
$\varepsilon_{\Psi}=59.3$, $\sigma_{\Psi}=1.54$ {[}S/m{]}, $SNR=100$
{[}dB{]}) - Profiles of (\emph{a})(\emph{c}) the relative permittivity
and (\emph{b})(\emph{d}) the conductivity of the (\emph{a})(\emph{b})
actual and (\emph{c})(\emph{d}) the \emph{SbD}-retrieved scenarios.
\item \textbf{Figure 6.} \emph{Numerical Results} (\emph{{}``Segmented''
Breast Phantom}, \emph{XD} \emph{Tissues} (Tab. I), $f=1.3$ {[}GHz{]},
$\varepsilon_{\Psi}=59.3$, $\sigma_{\Psi}=1.54$ {[}S/m{]}, $SNR\in\left[10,\,100\right]$
{[}dB{]}) - Plots of (\emph{a}) the total, (\emph{b}) the internal,
(\emph{c}) the external reconstruction errors, and (\emph{d}) the
execution time as a function of the \emph{SNR}.
\item \textbf{Figure 7.} \emph{Numerical Results} (\emph{{}``Segmented''
Breast Phantom}, \emph{XD} \emph{Tissues} (Tab. I), $f=1.3$ {[}GHz{]},
$\varepsilon_{\Psi}=59.3$, $\sigma_{\Psi}=1.54$ {[}S/m{]}, $SNR=20$
{[}dB{]}) - Profiles of (\emph{a})(\emph{c})(\emph{e}) the relative
permittivity and (\emph{b})(\emph{d})(\emph{f}) the conductivity retrieved
by the (\emph{a})(\emph{b}) \emph{SbD}, (\emph{c})(\emph{d}) the \emph{EA},
and (\emph{e})(\emph{f}) the \emph{DA} methods.
\item \textbf{Figure 8.} \emph{Numerical Results} (\emph{{}``Segmented''
Breast Phantom}, $f=1.3$ {[}GHz{]}, $\varepsilon_{\Psi}=59.3$, $\sigma_{\Psi}=1.54$
{[}S/m{]}, $SNR=100$ {[}dB{]}; \emph{SbD} inversion) - Total, internal,
and external reconstruction errors.
\item \textbf{Figure 9.} \emph{Numerical Results} (\emph{{}``Segmented''
Breast Phantom}, $f=1.3$ {[}GHz{]}, $\varepsilon_{\Psi}=59.3$, $\sigma_{\Psi}=1.54$
{[}S/m{]}, $SNR=100$ {[}dB{]}) - \textbf{}Maps of (\emph{a})(\emph{c})(\emph{e})(\emph{g})(\emph{i})(\emph{m})
the actual and (\emph{b})(\emph{d})(\emph{f})(\emph{h})(\emph{l})(\emph{n})
the retrieved profiles of the (\emph{a})(\emph{b})(\emph{e})(\emph{f})(\emph{i})(\emph{l})
relative permittivity and (\emph{c})(\emph{d})(\emph{g})(\emph{h})(\emph{m})(\emph{n})
conductivity when imaging breast phantoms belonging to the (\emph{a})-(\emph{d})
{}``fatty'', (\emph{e})-(\emph{h}) {}``scattered'', and (\emph{i})-(\emph{n})
\emph{{}``HD''} categories.
\item \textbf{Figure 10.} \emph{Numerical Results} (\emph{{}``Segmented''
Breast Phantom}, \emph{XD} \emph{Tissues} (Tab. I), $f=1.3$ {[}GHz{]},
$\varepsilon_{\Psi}^{\left(nom\right)}=59.3$, $\sigma_{\Psi}^{\left(nom\right)}=1.54$
{[}S/m{]}, $SNR=100$ {[}dB{]}) - Total, internal, and external reconstruction
error versus $\delta$.
\item \textbf{Figure 11.} \emph{Numerical Results} (\emph{{}``Segmented''
Breast Phantom}, \emph{XD} \emph{Tissues} (Tab. I), $f=1.3$ {[}GHz{]},
$\varepsilon_{\Psi}^{\left(nom\right)}=59.3$, $\sigma_{\Psi}^{\left(nom\right)}=1.54$
{[}S/m{]}, $SNR=100$ {[}dB{]}) - Maps of (\emph{a})(\emph{b})(\emph{e})(\emph{f})
the actual and (\emph{c})(\emph{d})(\emph{g})(\emph{h}) the reconstructed
profiles of the (\emph{a})-(\emph{d}) relative permittivity and (\emph{e})-(\emph{h})
conductivity when (\emph{a})(\emph{c})(\emph{e})(\emph{g}) $\delta=-10\%$
and (\emph{b})(\emph{d})(\emph{f})(\emph{h}) $\delta=+10\%$.
\item \textbf{Figure 12.} \emph{Numerical Results} (\emph{{}``Full'' Breast
Phantom}, \emph{Scattered} \emph{Tissues}) - Maps of (\emph{a})(\emph{c})(\emph{e})
the relative permittivity and (\emph{b})(\emph{d})(\emph{f}) the conductivity
profiles for the (\emph{a})(\emph{b}) full, (\emph{c})(\emph{d}) segmented,
and (\emph{e})(\emph{f}) constant \emph{prior} knowledge on the reference
scenario.
\item \textbf{Figure 13.} \emph{Numerical Results} (\emph{{}``Full'' Breast
Phantom}, \emph{Scattered} \emph{Tissues} (Tab. I), $f=1.3$ {[}GHz{]},
$\varepsilon_{\Psi}=59.3$, $\sigma_{\Psi}=1.54$ {[}S/m{]}, $SNR=100$
{[}dB{]}) - Maps of (\emph{a})(\emph{b}) the actual and (\emph{c})-(\emph{h})
the retrieved (\emph{a})(\emph{c})(\emph{e})(\emph{g}) relative permittivity
and (\emph{b})(\emph{d})(\emph{f})(\emph{h}) conductivity profiles
when considering the (\emph{c})(\emph{d}) full (\emph{FP}), (\emph{e})(\emph{f})
segmented (\emph{SP}), or (\emph{g})(\emph{h}) constant (\emph{CP})
\emph{prior} knowledge on the reference scenario.
\item \textbf{Figure 14.} \emph{Experimental Results} (\emph{UM-BMID Dataset}
\cite{Reimer 2021}\cite{Reimer 2020}, $\rho_{\Psi}=1.50$ {[}cm{]},
$f=2.0$ {[}GHz{]}) - Maps of (\emph{a})(\emph{b}) the actual and
(\emph{c})-(\emph{h}) the retrieved (\emph{a})(\emph{c})(\emph{e})(\emph{g})
relative permittivity and (\emph{b})(\emph{d})(\emph{f})(\emph{h})
conductivity profiles by the (\emph{c})(\emph{d}) \emph{SbD}, (\emph{e})(\emph{f})
\emph{EA}, and (\emph{g})(\emph{h}) \emph{DA}.
\item \textbf{Figure 15.} \emph{Experimental Results} (\emph{UM-BMID Dataset}
\cite{Reimer 2021}\cite{Reimer 2020}, $\rho_{\Psi}=0.75$ {[}cm{]},
$f=2.0$ {[}GHz{]}) - Maps of (\emph{a})(\emph{b}) the actual and
(\emph{c})-(\emph{h}) the retrieved (\emph{a})(\emph{c})(\emph{e})(\emph{g})
relative permittivity and (\emph{b})(\emph{d})(\emph{f})(\emph{h})
conductivity profiles by the (\emph{c})(\emph{d}) \emph{SbD}, (\emph{e})(\emph{f})
\emph{EA}, and (\emph{g})(\emph{h}) \emph{DA}.
\end{itemize}

\section*{TABLE CAPTIONS}

\begin{itemize}
\item \textbf{Table I.} \emph{Numerical Results} (\emph{Segmented Breast
Phantom}, $f=1.3$ {[}GHz{]}) - Description of the \emph{MRI}-derived
piece-wise constant phantoms \cite{Golnabi 2019}.
\item \textbf{Table II.} \emph{Numerical Results} (\emph{{}``Full'' Breast
Phantom}, \emph{Scattered} \emph{Tissues}, $f=1.3$ {[}GHz{]}, $\varepsilon_{\Psi}=59.3$,
$\sigma_{\Psi}=1.54$ {[}S/m{]}, $SNR=100$ {[}dB{]}) - Reconstruction
errors and execution time.
\item \textbf{Table III.} \emph{Numerical Results} (\emph{{}``Full'' Breast
Phantom}, \emph{Scattered} \emph{Tissues}, $f=1.3$ {[}GHz{]}, $\varepsilon_{\Psi}=59.3$,
$\sigma_{\Psi}=1.54$ {[}S/m{]}, $SNR=100$ {[}dB{]}; \emph{SbD} inversion)
- Error metrics.
\item \textbf{Table IV.} \emph{Experimental Results} (\emph{UM-BMID Dataset}
\cite{Reimer 2021}\cite{Reimer 2020}, $f=2.0$ {[}GHz{]}) - Reconstruction
errors and execution time.
\item \textbf{Table V.} \emph{Experimental Results} (\emph{UM-BMID Dataset}
\cite{Reimer 2021}\cite{Reimer 2020}, $f=2.0$ {[}GHz{]}) - Breast
tumor detection/localization indexes.
\end{itemize}
\newpage
\begin{center}~\vfill\end{center}

\begin{center}\begin{tabular}{c}
\includegraphics[%
  width=0.80\textwidth,
  keepaspectratio]{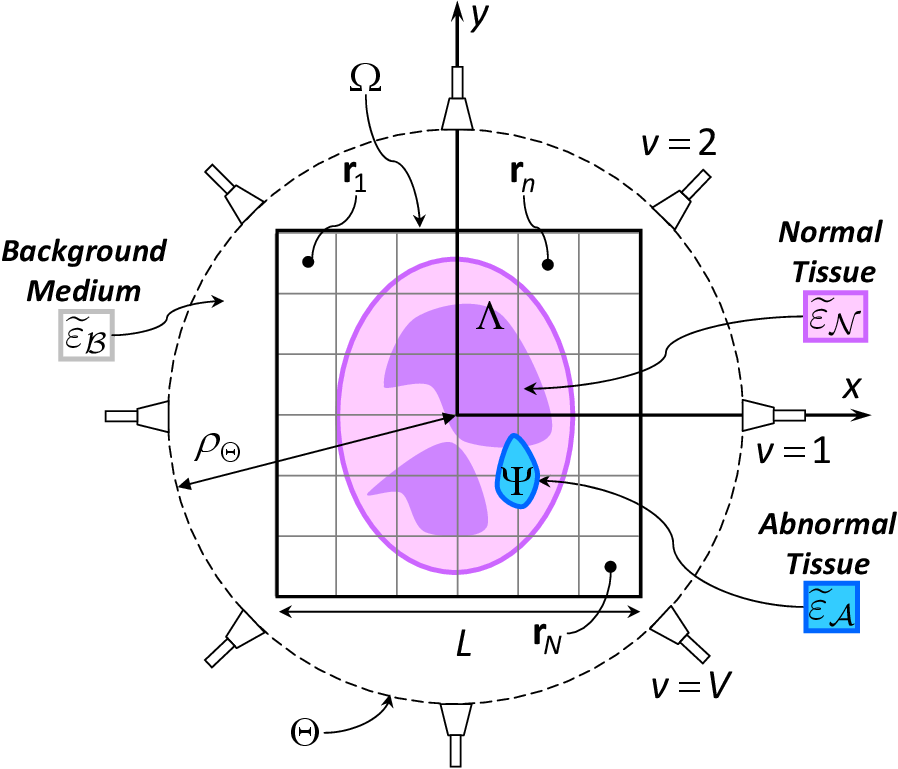}\tabularnewline
(\emph{a})\tabularnewline
\tabularnewline
\includegraphics[%
  width=0.95\textwidth]{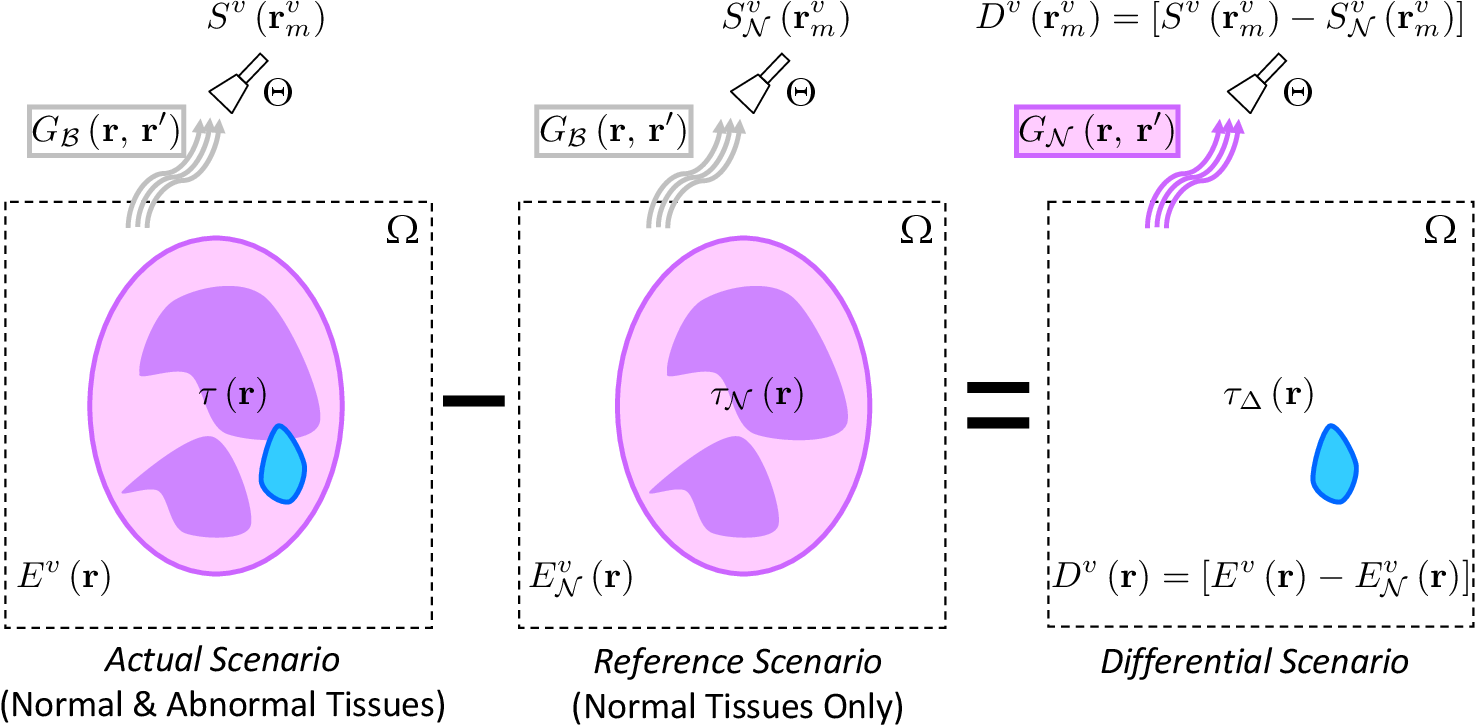}\tabularnewline
(\emph{b})\tabularnewline
\end{tabular}\end{center}

\begin{center}~\vfill\end{center}

\begin{center}\textbf{Fig. 1 - F. Zardi et} \textbf{\emph{al.}}\textbf{,}
\textbf{\emph{{}``}}A Physics-Driven \emph{AI} Approach ...''\end{center}

\newpage
\begin{center}~\vfill\end{center}

\begin{center}\begin{tabular}{cc}
\includegraphics[%
  width=0.45\columnwidth]{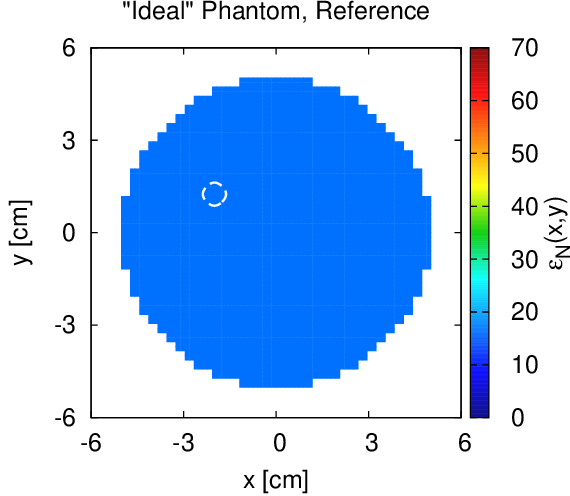}&
\includegraphics[%
  width=0.45\columnwidth]{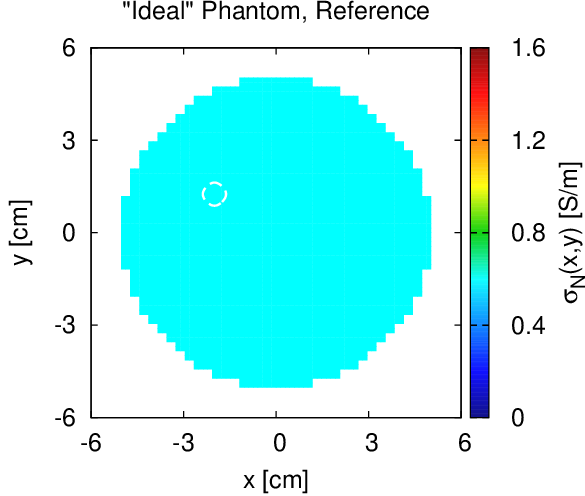}\tabularnewline
(\emph{a})&
(\emph{b})\tabularnewline
&
\tabularnewline
\includegraphics[%
  width=0.45\columnwidth]{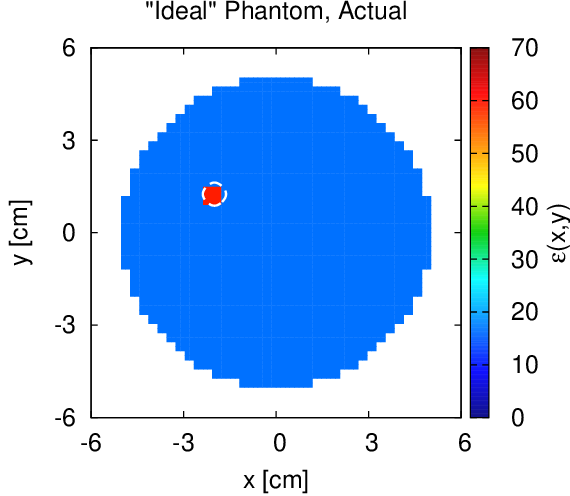}&
\includegraphics[%
  width=0.45\columnwidth]{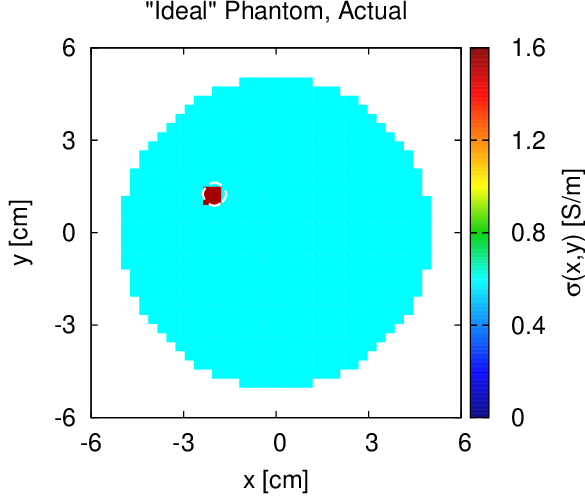}\tabularnewline
(\emph{c})&
(\emph{d})\tabularnewline
\end{tabular}\end{center}

\begin{center}~\vfill\end{center}

\begin{center}\textbf{Fig. 2 - F. Zardi et} \textbf{\emph{al.}}\textbf{,}
\textbf{\emph{{}``}}A Physics-Driven \emph{AI} Approach ...''\end{center}

\newpage
\begin{center}~\vfill\end{center}

\begin{center}\begin{tabular}{c}
\includegraphics[%
  width=0.75\textwidth]{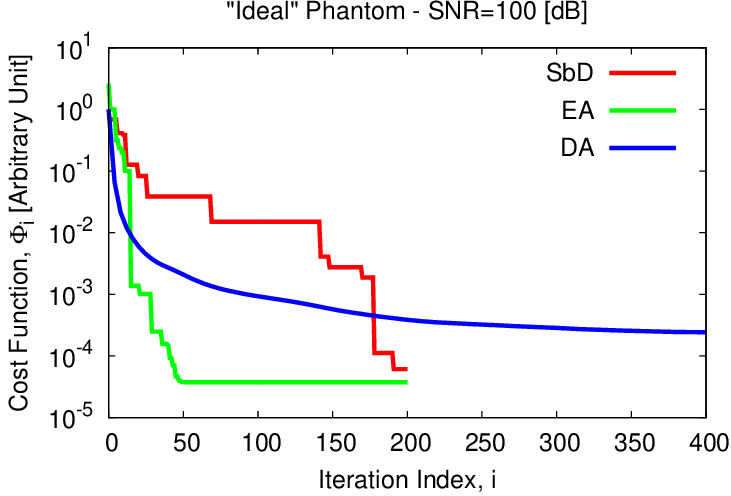}\tabularnewline
(\emph{a})\tabularnewline
\tabularnewline
\includegraphics[%
  width=0.75\textwidth]{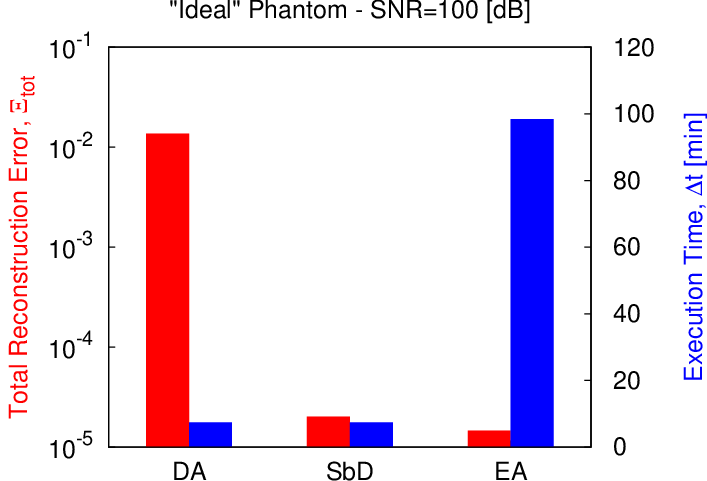}\tabularnewline
(\emph{b})\tabularnewline
\end{tabular}\end{center}

\begin{center}~\vfill\end{center}

\begin{center}\textbf{Fig. 3 - F. Zardi et} \textbf{\emph{al.}}\textbf{,}
\textbf{\emph{{}``}}A Physics-Driven \emph{AI} Approach ...''\end{center}

\newpage
\begin{center}~\vfill\end{center}

\begin{center}\begin{tabular}{ccc}
\begin{sideways}
\emph{~~~~~~~~~~~~~~~~~~~~~~~~SbD}%
\end{sideways}&
\includegraphics[%
  width=0.40\columnwidth]{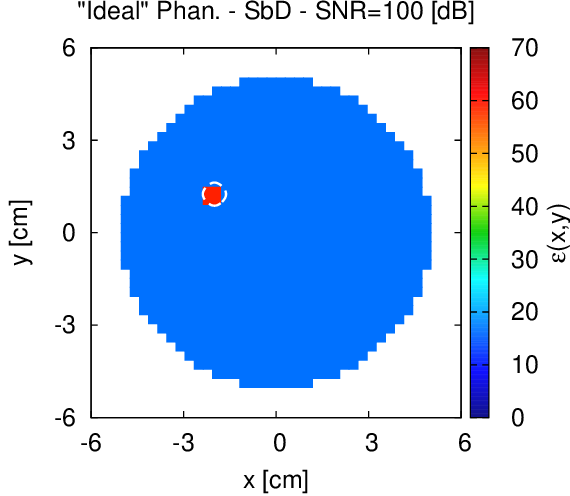}&
\includegraphics[%
  width=0.40\columnwidth]{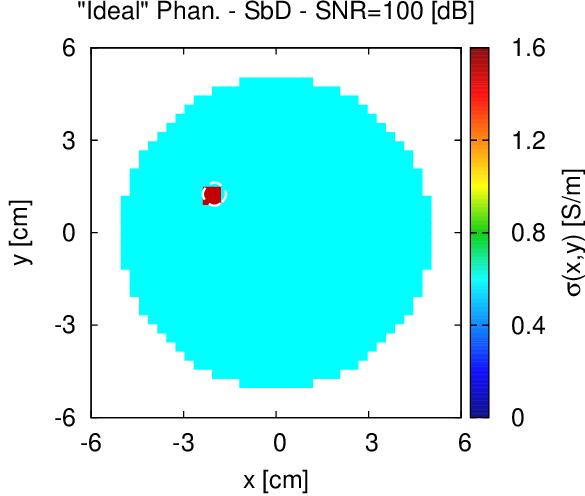}\tabularnewline
\begin{sideways}
\end{sideways}&
(\emph{a})&
(\emph{b})\tabularnewline
\begin{sideways}
\end{sideways}&
&
\tabularnewline
\begin{sideways}
\emph{~~~~~~~~~~~~~~~~~~~~~~~~~EA}%
\end{sideways}&
\includegraphics[%
  width=0.40\columnwidth]{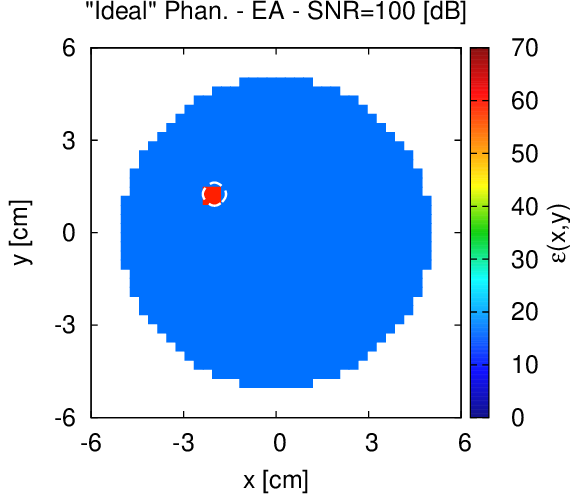}&
\includegraphics[%
  width=0.40\columnwidth]{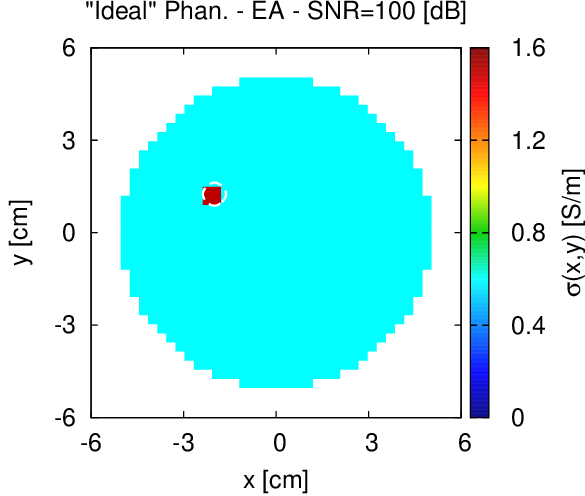}\tabularnewline
\begin{sideways}
\end{sideways}&
(\emph{c})&
(\emph{d})\tabularnewline
\begin{sideways}
\end{sideways}&
&
\tabularnewline
\begin{sideways}
\emph{~~~~~~~~~~~~~~~~~~~~~~~~~DA}%
\end{sideways}&
\includegraphics[%
  width=0.40\columnwidth]{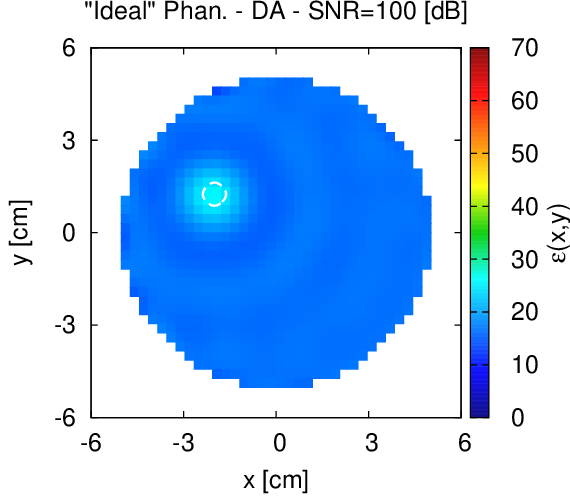}&
\includegraphics[%
  width=0.40\columnwidth]{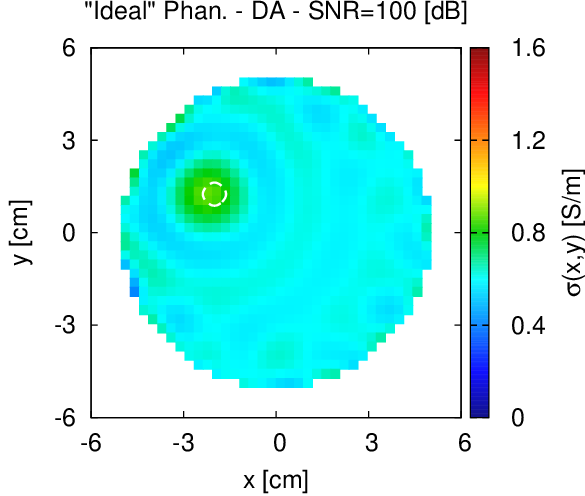}\tabularnewline
\begin{sideways}
\end{sideways}&
(\emph{e})&
(\emph{f})\tabularnewline
\end{tabular}\end{center}

\begin{center}~\vfill\end{center}

\begin{center}\textbf{Fig. 4 - F. Zardi et} \textbf{\emph{al.}}\textbf{,}
\textbf{\emph{{}``}}A Physics-Driven \emph{AI} Approach ...''\end{center}

\newpage
\begin{center}~\vfill\end{center}

\begin{center}\begin{tabular}{cc}
\includegraphics[%
  width=0.45\columnwidth]{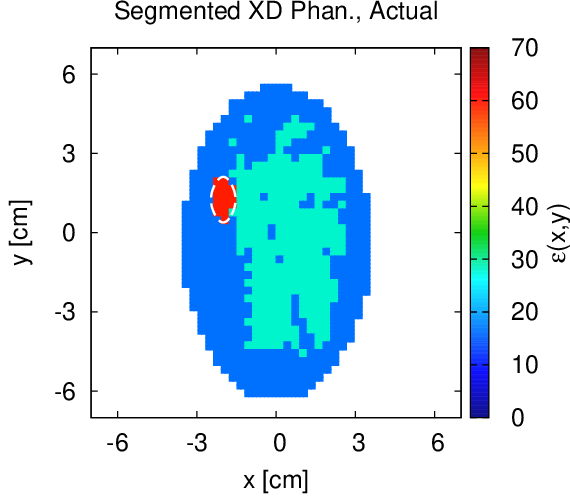}&
\includegraphics[%
  width=0.45\columnwidth]{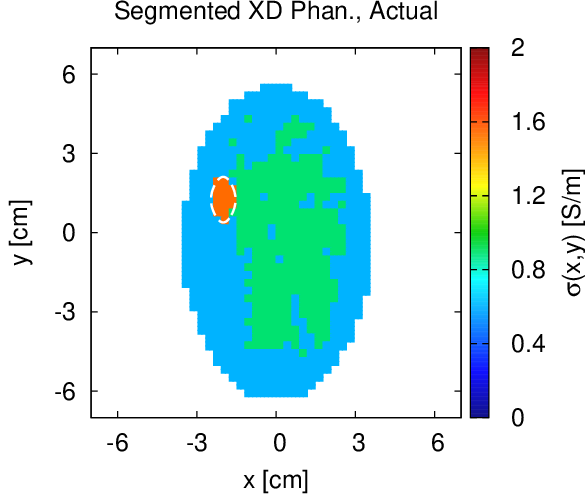}\tabularnewline
(\emph{a})&
(\emph{b})\tabularnewline
&
\tabularnewline
\includegraphics[%
  width=0.45\columnwidth]{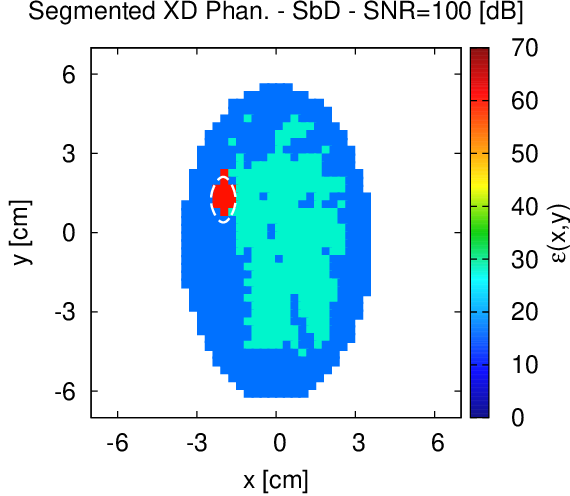}&
\includegraphics[%
  width=0.45\columnwidth]{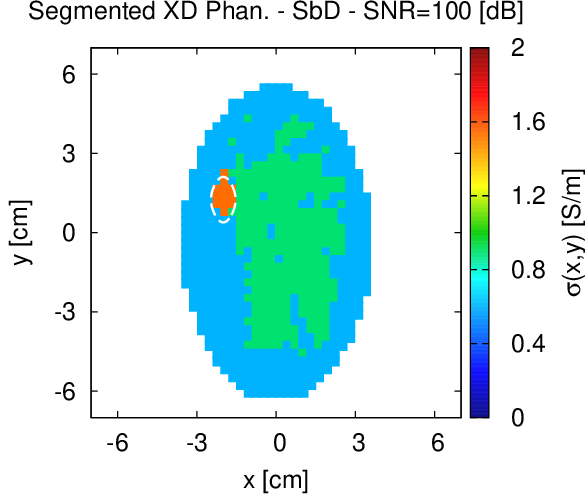}\tabularnewline
(\emph{c})&
(\emph{d})\tabularnewline
\end{tabular}\end{center}

\begin{center}~\vfill\end{center}

\begin{center}\textbf{Fig. 5 - F. Zardi et} \textbf{\emph{al.}}\textbf{,}
\textbf{\emph{{}``}}A Physics-Driven \emph{AI} Approach ...''\end{center}

\newpage
\begin{center}~\vfill\end{center}

\begin{center}\begin{tabular}{cc}
\includegraphics[%
  width=0.45\columnwidth]{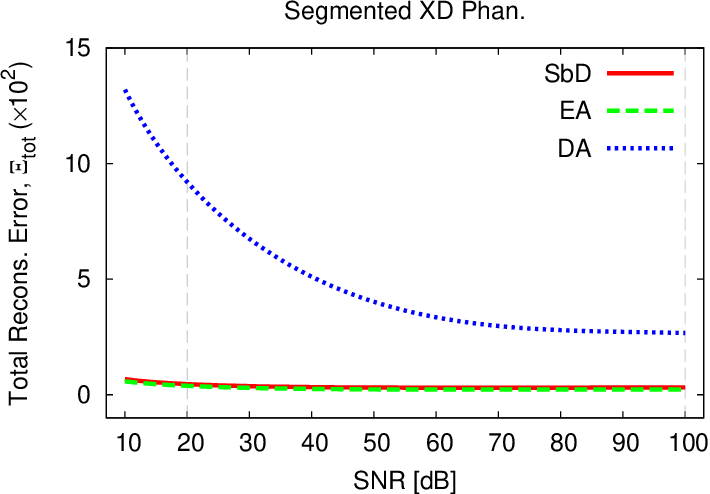}&
\includegraphics[%
  width=0.45\columnwidth]{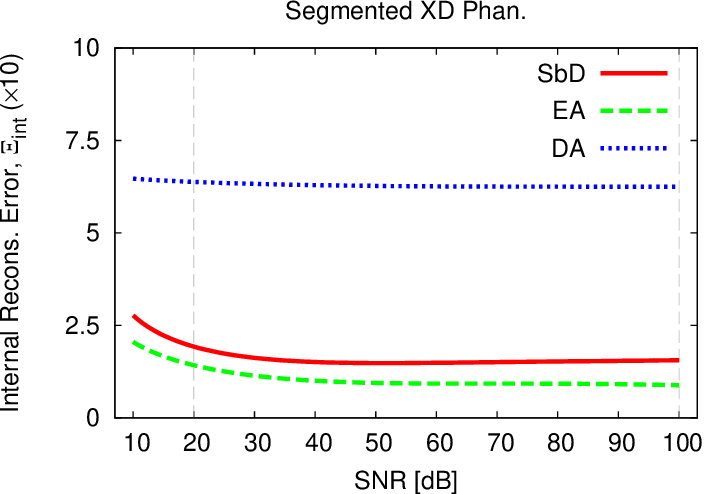}\tabularnewline
(\emph{a})&
(\emph{b})\tabularnewline
&
\tabularnewline
\includegraphics[%
  width=0.45\columnwidth]{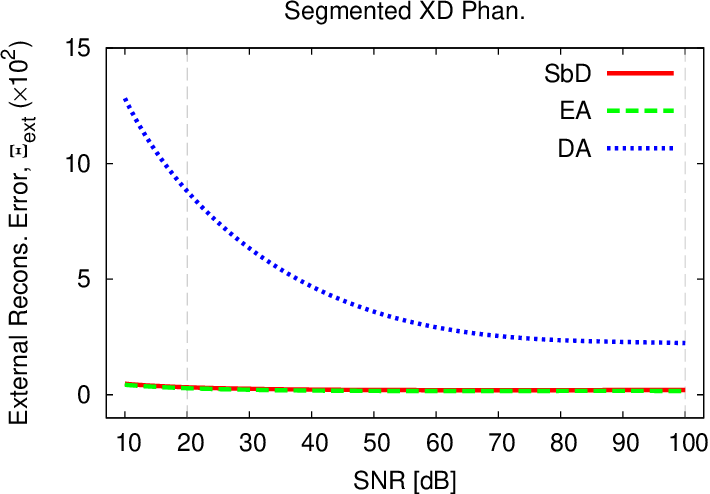}&
\includegraphics[%
  width=0.45\columnwidth]{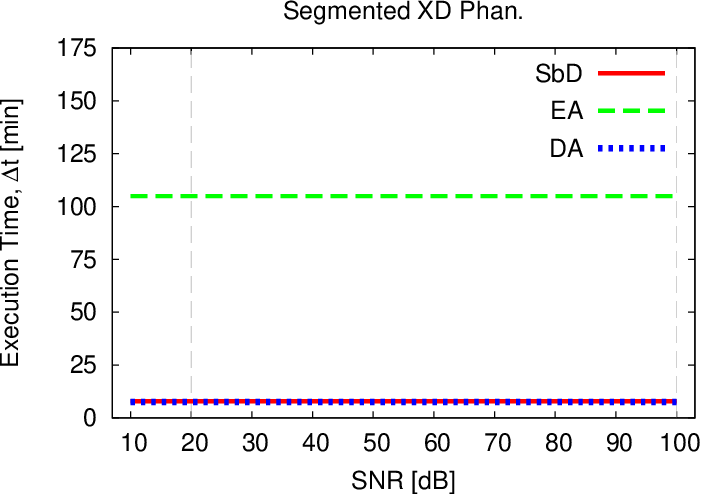}\tabularnewline
(\emph{c})&
(\emph{d})\tabularnewline
\end{tabular}\end{center}

\begin{center}~\vfill\end{center}

\begin{center}\textbf{Fig. 6 - F. Zardi et} \textbf{\emph{al.}}\textbf{,}
\textbf{\emph{{}``}}A Physics-Driven \emph{AI} Approach ...''\end{center}

\newpage
\begin{center}~\vfill\end{center}

\begin{center}\begin{tabular}{ccc}
\begin{sideways}
\emph{~~~~~~~~~~~~~~~~~~~~~~~~SbD}%
\end{sideways}&
\includegraphics[%
  width=0.40\columnwidth]{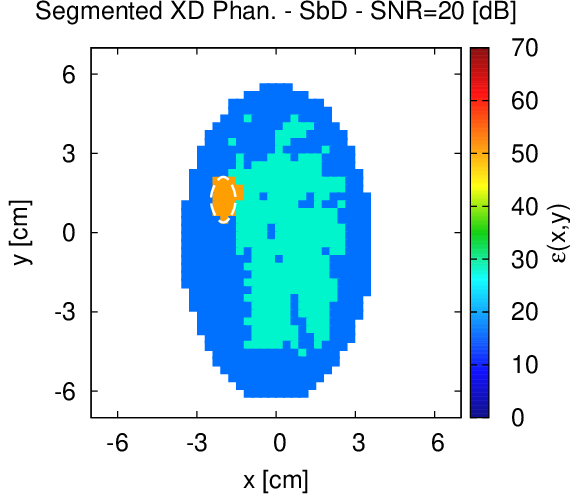}&
\includegraphics[%
  width=0.40\columnwidth]{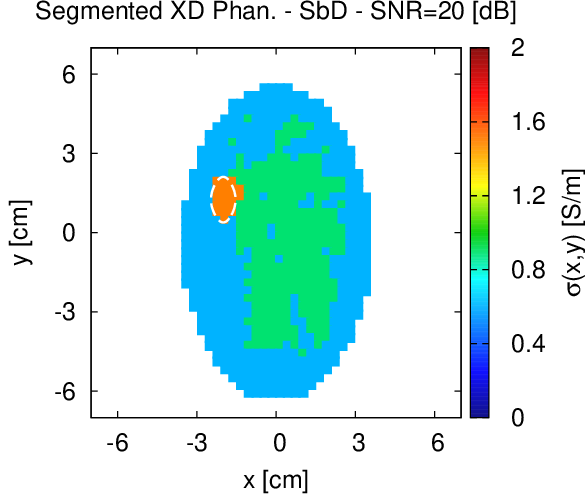}\tabularnewline
\begin{sideways}
\end{sideways}&
(\emph{a})&
(\emph{b})\tabularnewline
\begin{sideways}
\end{sideways}&
&
\tabularnewline
\begin{sideways}
\emph{~~~~~~~~~~~~~~~~~~~~~~~~~EA}%
\end{sideways}&
\includegraphics[%
  width=0.40\columnwidth]{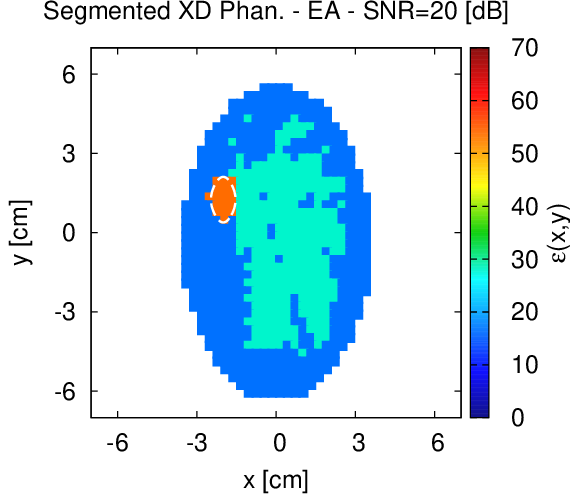}&
\includegraphics[%
  width=0.40\columnwidth]{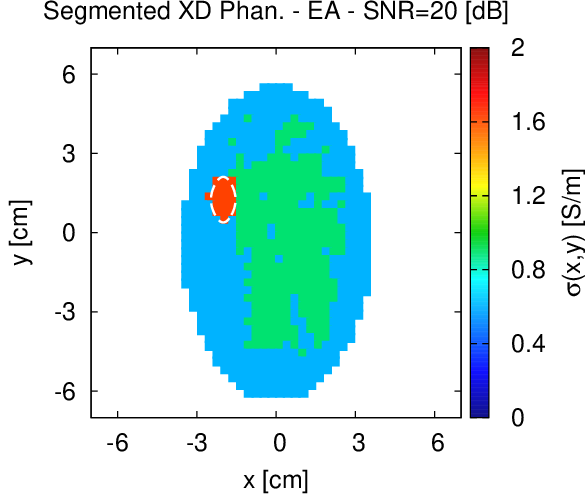}\tabularnewline
\begin{sideways}
\end{sideways}&
(\emph{c})&
(\emph{d})\tabularnewline
\begin{sideways}
\end{sideways}&
&
\tabularnewline
\begin{sideways}
\emph{~~~~~~~~~~~~~~~~~~~~~~~~~DA}%
\end{sideways}&
\includegraphics[%
  width=0.40\columnwidth]{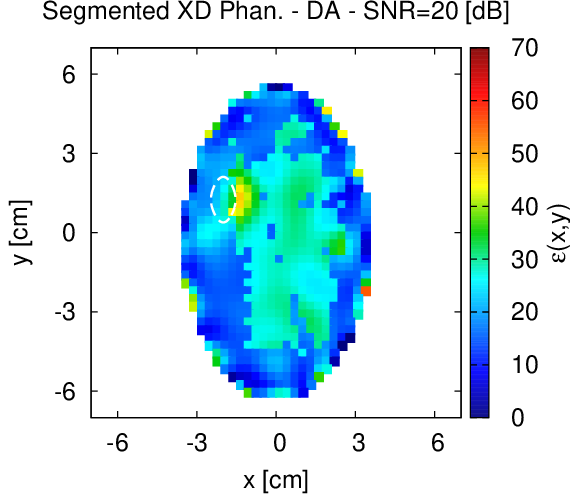}&
\includegraphics[%
  width=0.40\columnwidth]{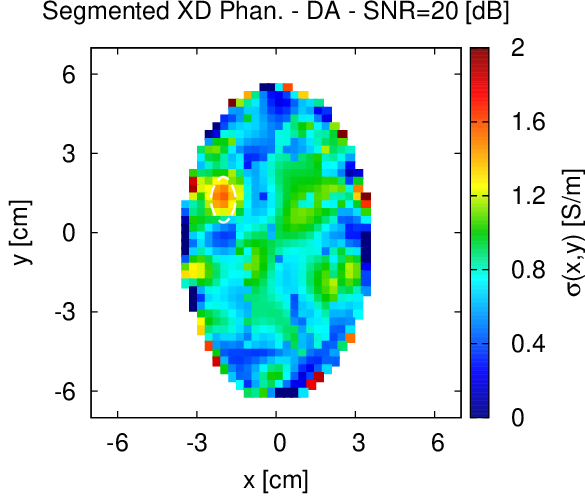}\tabularnewline
\begin{sideways}
\end{sideways}&
(\emph{e})&
(\emph{f})\tabularnewline
\end{tabular}\end{center}

\begin{center}~\vfill\end{center}

\begin{center}\textbf{Fig. 7 - F. Zardi et} \textbf{\emph{al.}}\textbf{,}
\textbf{\emph{{}``}}A Physics-Driven \emph{AI} Approach ...''\end{center}

\newpage
\begin{center}~\vfill\end{center}

\begin{center}\includegraphics[%
  width=0.75\columnwidth]{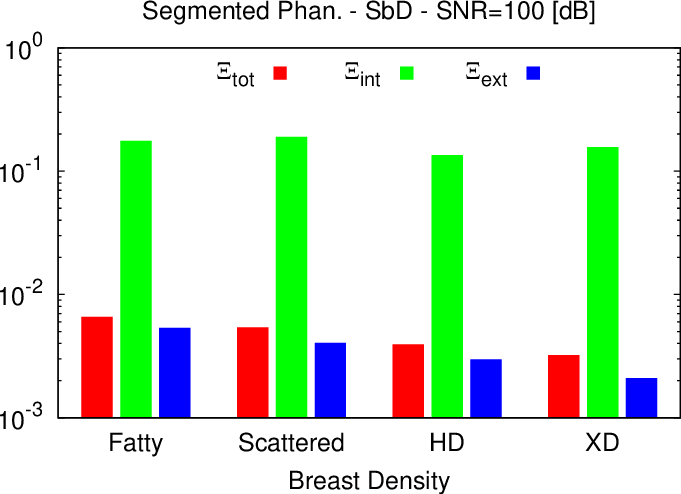}\end{center}

\begin{center}~\vfill\end{center}

\begin{center}\textbf{Fig. 8 - F. Zardi et} \textbf{\emph{al.}}\textbf{,}
\textbf{\emph{{}``}}A Physics-Driven \emph{AI} Approach ...''\end{center}

\newpage
\begin{center}~\vfill\end{center}

\begin{center}\begin{sideways}
\begin{tabular}{ccccc}
\begin{sideways}
\emph{~~~~~~~~~~~~~~~~Fatty}%
\end{sideways}&
\includegraphics[%
  width=0.30\columnwidth]{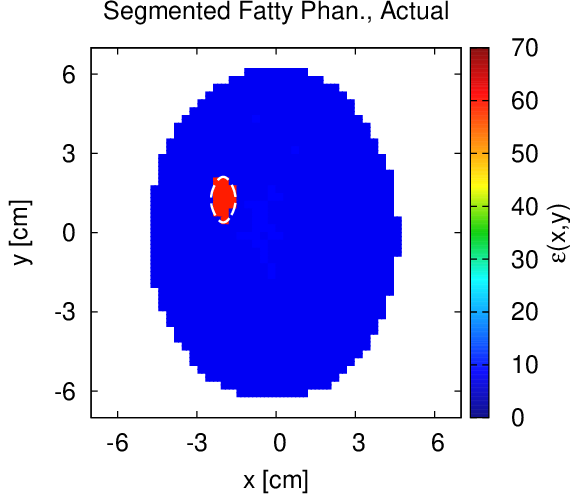}&
\includegraphics[%
  width=0.30\columnwidth]{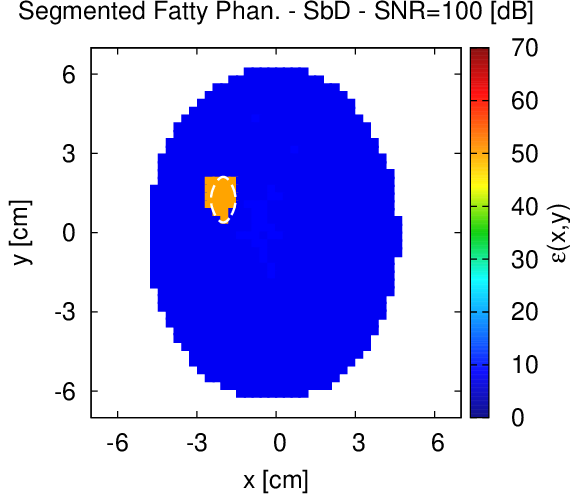}&
\includegraphics[%
  width=0.30\columnwidth]{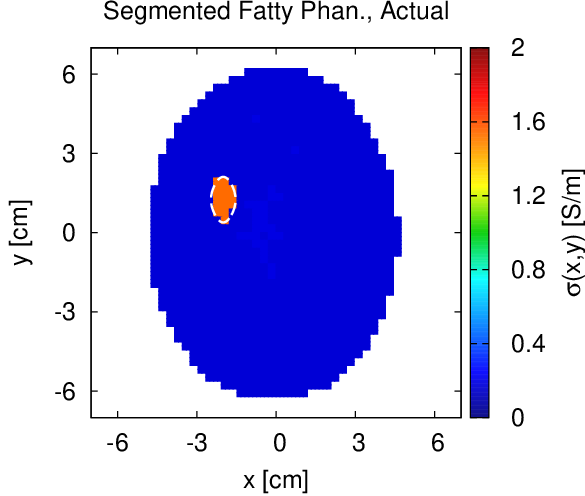}&
\includegraphics[%
  width=0.30\columnwidth]{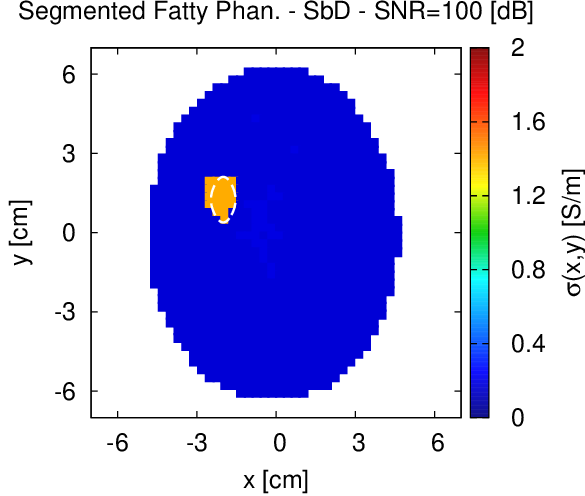}\tabularnewline
\begin{sideways}
\end{sideways}&
(\emph{a})&
(\emph{b})&
(\emph{c})&
(\emph{d})\tabularnewline
\begin{sideways}
\emph{~~~~~~~~~~~~~Scattered}%
\end{sideways}&
\includegraphics[%
  width=0.30\columnwidth]{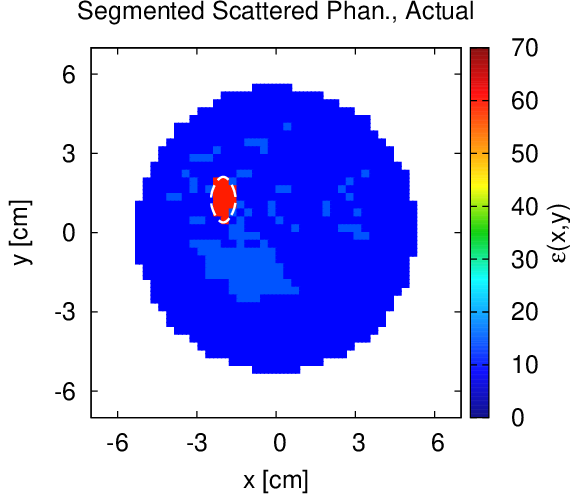}&
\includegraphics[%
  width=0.30\columnwidth]{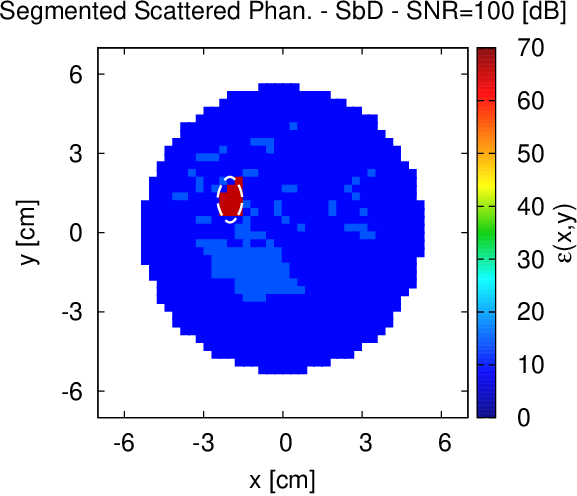}&
\includegraphics[%
  width=0.30\columnwidth]{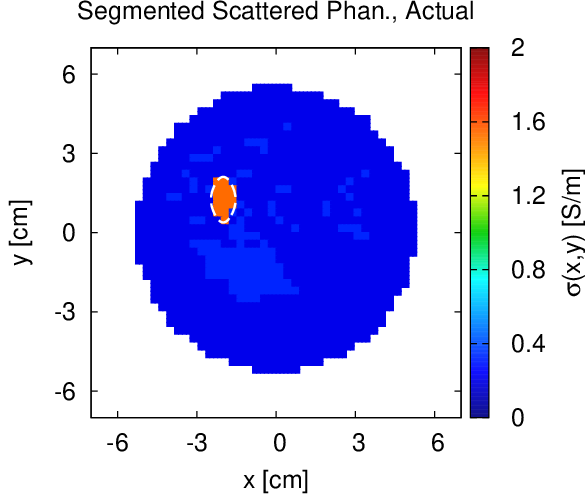}&
\includegraphics[%
  width=0.30\columnwidth]{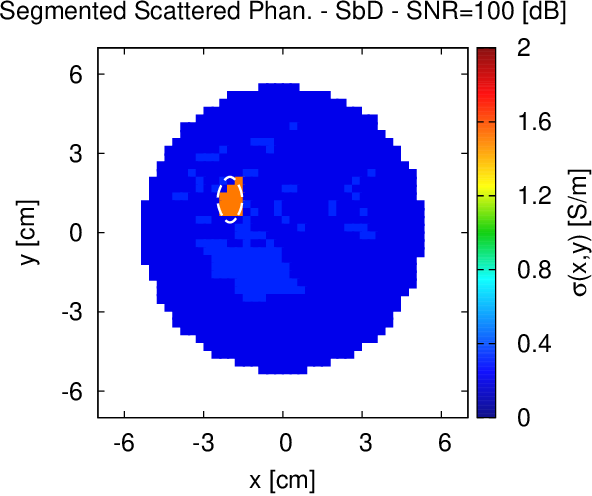}\tabularnewline
\begin{sideways}
\end{sideways}&
(\emph{e})&
(\emph{f})&
(\emph{g})&
(\emph{h})\tabularnewline
\begin{sideways}
\emph{~~~~~~~~~~~~~~~~~~~HD}%
\end{sideways}&
\includegraphics[%
  width=0.30\columnwidth]{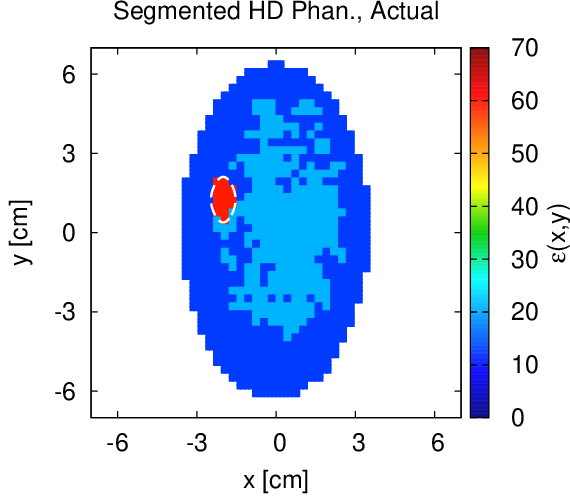}&
\includegraphics[%
  width=0.30\columnwidth]{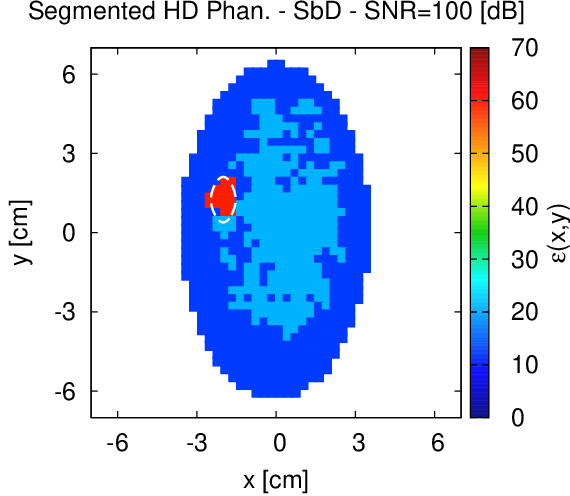}&
\includegraphics[%
  width=0.30\columnwidth]{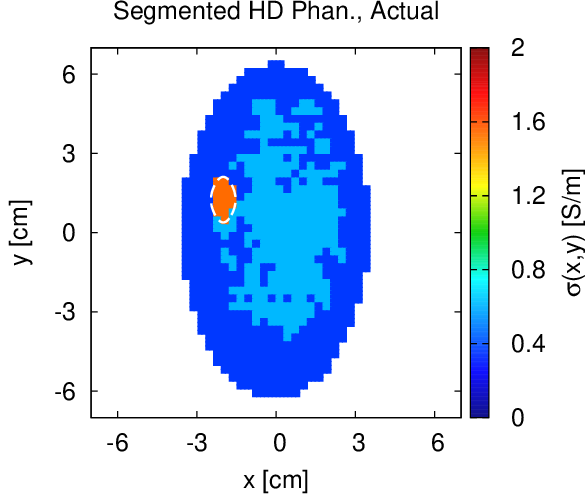}&
\includegraphics[%
  width=0.30\columnwidth]{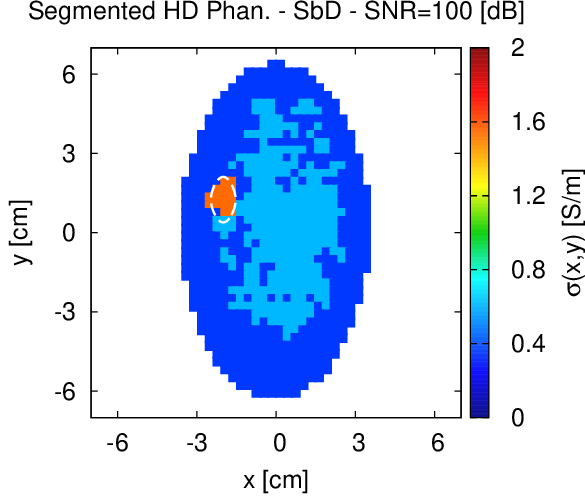}\tabularnewline
\begin{sideways}
\end{sideways}&
(\emph{i})&
(\emph{l})&
(\emph{m})&
(\emph{n})\tabularnewline
\end{tabular}
\end{sideways}\end{center}

\begin{center}~\vfill\end{center}

\begin{center}\textbf{Fig. 9 - F. Zardi et} \textbf{\emph{al.}}\textbf{,}
\textbf{\emph{{}``}}A Physics-Driven \emph{AI} Approach ...''\end{center}

\newpage
\begin{center}~\vfill\end{center}

\begin{center}\begin{tabular}{c}
\includegraphics[%
  width=0.80\textwidth]{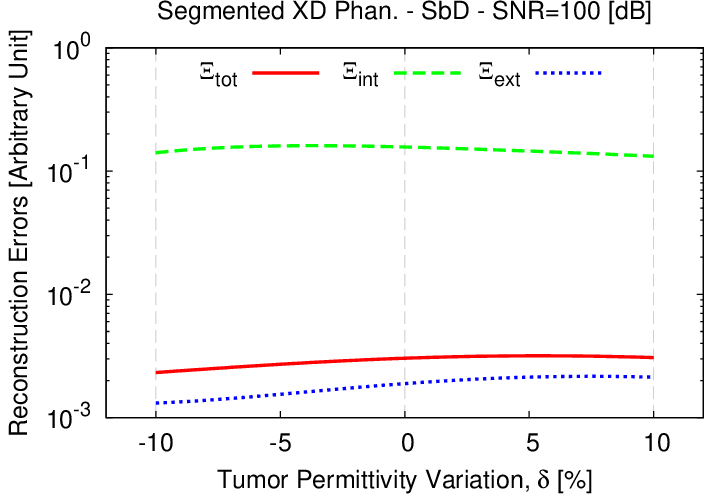}\tabularnewline
\end{tabular}\end{center}

\begin{center}~\vfill\end{center}

\begin{center}\textbf{Fig. 10 - F. Zardi et} \textbf{\emph{al.}}\textbf{,}
\textbf{\emph{{}``}}A Physics-Driven \emph{AI} Approach ...''\end{center}

\newpage
\begin{center}~\vfill\end{center}

\begin{center}\begin{tabular}{cc}
$\delta=-10\%$&
$\delta=+10\%$\tabularnewline
\includegraphics[%
  width=0.30\columnwidth]{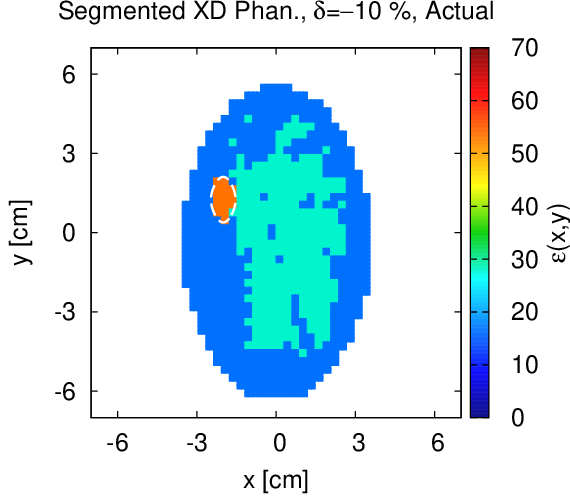}&
\includegraphics[%
  width=0.30\columnwidth]{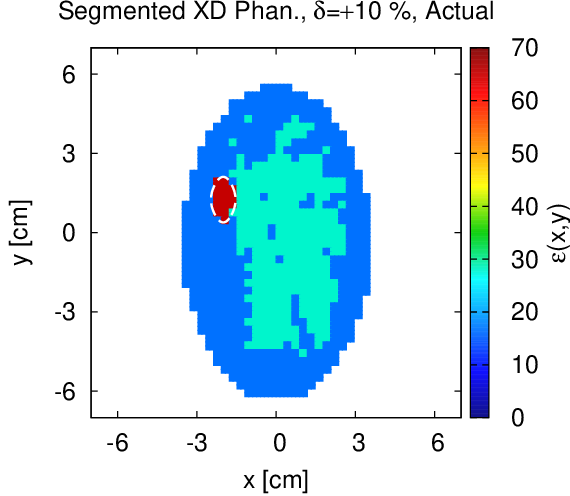}\tabularnewline
(\emph{a})&
(\emph{b})\tabularnewline
\includegraphics[%
  width=0.30\columnwidth]{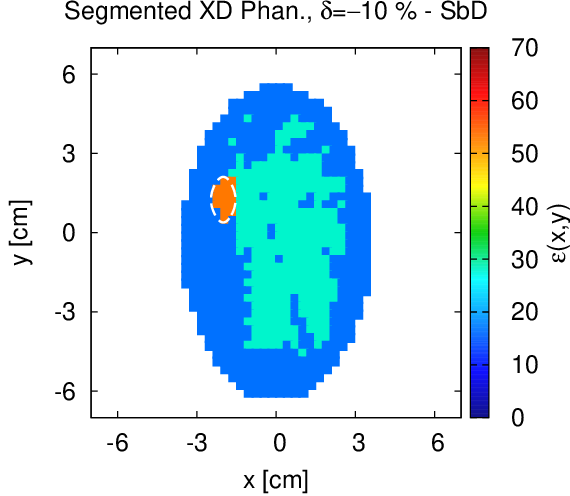}&
\includegraphics[%
  width=0.30\columnwidth]{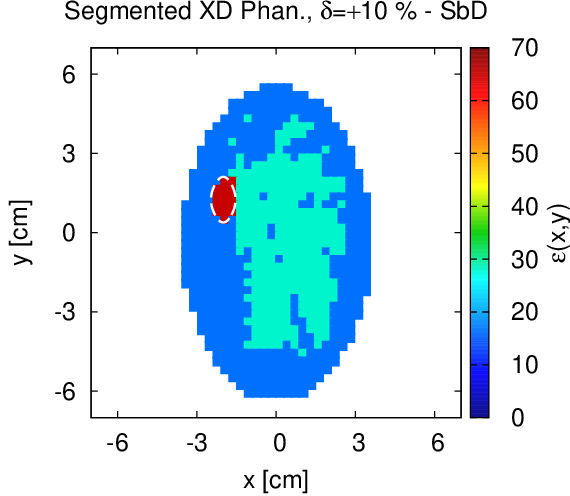}\tabularnewline
(\emph{c})&
(\emph{d})\tabularnewline
\includegraphics[%
  width=0.30\columnwidth]{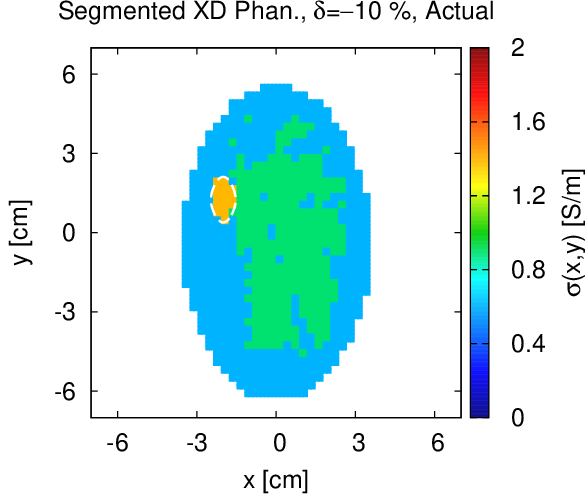}&
\includegraphics[%
  width=0.30\columnwidth]{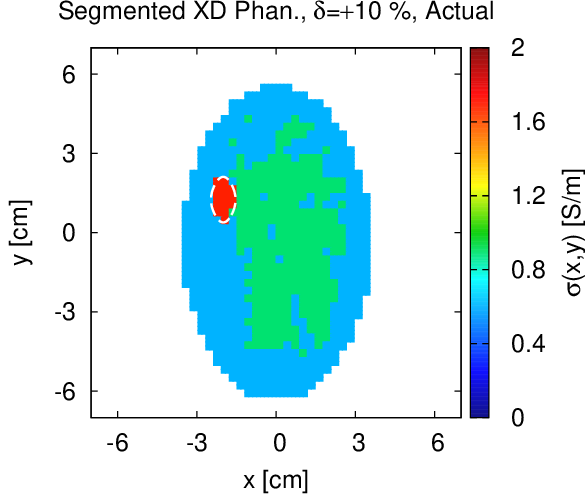}\tabularnewline
(\emph{e})&
(\emph{f})\tabularnewline
\includegraphics[%
  width=0.30\columnwidth]{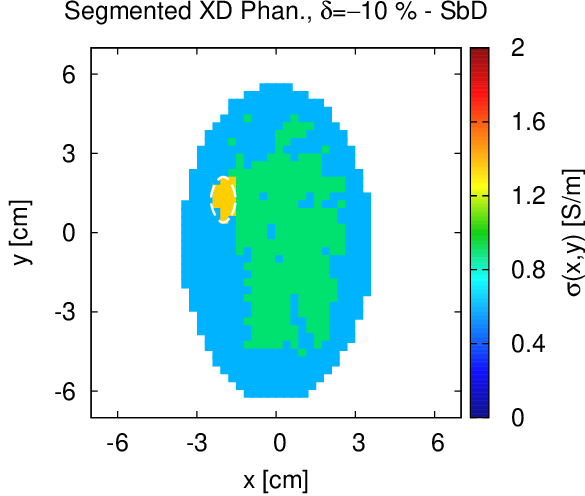}&
\includegraphics[%
  width=0.30\columnwidth]{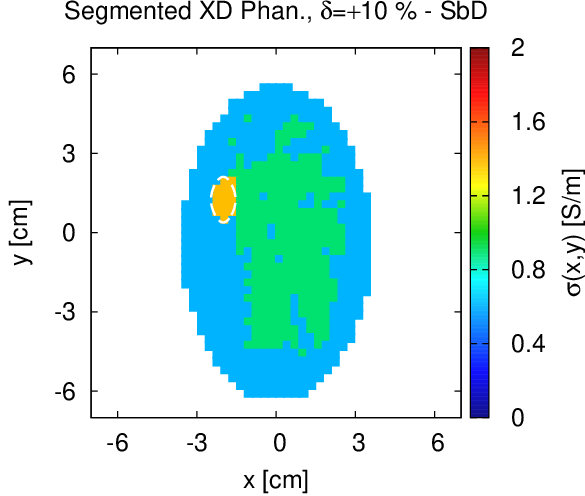}\tabularnewline
(\emph{g})&
(\emph{h})\tabularnewline
\end{tabular}\end{center}

\begin{center}~\vfill\end{center}

\begin{center}\textbf{Fig. 11 - F. Zardi et} \textbf{\emph{al.}}\textbf{,}
\textbf{\emph{{}``}}A Physics-Driven \emph{AI} Approach ...''\end{center}

\newpage
\begin{center}~\vfill\end{center}

\begin{center}\begin{tabular}{ccc}
\begin{sideways}
\emph{~~~~~~~~~~~~~~~~~~~~~~~Full}%
\end{sideways}&
\includegraphics[%
  width=0.40\columnwidth]{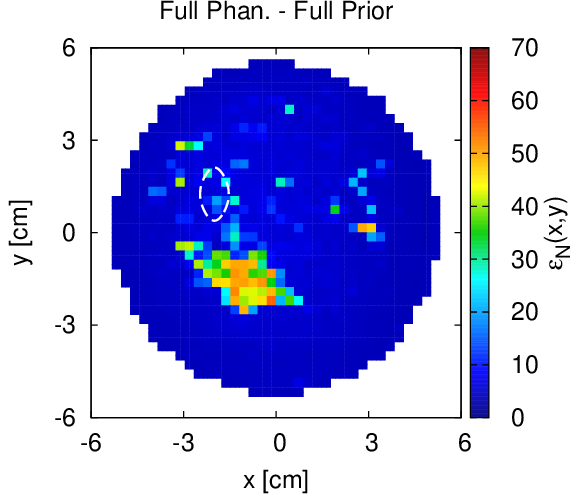}&
\includegraphics[%
  width=0.40\columnwidth]{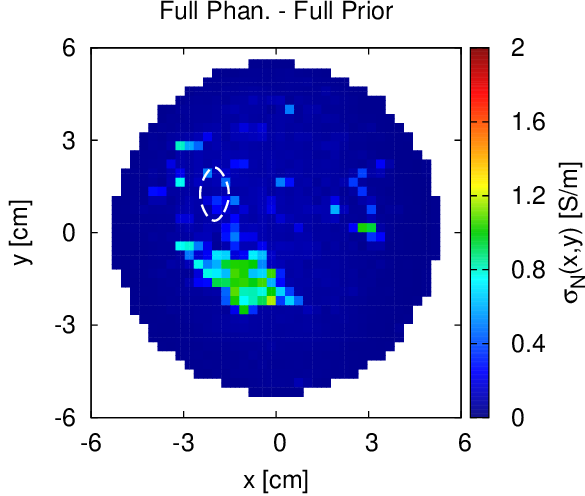}\tabularnewline
\begin{sideways}
\end{sideways}&
(\emph{a})&
(\emph{b})\tabularnewline
\begin{sideways}
\end{sideways}&
&
\tabularnewline
\begin{sideways}
\emph{~~~~~~~~~~~~~~~~~~Segmented}%
\end{sideways}&
\includegraphics[%
  width=0.40\columnwidth]{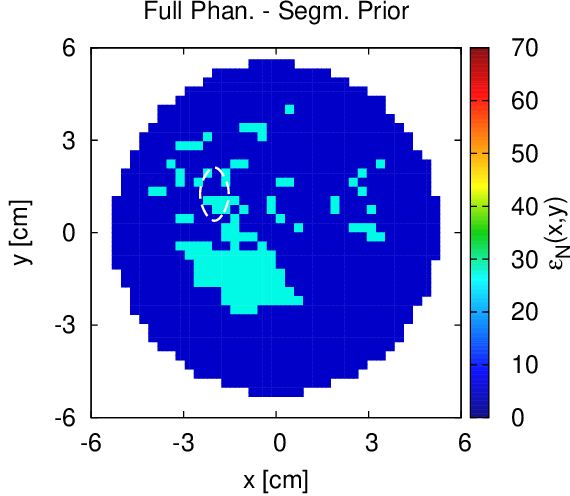}&
\includegraphics[%
  width=0.40\columnwidth]{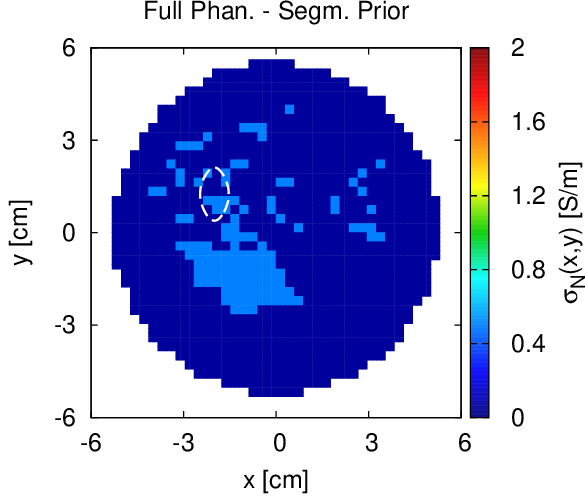}\tabularnewline
\begin{sideways}
\end{sideways}&
(\emph{c})&
(\emph{d})\tabularnewline
\begin{sideways}
\end{sideways}&
&
\tabularnewline
\begin{sideways}
\emph{~~~~~~~~~~~~~~~~~~~~Constant}%
\end{sideways}&
\includegraphics[%
  width=0.40\columnwidth]{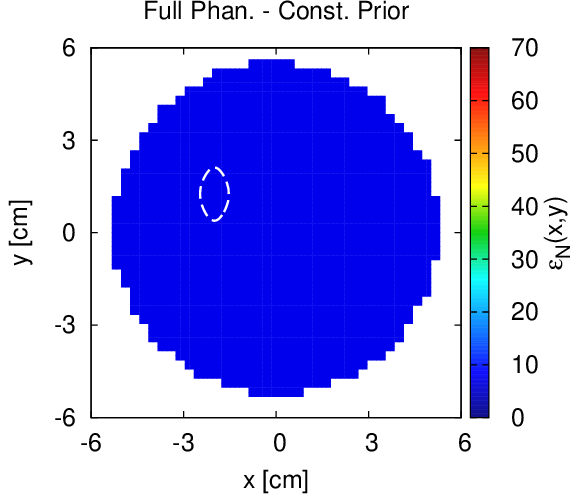}&
\includegraphics[%
  width=0.40\columnwidth]{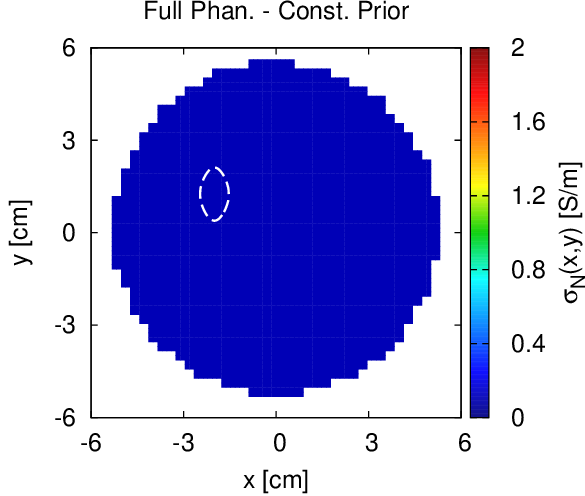}\tabularnewline
\begin{sideways}
\end{sideways}&
(\emph{e})&
(\emph{f})\tabularnewline
\end{tabular}\end{center}

\begin{center}~\vfill\end{center}

\begin{center}\textbf{Fig. 12 - F. Zardi et} \textbf{\emph{al.}}\textbf{,}
\textbf{\emph{{}``}}A Physics-Driven \emph{AI} Approach ...''\end{center}

\newpage
\begin{center}~\vfill\end{center}

\begin{center}\begin{tabular}{ccc}
\begin{sideways}
\emph{~~~~~~~~~~~~~~~~~~Actual}%
\end{sideways}&
\includegraphics[%
  width=0.30\columnwidth]{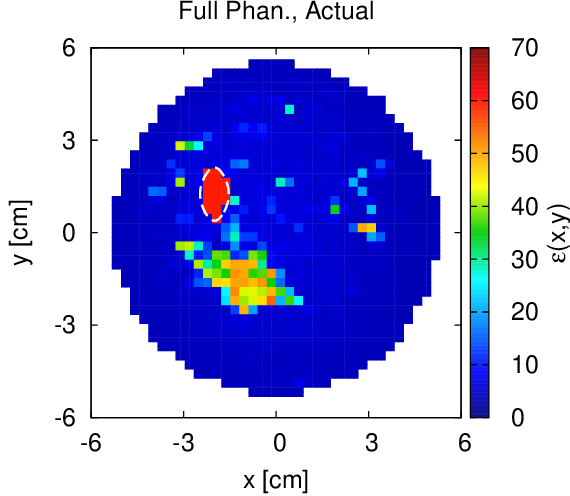}&
\includegraphics[%
  width=0.30\columnwidth]{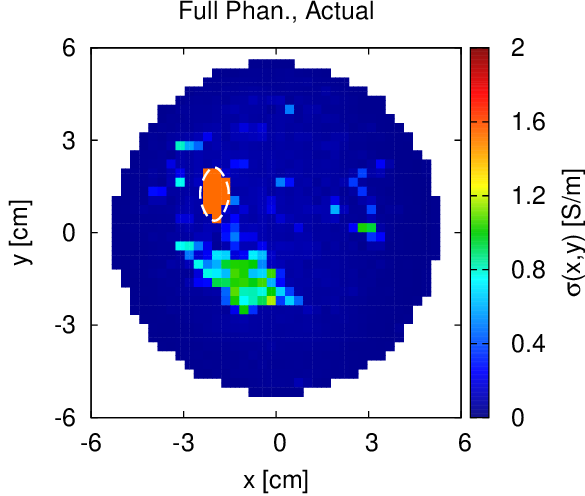}\tabularnewline
\begin{sideways}
\end{sideways}&
(\emph{a})&
(\emph{b})\tabularnewline
\begin{sideways}
\emph{~~~~~~~~~~~~~~~Full Prior}%
\end{sideways}&
\includegraphics[%
  width=0.30\columnwidth]{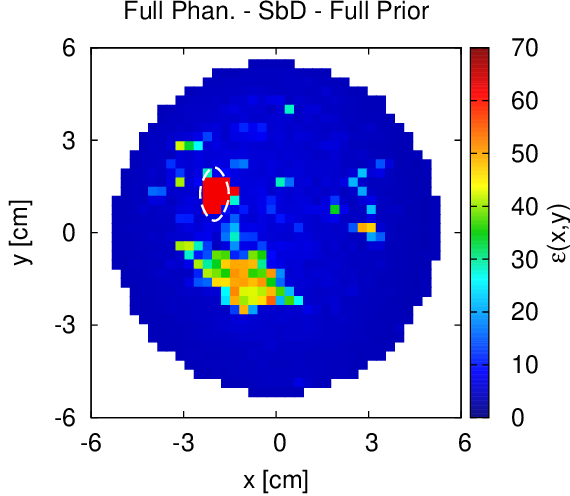}&
\includegraphics[%
  width=0.30\columnwidth]{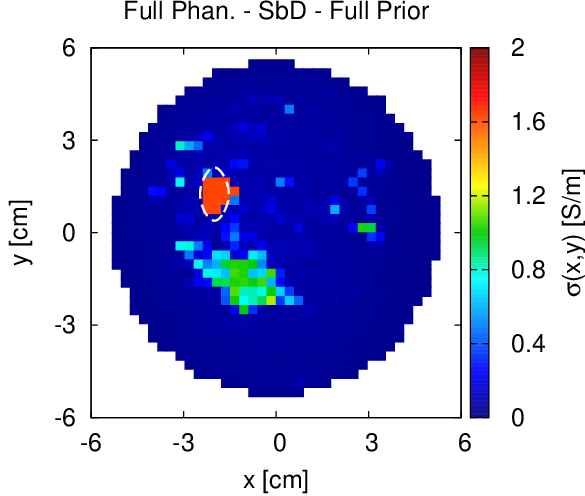}\tabularnewline
\begin{sideways}
\end{sideways}&
(\emph{c})&
(\emph{d})\tabularnewline
\begin{sideways}
\emph{~~~~~~~~~~Segmented Prior}%
\end{sideways}&
\includegraphics[%
  width=0.30\columnwidth]{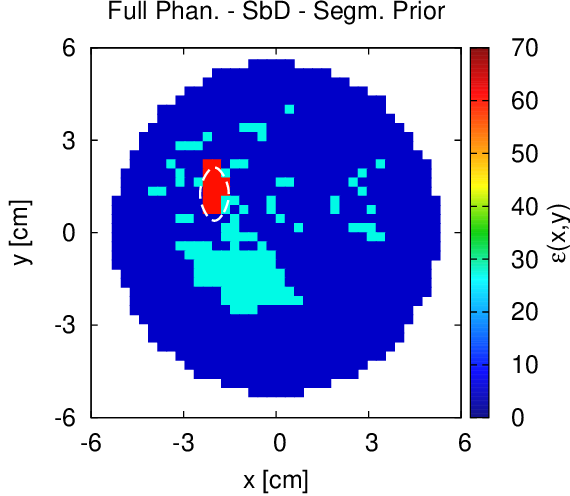}&
\includegraphics[%
  width=0.30\columnwidth]{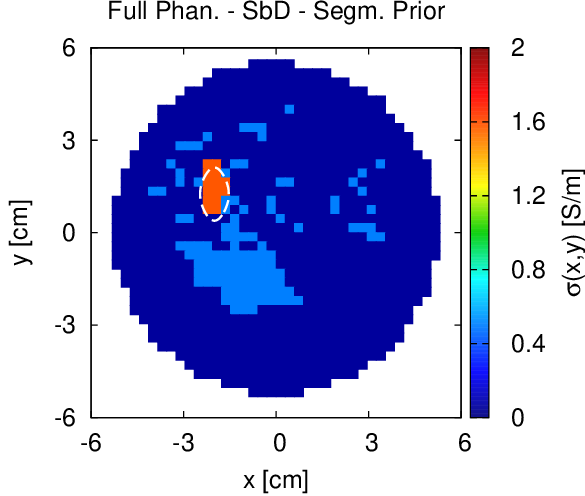}\tabularnewline
\begin{sideways}
\end{sideways}&
(\emph{e})&
(\emph{f})\tabularnewline
\begin{sideways}
\emph{~~~~~~~~~~Constant Prior}%
\end{sideways}&
\includegraphics[%
  width=0.30\columnwidth]{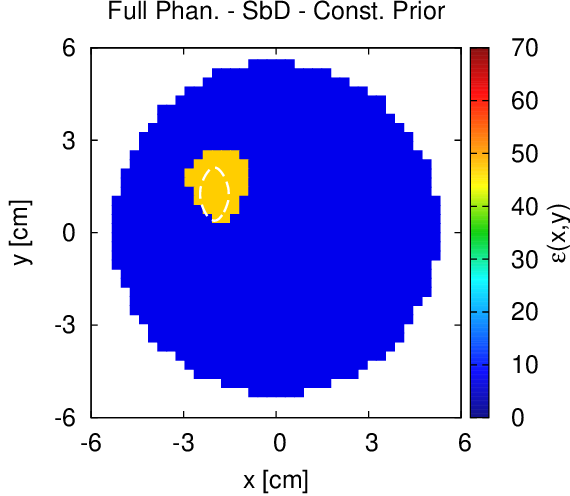}&
\includegraphics[%
  width=0.30\columnwidth]{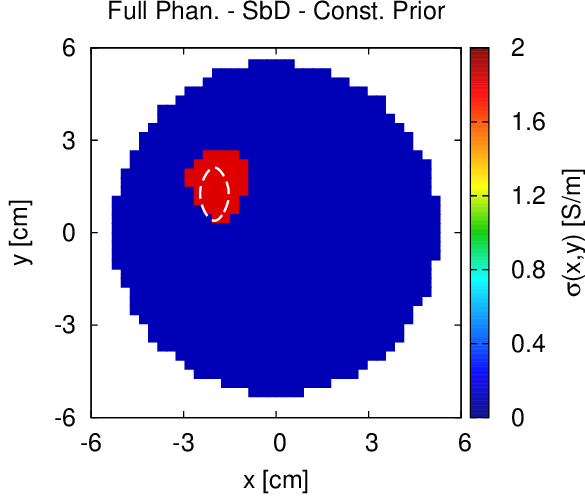}\tabularnewline
\begin{sideways}
\end{sideways}&
(\emph{g})&
(\emph{h})\tabularnewline
\end{tabular}\end{center}

\begin{center}~\vfill\end{center}

\begin{center}\textbf{Fig. 13 - F. Zardi et} \textbf{\emph{al.}}\textbf{,}
\textbf{\emph{{}``}}A Physics-Driven \emph{AI} Approach ...''\end{center}

\newpage
\begin{center}~\vfill\end{center}

\begin{center}\begin{tabular}{ccc}
\begin{sideways}
\emph{~~~~~~~~~~~~~~~Actual}%
\end{sideways}&
\includegraphics[%
  width=0.30\columnwidth]{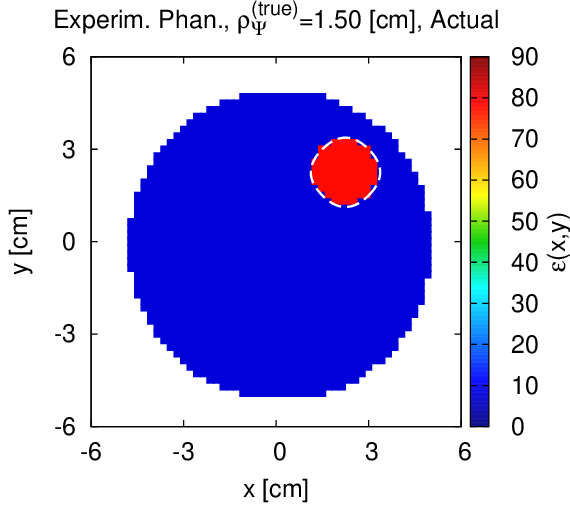}&
\includegraphics[%
  width=0.30\columnwidth]{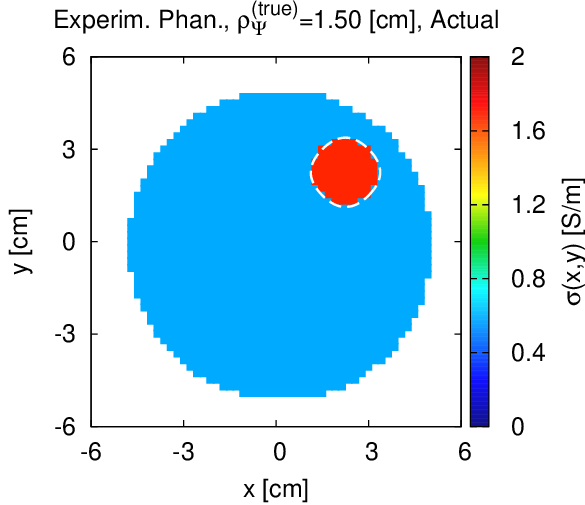}\tabularnewline
\begin{sideways}
\end{sideways}&
(\emph{a})&
(\emph{b})\tabularnewline
\begin{sideways}
\emph{~~~~~~~~~~~~~~~~~~SbD}%
\end{sideways}&
\includegraphics[%
  width=0.30\columnwidth]{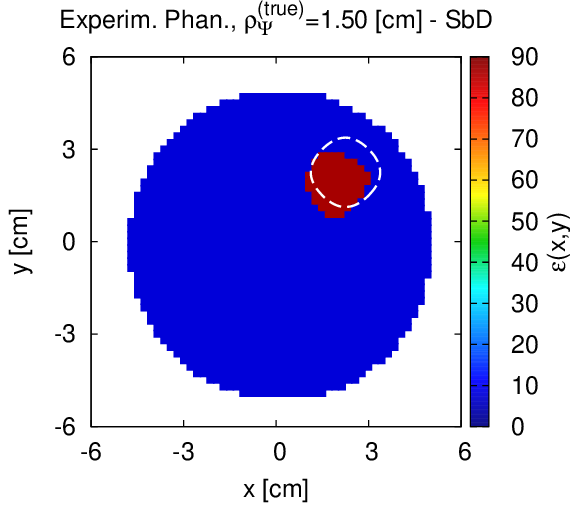}&
\includegraphics[%
  width=0.30\columnwidth]{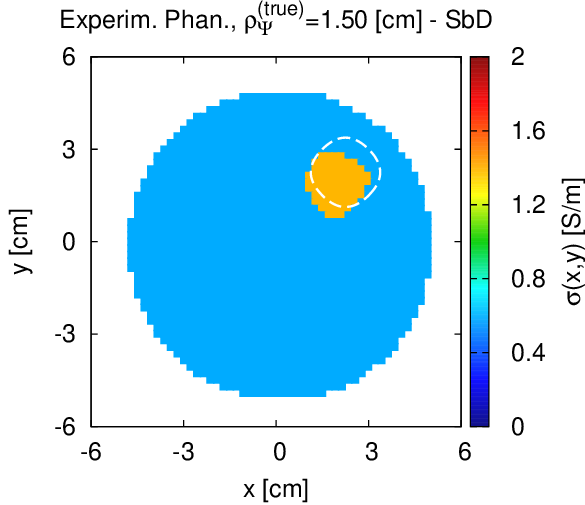}\tabularnewline
\begin{sideways}
\end{sideways}&
(\emph{c})&
(\emph{d})\tabularnewline
\begin{sideways}
\emph{~~~~~~~~~~~~~~~~~~EA}%
\end{sideways}&
\includegraphics[%
  width=0.30\columnwidth]{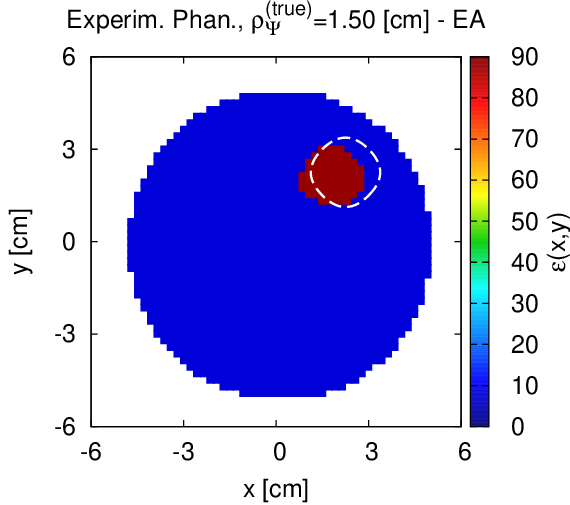}&
\includegraphics[%
  width=0.30\columnwidth]{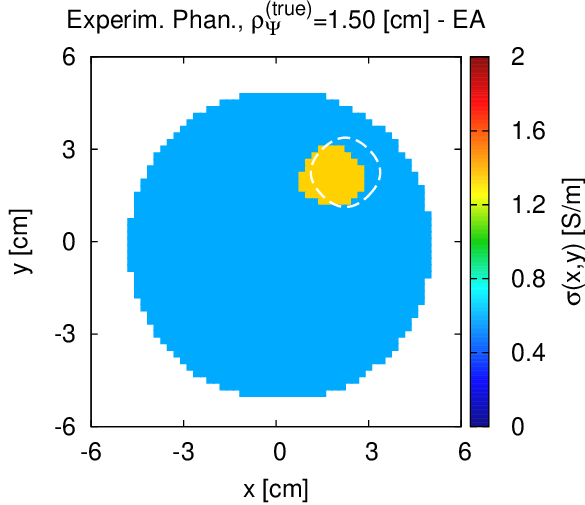}\tabularnewline
\begin{sideways}
\end{sideways}&
(\emph{e})&
(\emph{f})\tabularnewline
\begin{sideways}
\emph{~~~~~~~~~~~~~~~~~~~DA}%
\end{sideways}&
\includegraphics[%
  width=0.30\columnwidth]{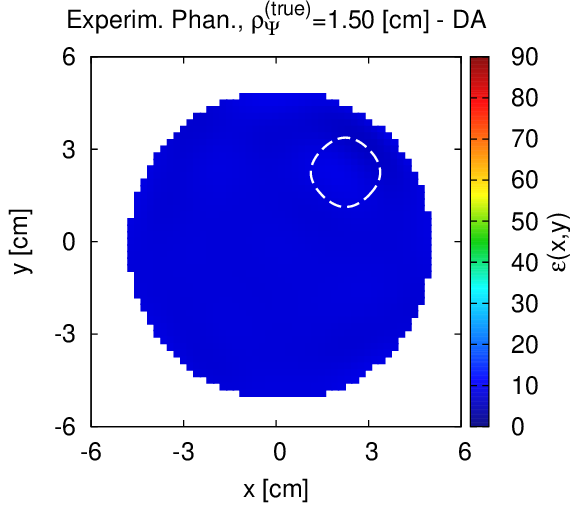}&
\includegraphics[%
  width=0.30\columnwidth]{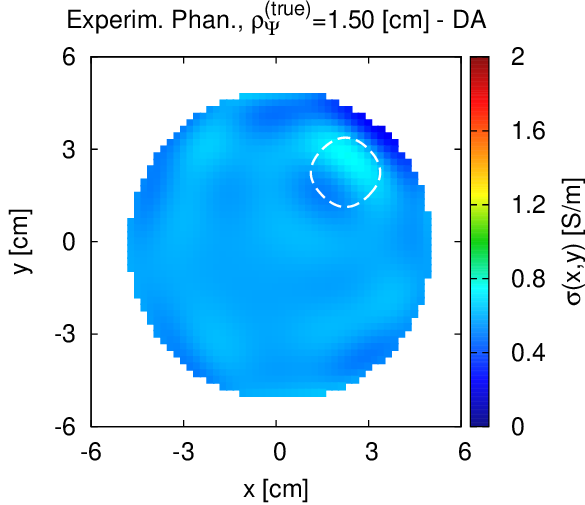}\tabularnewline
\begin{sideways}
\end{sideways}&
(\emph{g})&
(\emph{h})\tabularnewline
\end{tabular}\end{center}

\begin{center}~\vfill\end{center}

\begin{center}\textbf{Fig. 14 - F. Zardi et} \textbf{\emph{al.}}\textbf{,}
\textbf{\emph{{}``}}A Physics-Driven \emph{AI} Approach ...''\end{center}

\newpage
\begin{center}~\vfill\end{center}

\begin{center}\begin{tabular}{ccc}
\begin{sideways}
\emph{~~~~~~~~~~~~~~~Actual}%
\end{sideways}&
\includegraphics[%
  width=0.30\columnwidth]{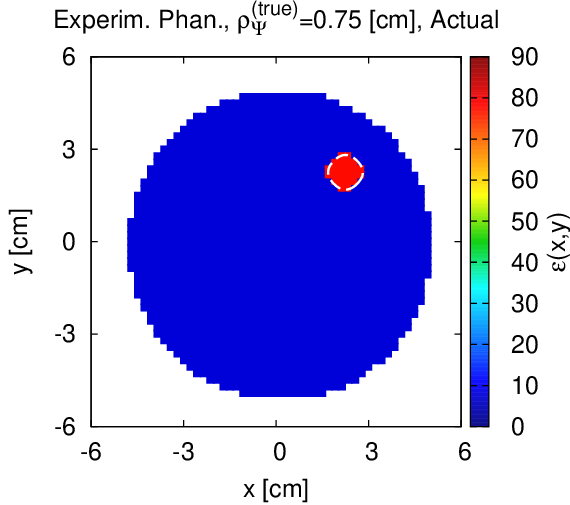}&
\includegraphics[%
  width=0.30\columnwidth]{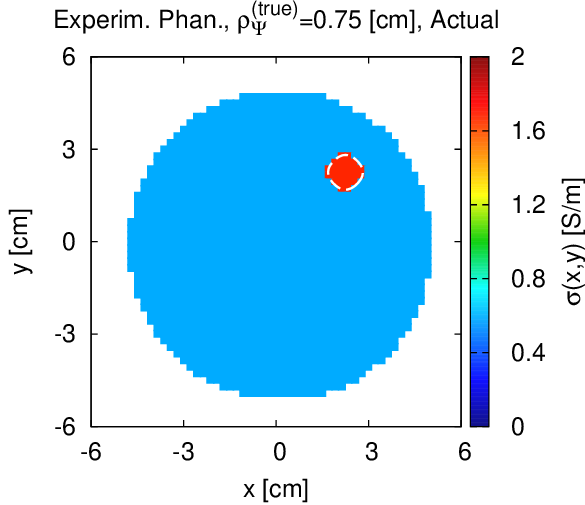}\tabularnewline
\begin{sideways}
\end{sideways}&
(\emph{a})&
(\emph{b})\tabularnewline
\begin{sideways}
\emph{~~~~~~~~~~~~~~~~~~SbD}%
\end{sideways}&
\includegraphics[%
  width=0.30\columnwidth]{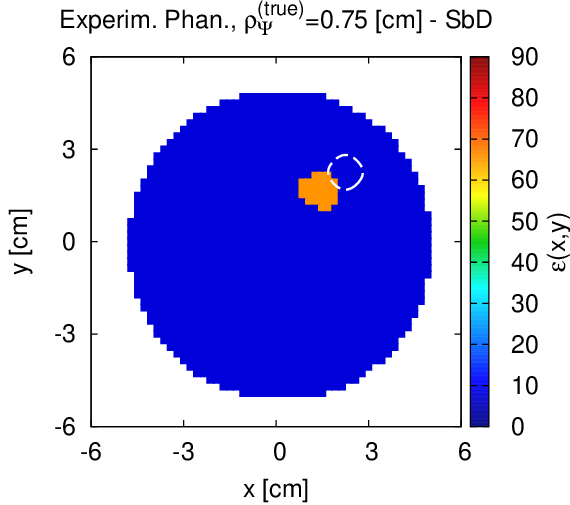}&
\includegraphics[%
  width=0.30\columnwidth]{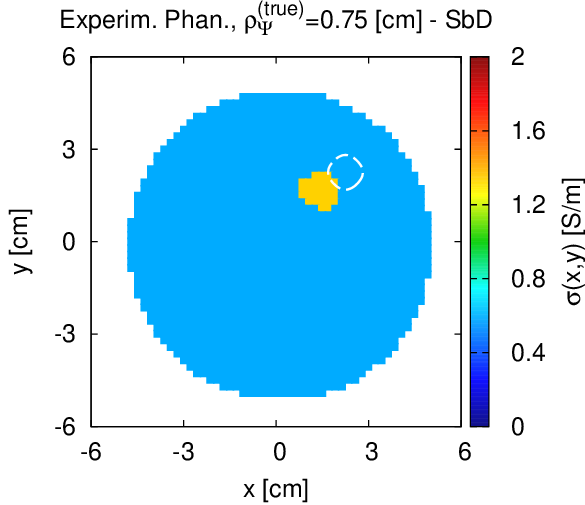}\tabularnewline
\begin{sideways}
\end{sideways}&
(\emph{c})&
(\emph{d})\tabularnewline
\begin{sideways}
\emph{~~~~~~~~~~~~~~~~~~EA}%
\end{sideways}&
\includegraphics[%
  width=0.30\columnwidth]{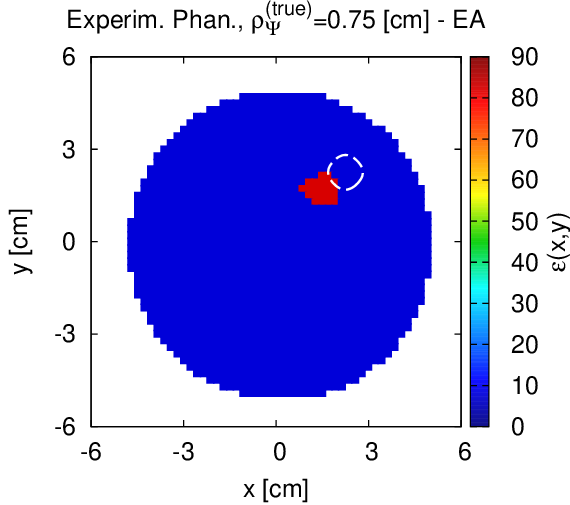}&
\includegraphics[%
  width=0.30\columnwidth]{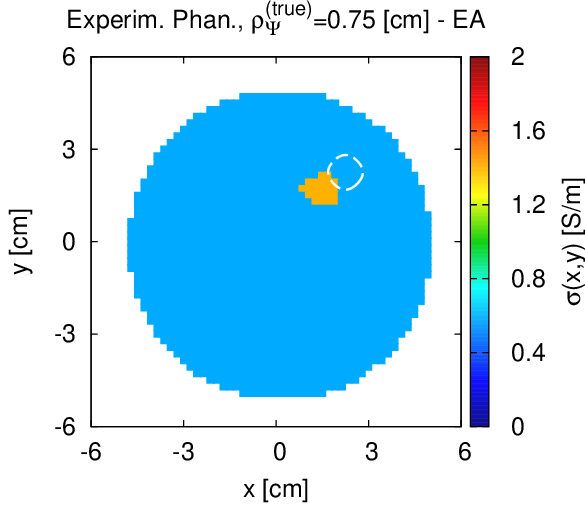}\tabularnewline
\begin{sideways}
\end{sideways}&
(\emph{e})&
(\emph{f})\tabularnewline
\begin{sideways}
\emph{~~~~~~~~~~~~~~~~~~~DA}%
\end{sideways}&
\includegraphics[%
  width=0.30\columnwidth]{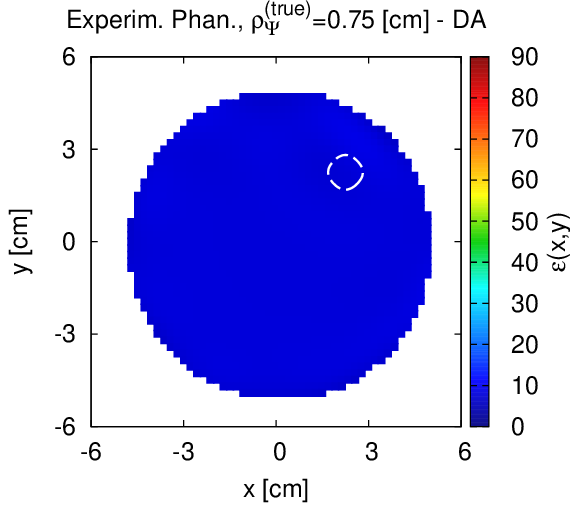}&
\includegraphics[%
  width=0.30\columnwidth]{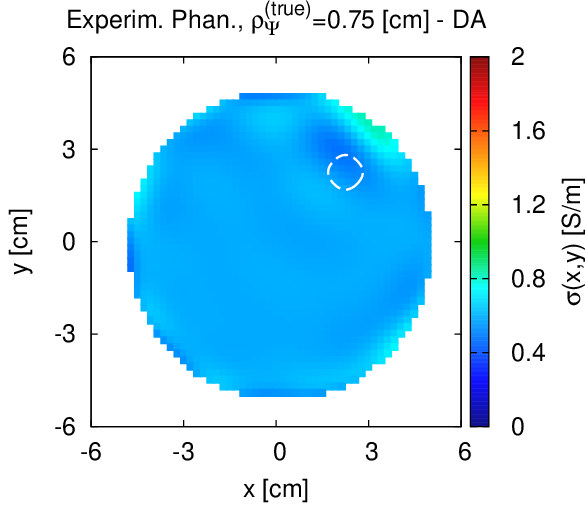}\tabularnewline
\begin{sideways}
\end{sideways}&
(\emph{g})&
(\emph{h})\tabularnewline
\end{tabular}\end{center}

\begin{center}~\vfill\end{center}

\begin{center}\textbf{Fig. 15 - F. Zardi et} \textbf{\emph{al.}}\textbf{,}
\textbf{\emph{{}``}}A Physics-Driven \emph{AI} Approach ...''\end{center}

\newpage
\begin{center}~\vfill\end{center}

\begin{center}\begin{tabular}{|c||c|c|c|c|c|c|c|}
\hline 
\emph{UWCEM}&
\emph{Breast}&
\multicolumn{2}{c|}{\emph{Adipose}}&
\multicolumn{2}{c|}{\emph{Fibroglandular}}&
\multicolumn{2}{c|}{\emph{Background }}\tabularnewline
\cline{3-4} \cline{5-6} \cline{7-8} 
\emph{ID}&
\emph{Density}&
\emph{$\varepsilon_{\mathcal{N}}$}&
\emph{$\sigma_{\mathcal{N}}$} {[}S/m{]}&
\emph{$\varepsilon_{\mathcal{N}}$}&
\emph{$\sigma_{\mathcal{N}}$} {[}S/m{]}&
\emph{$\varepsilon_{\mathcal{B}}$}&
\emph{$\sigma_{\mathcal{B}}$} {[}S/m{]}\tabularnewline
\hline
\hline 
$012304$&
\emph{XD}&
$16.5$&
$0.60$&
$28.0$&
$0.89$&
$22.4$&
$1.26$\tabularnewline
\hline 
$080304$&
\emph{HD}&
$12.8$&
$0.36$&
$21.0$&
$0.61$&
$22.4$&
$1.26$\tabularnewline
\hline 
$012204$&
Scattered&
$9.0$&
$0.21$&
$14.5$&
$0.32$&
$13.6$&
$0.87$\tabularnewline
\hline 
$071904$&
Fatty&
$8.0$&
$0.17$&
$8.8$&
$0.195$&
$13.6$&
$0.87$\tabularnewline
\hline
\end{tabular}\end{center}

\begin{center}\vfill\end{center}

\begin{center}\textbf{Table I - F. Zardi et} \textbf{\emph{al.}},
{}``A Physics-Driven \emph{AI} Approach ...''\end{center}

\newpage
\begin{center}~\vfill\end{center}

\begin{center}\begin{tabular}{|c||c|c|c|c|}
\hline 
\emph{Method}&
\textbf{$\Xi_{tot}$}&
\textbf{$\Xi_{int}$}&
\textbf{$\Xi_{ext}$}&
\textbf{$\Delta t$} {[}min{]}\tabularnewline
\hline
\hline 
\emph{SbD}&
$3.48\times10^{-3}$&
$1.89\times10^{-1}$&
$3.72\times10^{-3}$&
$12$\tabularnewline
\hline 
\emph{EA}&
$3.46\times10^{-3}$&
$1.29\times10^{-1}$&
$3.69\times10^{-3}$&
$164$\tabularnewline
\hline 
\emph{DA}&
$7.66\times10^{-2}$&
$7.53\times10^{-1}$&
$7.17\times10^{-2}$&
$12$\tabularnewline
\hline
\end{tabular}\end{center}

\begin{center}\vfill\end{center}

\begin{center}\textbf{Table II - F. Zardi et} \textbf{\emph{al.}},
{}``A Physics-Driven \emph{AI} Approach ...''\end{center}

\newpage
\begin{center}~\vfill\end{center}

\begin{center}\begin{tabular}{|c||c|c|c|}
\hline 
\emph{Prior Knowledge}&
\textbf{$\Xi_{tot}$}&
\textbf{$\Xi_{int}$}&
\textbf{$\Xi_{ext}$}\tabularnewline
\hline
\hline 
Full (\emph{FP})&
$3.08\times10^{-3}$&
$1.59\times10^{-1}$&
$1.72\times10^{-3}$\tabularnewline
\hline 
Segmented (\emph{SP})&
$4.16\times10^{-3}$&
$1.71\times10^{-1}$&
$3.31\times10^{-3}$\tabularnewline
\hline 
Constant (\emph{CP})&
$3.63\times10^{-2}$&
$1.90\times10^{-1}$&
$3.52\times10^{-2}$\tabularnewline
\hline
\end{tabular}\end{center}

\begin{center}\vfill\end{center}

\begin{center}\textbf{Table III - F. Zardi et} \textbf{\emph{al.}},
{}``A Physics-Driven \emph{AI} Approach ...''\end{center}

\newpage
\begin{center}~\vfill\end{center}

\begin{center}\begin{tabular}{|c||c|c|c|c|}
\hline 
\emph{Method}&
$\Xi_{tot}$&
$\Xi_{int}$&
$\Xi_{ext}$&
$\Delta t$ {[}min{]}\tabularnewline
\hline
\hline 
&
\multicolumn{4}{c|}{$\rho_{\Psi}^{\left(true\right)}=1.5$ {[}cm{]}}\tabularnewline
\hline 
\emph{SbD}&
$2.23\times10^{-2}$&
$3.19\times10^{-1}$&
$1.80\times10^{-2}$&
$21$\tabularnewline
\hline 
\emph{EA}&
$2.06\times10^{-2}$&
$3.50\times10^{-1}$&
$1.58\times10^{-2}$&
$304$\tabularnewline
\hline 
\emph{DA}&
$2.51\times10^{-2}$&
$8.86\times10^{-1}$&
$1.27\times10^{-2}$&
$21$\tabularnewline
\hline
\hline 
&
\multicolumn{4}{c|}{$\rho_{\Psi}^{\left(true\right)}=0.75$ {[}cm{]}}\tabularnewline
\hline 
\emph{SbD}&
$1.92\times10^{-2}$&
$4.96\times10^{-1}$&
$1.75\times10^{-2}$&
$21$\tabularnewline
\hline 
\emph{EA}&
$1.86\times10^{-2}$&
$4.72\times10^{-1}$&
$1.68\times10^{-2}$&
$295$\tabularnewline
\hline 
\emph{DA}&
$2.06\times10^{-2}$&
$8.91\times10^{-1}$&
$8.24\times10^{-3}$&
$21$\tabularnewline
\hline
\end{tabular}\end{center}

\begin{center}\vfill\end{center}

\begin{center}\textbf{Table IV - F. Zardi et} \textbf{\emph{al.}},
{}``A Physics-Driven \emph{AI} Approach ...''\end{center}

\newpage
\begin{center}~\vfill\end{center}

\begin{center}\begin{tabular}{|c||c|c||c|c|}
\hline 
&
\multicolumn{2}{c||}{$\rho_{\Psi}=1.5$ {[}cm{]}}&
\multicolumn{2}{c|}{$\rho_{\Psi}=0.75$ {[}cm{]}}\tabularnewline
\hline 
\emph{Method}&
$\zeta$ {[}cm{]}&
\emph{Detection}&
$\zeta$ {[}cm{]}&
\emph{Detection}\tabularnewline
\hline
\hline 
\emph{SbD}&
$0.48$&
Yes&
$1.03$&
Yes\tabularnewline
\hline 
\emph{EA}&
$0.45$&
Yes&
$0.96$&
Yes\tabularnewline
\hline 
\emph{DA}&
$1.83$&
Yes&
$2.47$&
No\tabularnewline
\hline 
\emph{DAS} \cite{Reimer 2021}&
$1.00$&
Yes&
$3.76$&
No\tabularnewline
\hline 
\emph{DMAS} \cite{Reimer 2021}&
$1.01$&
Yes&
$3.76$&
No\tabularnewline
\hline 
\emph{ORR} \cite{Reimer 2021}&
$1.21$&
Yes&
$3.62$&
No\tabularnewline
\hline
\end{tabular}\end{center}

\begin{center}\vfill\end{center}

\begin{center}\textbf{Table V - F. Zardi et} \textbf{\emph{al.}},
{}``A Physics-Driven \emph{AI} Approach ...''\end{center}

\newpage

\begin{thebibliography}{10}
\bibitem{WHO 2022}World Health Organization, {}``Breast cancer''. Accessed: Nov. 4,
2022. {[}Online{]}. Available: https://www.who.int/news-room/fact-sheets/detail/breast-cancer
\bibitem{Tabar 2019}L. Tabar, P. B. Dean, T. H. Chen, A. M. Yen, S. L. Chen, J. C. Fann,
S. Y. Chiu, M. M. Ku, W. Y. Wu, C. Hsu, Y. Chen, K. Beckmann, R. A.
Smith, and S. W. Duffy, {}``The incidence of fatal breast cancer
measures the increased effectiveness of therapy in women participating
in mammography screening,'' \emph{Cancer}, vol. 125, no. 4, pp. 515-523,
Feb. 2019.
\bibitem{Nikolova 2011}N. Nikolova, {}``Microwave imaging for breast cancer,'' \emph{IEEE
Microw. Mag.}, vol. 12, no. 7, pp. 78-94, Dec. 2011.
\bibitem{Mirbeik-Sabzevari 2019}A. Mirbeik-Sabzevari and N. Tavassolian, {}``Tumor detection using
millimeter-wave technology: Differentiating between benign lesions
and cancer tissues,'' \emph{IEEE Microw. Mag.}, vol. 20, no. 8, pp.
30-43, Aug. 2019.
\bibitem{Bassi 2013}M. Bassi, M. Caruso, M. S. Khan, A. Bevilacqua, A.-D. Capobianco,
and A. Neviani, {}``An integrated microwave imaging radar with planar
antennas for breast cancer detection,'' \emph{IEEE Trans. Microw.
Theory Techn.}, vol. 61, no. 5, pp. 2108-2118, May 2013.
\bibitem{Porter 2016 b}E. Porter, H. Bahrami, A. Santorelli, B. Gosselin, L. A. Rusch, and
M. Popovic, {}``A wearable microwave antenna array for time-domain
breast tumor screening,'' \emph{IEEE Trans. Med. Imaging}, vol. 35,
no. 6, pp. 1501-1509, Jun. 2016.
\bibitem{O'Loughlin 2018}D. O'Loughlin, M. O'Halloran, B. M. Moloney, M. Glavin, E. Jones,
and M. A. Elahi, {}``Microwave breast imaging: Clinical advances
and remaining challenges,'' \emph{IEEE Trans. Biomed. Eng.}, vol.
65, no. 11, pp. 2580-2590, Nov. 2018.
\bibitem{Chen 2018}X. Chen, \emph{Computational Methods for Electromagnetic Inverse Scattering.}
Hoboken, NJ, USA: Wiley, 2018.
\bibitem{Caorsi 2004}S. Caorsi, A. Massa, M. Pastorino, and M. Donelli, {}``Improved microwave
imaging procedure for nondestructive evaluations of two-dimensional
structures,'' \emph{IEEE Trans. Antennas Propag.}, vol. 52, no. 6,
pp. 1386-1397, Jun. 2004.
\bibitem{Xu 2018b}K. Xu, Y. Zhong, X. Chen, and D. Lesselier, {}``A fast integral equation-based
method for solving electromagnetic inverse scattering problems with
inhomogeneous background,'' \emph{IEEE Trans. Antennas Propag.},
vol. 66, no. 8, pp. 4228-4239, Aug. 2018.
\bibitem{Kurrant 2013}D. Kurrant and E. Fear, {}``Defining regions of interest for microwave
imaging using near-field reflection data,'' \emph{IEEE Trans. Microw.
Theory Techn.}, vol. 61, no. 5, pp. 2137-2145, May 2013.
\bibitem{Colgan 2015}T. J. Colgan, S. C. Hagness, and B. D. Van Veen, {}``A 3-D level
set method for microwave breast imaging,'' \emph{IEEE Trans. Biomed.
Eng.}, vol. 62, no. 10, pp. 2526-2534, Oct. 2015.
\bibitem{Abdollahi 2020}N. Abdollahi, I. Jeffrey, and J. LoVetri, {}``Improved tumor detection
via quantitative microwave breast imaging using eigenfunction-based
prior,'' \emph{IEEE Trans. Comput. Imaging}, vol. 6, pp. 1194-1202,
Jul. 2020.
\bibitem{Porter 2016 a}E. Porter, M. Coates, and M. Popovic, {}``An early clinical study
of time-domain microwave radar for breast health monitoring,'' \emph{IEEE
Trans. Biomed. Eng.}, vol. 63, no. 3, pp. 530-539, Mar. 2016.
\bibitem{Smith 2022}K. Smith, J. Bourqui, D. Garrett, S. Zarnke, M. Owjimehr, D. Deutscher,
T. Fung, and E. Fear, {}``Microwave imaging of the breast: Consistency
of measurements over time,'' \emph{IEEE J. Electromagn. RF Microw.
Med. Biol.}, vol. 6, no. 1, pp. 61-67, Mar. 2022.
\bibitem{Gubern-Merida 2015}A. Gubern-Merida, M. Kallenberg, R. M. Mann, R. Marti, and N. Karssemeijer,
{}``Breast segmentation and density estimation in breast MRI: A fully
automatic framework,'' \emph{IEEE J. Biomed. Health Inform}., vol.
19, no. 1, pp. 349-357, Jan. 2015.
\bibitem{Golnabi 2019}A. H. Golnabi, P. M. Meaney, S. D. Geimer, and K. D. Paulsen, {}``3-D
microwave tomography using the soft prior regularization technique:
Evaluation in anatomically realistic MRI-derived numerical breast
phantoms,'' \emph{IEEE Trans. Biomed. Eng.}, vol. 66, no. 9, pp.
2566-2575, Sep. 2019.
\bibitem{Abdollahi 2019}N. Abdollahi, D. Kurrant, P. Mojabi, M. Omer, E. Fear, and J. LoVetri,
{}``Incorporation of ultrasonic prior information for improving quantitative
microwave imaging of breast,'' \emph{IEEE J. Multiscale Multiphys.
Comput. Tech.}, vol. 4, pp. 98-110, Mar. 2019.
\bibitem{Mojabi 2019}P. Mojabi and J. LoVetri, {}``Experimental evaluation of composite
tissue-type ultrasound and microwave imaging,'' \emph{IEEE J. Multiscale
Multiphys. Comput. Tech.}, vol. 4, pp. 119-132, Mar. 2019.
\bibitem{Qin 2020}Y. Qin, T. Rodet, M. Lambert, and D. Lesselier, {}``Microwave breast
imaging with prior ultrasound information,'' \emph{IEEE Open J. Antennas
Propag.}, vol. 1, pp. 472-482, Aug. 2020.
\bibitem{Lim 2008}H. B. Lim, N. T. T. Nhung, E.-P. Li, and N. D. Thang, {}``Confocal
microwave imaging for breast cancer detection: Delay-multiply-and-sum
image reconstruction algorithm,'' \emph{IEEE Trans. Biomed. Eng.},
vol. 55, no. 6, pp. 1697-1704, June 2008.
\bibitem{Fear 2013}E. C. Fear, J. Bourqui, C. Curtis, D. Mew, B. Docktor, and C. Romano,
{}``Microwave breast imaging with a monostatic radar-based system:
A study of application to patients,'' \emph{IEEE Trans. Microw. Theory
Techn.}, vol. 61, no. 5, pp. 2119-2128, May 2013.
\bibitem{Gholipur 2018}T. Gholipur, M. Nakhkash, and M. Zoofaghari, {}``A linear synthetic
focusing method for microwave imaging of 2-D objects,'' \emph{IEEE
Trans. Microw. Theory Techn.}, pp. 5042-5050, Nov. 2018.
\bibitem{Mukherjee 2019}S. Mukherjee, L. Udpa, S. Udpa, E. J. Rothwell, and Y. Deng, {}``A
time reversal-based microwave imaging system for detection of breast
tumors,'' \emph{IEEE Trans. Microw. Theory Techn.}, vol. 67, no.
5, pp. 2062-2075, May 2019.
\bibitem{Reimer 2021}T. Reimer and S. Pistorius, {}``An optimization-based approach to
radar image reconstruction in breast microwave sensing,'' \emph{Sensors},
vol. 21, no. 24, pp. 8172, Dec. 2021.
\bibitem{Casu 2017}M. R. Casu, M. Vacca, J. A. Tobon, A. Pulimeno, I. Sarwar, R. Solimene,
and F. Vipiana, {}``A COTS-based microwave imaging system for breast-cancer
detection,'' \emph{IEEE Trans. Biomed. Circuits Syst.}, vol. 11,
no. 4, pp. 804-814, Aug. 2017.
\bibitem{Poplack 2007}S. P. Poplack, T. D. Tosteson, W. A. Wells, B. W. Pogue, P. M. Meaney,
A. Hartov, C. A. Kogel, S. K. Soho, J. J. Gibson, and K. D. Paulsen,
{}``Electromagnetic breast imaging: Results of a pilot study in women
with abnormal mammograms,'' \emph{Radiology}, vol. 243, no. 2, pp.
350-359, May 2007.
\bibitem{Meaney 2013}P. M. Meaney, P. A. Kaufman, L. S. Muffly, M. Click, S. P. Poplack,
W. A. Wells, G. N. Schwartz, R. M. di Florio-Alexander, T. D. Tosteson,
Z. Li, S. D. Geimer, M. W. Fanning, T. Zhou, N. R. Epstein, and K.
D. Paulsen, {}``Microwave imaging for neoadjuvant chemotherapy monitoring:
Initial clinical experience,'' \emph{Breast Cancer Research}, vol.
15, no. 2, Apr. 2013.
\bibitem{Miao 2017}Z. Miao and P. Kosmas, {}``Multiple-frequency BBIM-TwIST algorithm
for microwave breast imaging,'' \emph{IEEE Trans. Antennas Propag.},
vol. 65, no. 5, pp. 2507-2516, May 2017.
\bibitem{Neira 2017}L. M. Neira, B. D. Van Veen, and S. C. Hagness, {}``High-resolution
microwave breast imaging using a 3-D inverse scattering algorithm
with a variable-strength spatial prior constraint,'' \emph{IEEE Trans.
Antennas Propag.}, vol. 65, no. 11, pp. 6002-6014, Nov. 2017.
\bibitem{Tajik 2022}D. Tajik, R. Kazemivala, and N. K. Nikolova, {}``Real-time imaging
with simultaneous use of Born and Rytov approximations in quantitative
microwave holography,'' \emph{IEEE Trans. Microw. Theory Techn.},
vol. 70, no. 3, pp. 1896-1909, Mar. 2022.
\bibitem{Gao.2015}F. Gao, B. D. Van Veen, and S. C. Hagness, {}``Sensitivity of the
distorted Born iterative method to the initial guess in microwave
breast imaging,'' \emph{IEEE Trans. Antennas Propag.}, vol. 63, no.
8, pp. 3540-3547, Aug. 2015.
\bibitem{Hosseinzadegan 2021}S. Hosseinzadegan, A. Fhager, M. Persson, S. D. Geimer, and P. M.
Meaney, {}``Discrete dipole approximation-based microwave tomography
for fast breast cancer imaging,'' \emph{IEEE Trans. Microw. Theory
Techn.}, vol. 69, no. 5, pp. 2741-2752, May 2021.
\bibitem{Liu 2002}Q. H. Liu, Z. Q. Zhang, T. T. Wang, J. A. Bryan, G. A. Ybarra, L.
W. Nolte, and W. T. Joineset, {}``Active microwave imaging. I. 2-D
forward and inverse scattering methods,'' \emph{IEEE Trans. Microw.
Theory Techn.}, vol. 50, no. 1, pp. 123-133, Jan. 2002.
\bibitem{Rocca 2009}P. Rocca, M. Benedetti, M. Donelli, D. Franceschini, and A. Massa,
{}``Evolutionary optimization as applied to inverse scattering problems,''
\emph{Inverse Probl.}, vol. 25, no. 12, 123003, Dec. 2009.
\bibitem{Rocca 2011}P. Rocca, G. Oliveri, and A. Massa, {}``Differential evolution as
applied to electromagnetics,'' \emph{IEEE Antennas Propag. Mag.},
vol. 53, no. 1, pp. 38-49, Feb. 2011
\bibitem{Salucci 2022}M. Salucci, L. Poli, P. Rocca, and A. Massa, {}``Learned global optimization
for inverse scattering problems: Matching global search with computational
efficiency,'' \emph{IEEE Trans. Antennas Propag.}, vol. 70, no. 8,
pp. 6240-6255, Aug. 2022.
\bibitem{Pastorino 2007}M. Pastorino, {}``Stochastic optimization methods applied to microwave
imaging: A review,'' \emph{IEEE Trans. Antennas Propag.}, vol. 55,
no. 3, pp. 538-548, Mar. 2007.
\bibitem{Goudos 2021}S. Goudos, \emph{Emerging Evolutionary Algorithms for Antennas and
Wireless Communications}. SciTech/IET, 2021 (ISBN-13: 978-1-78561-552-8).
\bibitem{Salucci 2022b}M. Salucci, M. Arrebola, T. Shan, and M. Li, {}``Artificial intelligence:
New frontiers in real-time inverse scattering and electromagnetic
imaging,'' \emph{IEEE Trans. Antennas Propag.}, vol. 70, no. 8, pp.
6349-6364, Aug. 2022
\bibitem{Massa 2018}A. Massa, G. Oliveri, M. Salucci, N. Anselmi, and P. Rocca, {}``Learning-by-examples
techniques as applied to electromagnetics,'' \emph{J. Electromagn.
Waves Appl.}, vol. 32, no. 4, pp. 516-541, 2018.
\bibitem{Zhou 2021}Y. Zhou, Y. Zhong, Z. Wei, T. Yin, and X. Chen, {}``An improved deep
learning scheme for solving 2-D and 3-D inverse scattering problems,''
\emph{IEEE Trans. Antennas Propag.}, vol. 69, no. 5, pp. 2853-2863,
May 2021.
\bibitem{Friedman 2001}J. Friedman et \emph{al.} \emph{The elements of statistical learning}.
Springer series in statistics New York, 2001, vol. 1, no. 10.
\bibitem{Massa 2022}A. Massa and M. Salucci, {}``On the design of complex EM devices
and systems through the System-by-Design paradigm: A framework for
dealing with the computational complexity,'' \emph{IEEE Trans. Antennas
Propag.}, vol. 70, no. 2, pp. 1328-1343, Feb. 2022.
\bibitem{Burfeindt 2012}M. J. Burfeindt, T. J. Colgan, R. O. Mays, J. D. Shea, N. Behdad,
B. D. Van Veen, and S. C. Hagness, {}``MRI-derived 3-D-printed breast
phantom for microwave breast imaging validation,'' \emph{IEEE Antennas
Wireless Propag. Lett.}, vol. 11, pp. 1610-1613, Dec. 2012.
\bibitem{Solis-Nepote 2019}M. Solis-Nepote, T. Reimer, and S. Pistorius, {}``An air-operated
bistatic system for breast microwave radar imaging: Pre-clinical validation,''
\emph{2019 41st IEEE Annu. Int. Conf. Eng. Med. Biol. Soc. (EMBC),
2019}, pp. 1859-1862.
\bibitem{Forrester 2008}A. I. J. Forrester, A. Sobester, and A. J. Keane, \emph{Engineering
Design via Surrogate Modelling: A Practical Guide}. Hoboken, NJ, USA:
Wiley, 2008.
\bibitem{Lophaven 2002}S. N. Lophaven, H. B. Nielsen, and J. Sondergaard, {}``Dace: A MATLAB
Kriging toolbox,'' Dept. Informat. Math. Model., Tech. Univ. Denmark,
Lyngby, Denmark, Tech. Rep. IMM-TR-2002-12, 2002.
\bibitem{D'Orsi 2013}C. J. D'Orsi, E. A. Sickles, E. B. Mendelson, and E. A. Morris, \emph{ACR
BI-RADS Atlas: Breast Imaging Reporting and Data System}. American
College of Radiology, 2013.
\bibitem{O'Loughlin 2019}D. O'Loughlin, B. L. Oliveira, A. Santorelli, E. Porter, M. Glavin,
E. Jones, M. Popovic, and M. O'Halloran, {}``Sensitivity and specificity
estimation using patient-specific microwave imaging in diverse experimental
breast phantoms,'' \emph{IEEE Trans. Med. Imaging}, vol. 38, no.
1, pp. 303-311, Jan. 2019.
\bibitem{Martellosio 2017}A. Martellosio, M. Bellomi, M. Pasian, M. Bozzi, L. Perregrini, A.
Mazzanti, F. Svelto, P. E. Summers, G. Renne, and L. Preda, {}``Dielectric
properties characterization from 0.5 to 50 GHz of breast cancer tissues,''
\emph{IEEE Trans. Microw. Theory Techn.}, vol. 65, no. 3, pp. 998-1011,
Mar. 2017.
\bibitem{Zastrow 2008}E. Zastrow, S. K. Davis, M. Lazebnik, F. Kelcz, B. D. Van Veen, and
S. C. Hagness, {}``Development of anatomically realistic numerical
breast phantoms with accurate dielectric properties for modeling microwave
interactions with the human breast,'' \emph{IEEE Trans. Biomed. Eng.},
vol. 55, no. 12, pp. 2792-2800, Dec. 2008.
\bibitem{A-Info 2022}A-Info, {}``LB-20200 Broadband Horn Antenna Specifications''. Accessed:
Oct. 12, 2022. {[}Online{]}. Available: http://ainfoinc.com.cn/en/pro\_pdf/new\_products/antenna/Broadband\%20Horn\%20Antenna/tr\_LB-20200.pdf
\bibitem{Veronesi 1990}U. Veronesi, B. Salvadori, A. Luini, A. Banfi, R. Zucali, M. Del Vecchio,
R. Saccozzi, E. Beretta, P. Boracchi, G. Farante, V. Galimberti, G.
Mezzanotte, V. Sacchini, S. Tana, and E. Marubini, {}``Conservative
treatment of early breast cancer. Long-term results of 1232 cases
treated with quadrantectomy, axillary dissection, and radiotherapy,''
\emph{Ann. Surg.}, vol. 211, no. 3, pp. 250-259, Mar. 1990. 
\bibitem{Reimer 2022 a}T. Reimer and S. Pistorius, {}``The diagnostic performance of machine
learning in breast microwave sensing on an experimental dataset,''
\emph{IEEE J. Electromagn. RF Microw. Med. Biol.}, vol. 6, no. 1,
pp. 139-145, Mar. 2022.
\bibitem{Garud 2017}S. S. Garud, I. A. Karimi, and M. Kraft, ''Design of computer experiments:
a review,'' \emph{Comput. Chem. Eng.}, vol\emph{.} 106, pp. 71-95,
May 2017.
\bibitem{van den Berg 1997}P. van den Berg and R. Kleinman, {}``A contrast source inversion
method,'' \emph{Inverse Probl.}, vol. 13, no. 6, pp. 1607-1620, Jul.
1997.
\bibitem{Reimer 2020}T. Reimer, J. Krenkevich, and S. Pistorius, {}``An open-access experimental
dataset for breast microwave imaging,'' \emph{2020 14th European
Conf. Antennas Propag.} (EuCAP), 2020, pp. 1-5.
\end{thebibliography}
\end{document}